\documentclass{aa}

\usepackage[varg]{txfonts}
\usepackage{natbib}
\usepackage{graphicx}
\usepackage{times}
\usepackage{color}
\usepackage{hyperref}
\hypersetup{
  hidelinks,
  colorlinks=true,
  linkcolor=blue,
  filecolor=blue,
  urlcolor=blue,
  citecolor=blue,
}
\usepackage{bm}
\usepackage{enumitem}   
\usepackage{scalerel}
\usepackage{multirow}

\usepackage{tabularx}
    \newcolumntype{C}{>{\centering\arraybackslash}X}
    \newcolumntype{L}{>{\raggedright\arraybackslash}X}
    \newcolumntype{R}{>{\raggedleft\arraybackslash}X}
\usepackage{array}

\bibpunct{(}{)}{;}{a}{}{,}
\newcommand{\de}{$\delta$~Ori }

\newcommand{\korel}{{\tt KOREL} }
\newcommand{\korele}{{\tt KOREL}}

\newcommand{\pyt}{{\tt PYTERPOL} }
\newcommand{\pyte}{{\tt PYTERPOL}}

\newcommand{\ubv}{\hbox{$U\!B{}V$}}

\newcommand{\st}{$^\circ$}

\newcommand{\ks}{km\,s$^{-1}$}

\newcommand{\vsi}{$v \sin i$}
\newcommand{\tef}{$T_{\rm eff}$}
\newcommand{\lgg}{{\rm log}\,$g$}

\newcommand{\Rnom}{\ifmmode{\rm R}_\odot\else R$_{\odot}$\fi}
\newcommand{\Mnom}{\ifmmode{\rm M}_\odot\else M$_{\odot}$\fi}

\usepackage{siunitx,booktabs}
\usepackage[version=4]{mhchem}
\usepackage{collcell}

\definecolor{gnu_green}{RGB}{0,158,115}
\definecolor{gnu_navy}{RGB}{0,0,128}
\definecolor{gnu_orange}{RGB}{255,165,0}
\definecolor{pygrey}{RGB}{128, 128, 128}
\definecolor{pybrown}{RGB}{50, 0, 0}
\definecolor{pymediumblue}{RGB}{0, 0, 205}
\definecolor{forestgreen}{RGB}{34, 139, 34}
\definecolor{pypurple}{RGB}{128,0,128}
\definecolor{brown}{RGB}{165,42,42}
\definecolor{green}{RGB}{0,230,0}
\definecolor{orange}{RGB}{255,165,0}

\begin{document}

\title{
Spectrum of the secondary component and new orbital elements of the massive triple star $\delta$~Ori~A
\thanks{
Based on spectroscopic CCD observations with a coud\'e spectrograph attached
to the 2m reflector of the Astronomical Institute AS~\v{C}R at Ond\v{r}ejov,
archival Haute Provence and ESO La Silla spectra,
ground-based \ubv\ photometry from Hvar, and 
data collected by the BRITE Constellation satellite mission,
designed, built, launched, operated, and supported by the Austrian Research Promotion 
Agency (FFG), the University of Vienna, the Technical University of Graz, 
the University of Innsbruck, the Canadian Space Agency (CSA), the University  
of  Toronto  Institute  for  Aerospace  Studies  (UTIAS),  the  Foundation  for 
Polish Science \& Technology (FNiTP MNiSW), and National Science Centre (NCN).}
\fnmsep\thanks{Tables \ref{oes} and \ref{el_fer} are also available 
in electronic form at the CDS via anonymous ftp to 
cdsarc.cds.unistra.fr (130.79.128.5)
or via https://cdsarc.cds.unistra.fr/…+A/
}
}

\titlerunning{Spectrum of the secondary component of $\delta$~Ori~A}

\author{A.~Opli\v{s}tilov\'a\inst{1}\and
        P.~Mayer\thanks{Pavel Mayer passed away on November 7, 2018.}\inst{1}\and
        P.~Harmanec\inst{1}\and
        M.~Bro\v{z}\inst{1}\and
        A.~Pigulski\inst{2}\and
        H.~Bo\v{z}i\'c\inst{3}\and
        P.~Zasche\inst{1}\and
        M.~\v{S}lechta\inst{4}\and
        H.~Pablo\inst{5}\and
        P.~A.~Ko\l aczek-Szyma\'nski\inst{2}\and
        A.~F.~J.~Moffat\inst{5}\and
        C.~C.~Lovekin\inst{6}\and     
        G.~A.~Wade\inst{7}\and
        K.~Zwintz\inst{8}\and       
        A.~Popowicz\inst{9} \and
        W.~W.~Weiss\inst{10}
}

\offprints{A.O., \email{betsimsim@seznam.cz}}

\institute{
   Charles University, Faculty of Mathematics and Physics, Astronomical Institute, 
   V~Hole\v{s}ovi\v{c}k\'ach~2, CZ-180 00 Praha~8-Tr\'oja, Czech Republic
  \and
   Uniwersytet Wroc\l awski, Instytut Astronomiczny,
   Kopernika 11, 51-622 Wroc\l aw, Poland
  \and
   Hvar Observatory, Faculty of Geodesy, Zagreb University, 
   Ka\v{c}ic\'eva 26, 10000 Zagreb, Croatia
  \and
   Czech Academy of Sciences, Astronomical Institute,
   CZ-25165 Ond\v{r}ejov, Czech Republic
  \and
   Universit\'e de Montr\'eal, D\'epartement de physique,
   C.P.6128, Succursale center-Ville, Montr\'eal, Qu\'ebec, H3C 3J7, Canada
  \and
   Department of Physics, Mount Allison University, Sackville, NB, E4L1E6, Canada
  \and
   Department of Physics and Space Science, Royal Military College of Canada, 
   Kingston, Ontario K7K 7B4, Canada
  \and
   Institut f\"ur Astro- und Teilchenphysik, Universit\"at Innsbruck, Technikerstra{\ss}e 25, A-6020 Innsbruck, Austria
  \and
   Department of Electronics, Electrical Engineering and Microelectronics, 
   Silesian University of Technology, Akademicka 16, \mbox{44-100} Gliwice, Poland.
  \and
   University of Vienna, Institute for Astrophysics, Türkenschanzstraße 17, 1180 Vienna, Austria
}

\date{Received \today}

\abstract{
$\delta$~Orionis is the closest massive multiple stellar system and
one of the brightest members of the Orion OB association.
The primary (Aa1) is a unique evolved O star.
In this work,
we applied a two-step disentangling method
to a series of spectra in the blue region (430 to 450\,nm),
and we detected spectral lines of the secondary (Aa2).
For the first time, 
we were able to constrain the orbit of the tertiary (Ab)
-- to $55\,450\,{\rm d}$ or 152\,yr -- using 
variable $\gamma$ velocities and new speckle interferometric measurements, 
which have been published in the Washington Double Star Catalogue.
In addition, 
the Gaia DR3 parallax of the faint component (Ca+Cb)
constrains the distance of the system to
$(381\pm 8)\,{\rm pc}$,
which is just in the centre of the Orion OB1b association,
at $(382\pm 1)\,{\rm pc}$.
Consequently, we found that 
the component masses according to the three-body model are
$17.8$,
$8.5$, and
$8.7$\,\Mnom,
for Aa1, Aa2, and Ab, respectively,
with the uncertainties of the order of $1$\,\Mnom.
We used new photometry from the BRITE satellites
together with
astrometry,
radial velocities,
eclipse timings,
eclipse duration,
spectral line profiles, and
spectral energy distribution
to refine radiative properties.
The components, classified as
O9.5\,II +
B2\,V +
B0\,IV,
have radii of
$13.1$,
$4.1$, and
$12.0\,\Rnom$,
which means that $\delta$~Ori~A is a pre-mass-transfer object.
The frequency of $0.478\,$cycles per day,
known from the Fourier analysis of the residual light curve and X-ray observations,
was identified as the rotation frequency of the tertiary.
$\delta$~Ori could be related to other bright stars in Orion,
in particular, $\zeta$~Ori,
which has a similar architecture,
or $\varepsilon$~Ori,
which is a single supergiant, and
possibly a post-mass-transfer object.
}


\maketitle


\section{Introduction}

\begin{figure}
\centering
\includegraphics[width=0.5\textwidth]{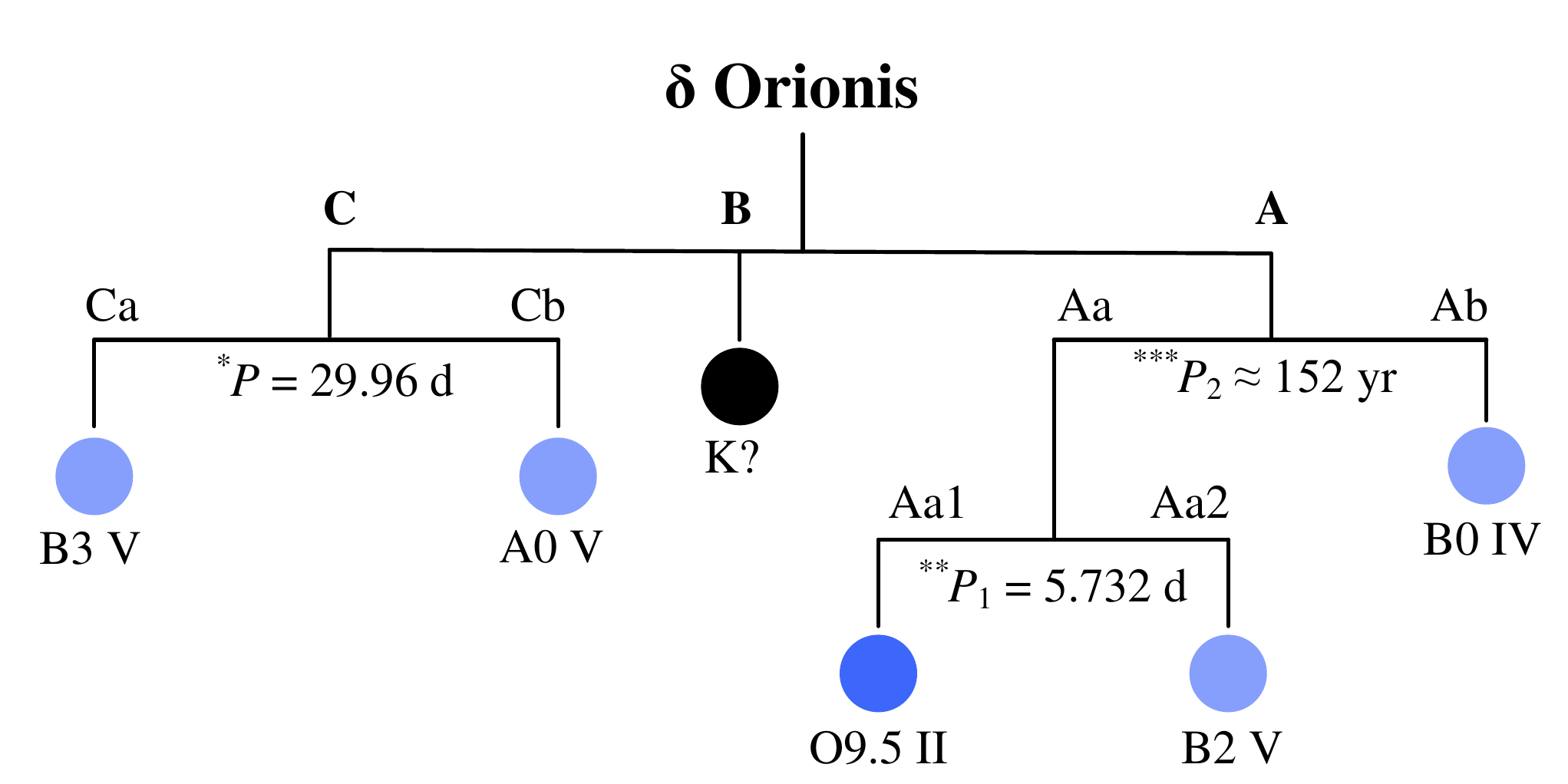}
\caption{Scheme of the multiple system $\delta$~Ori (HD~36486, ADS~4134, Mintaka).
Orbital periods were taken from
$^*$\,\citet{leone10},
$^{**}$\,\citet{mayer2010}, and
$^{***}$\,this paper.}
\label{schema}
\end{figure}

The bright star $\delta$~Ori 
(HR\,1852, HD\,36486, HIP\,25930, ADS\,4134)
is a multiple stellar system consisting of six components:
Aa1, Aa2, Ab, B, Ca, and Cb,
more specifically,
the eclipsing binary Aa1+Aa2,
the interferometric binary (Aa1+Aa2)+Ab,
the faint visual companion~B, and
the spectroscopic binary Ca+Cb
(see Fig.~\ref{schema}).
Their properties can be summarised as follows:

\begin{itemize}
\item Aa1+Aa2
($V_\mathrm{Aa1} = 2.55\,{\rm mag}$, 
$V_\mathrm{Aa2} \simeq 5.5\,{\rm mag}$)%
\footnote{$\alpha_{\mathrm{J2000}} = 5^\mathrm{h}\,32^\mathrm{m}\,0.398^\mathrm{s}$ and
$\delta_{\mathrm{J2000}} = -00^\circ\,17'\,56.69''$}
is a detached eclipsing binary with a negligible mass transfer, 
the orbital period $P_1 = 5.732436\,{\rm d}$ \citep{mayer2010},
a slightly eccentric orbit ($0.08$), and
apsidal motion ($1.45^\circ\,{\rm yr}^{-1}$) \citep{pablo2015}.

\item Ab ($V_\mathrm{Ab} = 3.7\,{\rm mag}$)
is a nearby companion, which forms an interferometric pair with Aa1+Aa2.
It was discovered by \citet{heintz80},
confirmed by speckle interferometry \citep{mason99} and
by {\em Hipparcos} astrometry \citep{esa97} of the (Aa1+Aa2)+Ab system.
The corresponding orbital period $P_2$ must be of the order of tens of thousands of days.

\item B ($V_\mathrm{B} \simeq 14\,{\rm mag}$) is 
a very faint distant companion%
\footnote{$\alpha_{\mathrm{J2000}} = 5^\mathrm{h}\,31^\mathrm{m}\,58.745^\mathrm{s}$ and
$\delta_{\mathrm{J2000}} = -00^\circ\,18'\,18.65''$}
that is probably not associated with the system. 
Assuming that the component is a main-sequence star,
its absolute magnitude of 6.7 mag corresponds to 
the spectral type K.

\item Ca+Cb ($V_\mathrm{Ca+Cb} = 6.85\,{\rm mag}$)
is another distant companion%
\footnote{$\alpha_{\mathrm{J2000}} = 5^\mathrm{h}\,32^\mathrm{m}\,00.406^\mathrm{s}$ and
$\delta_{\mathrm{J2000}} = -00^\circ\,17'\,04.38''$}
to (Aa1+Aa2)+Ab that is a spectroscopic, non-eclipsing binary 
with a period of $29.96\,{\rm d}$ and of spectral types B3\,V\,+\,A0\,V \citep{leone10}.
\end{itemize}

In the present paper, we focus on the triple sub-system $\delta$~Ori (Aa1+Aa2)+Ab,
with $V_{\mathrm{Aa1+Aa2+Ab}} = 2.223\,{\rm mag}$ 
(from the differential photometry at the Hvar Observatory),
$\alpha_\mathrm{J2000} = 5^\mathrm{h}\,32^\mathrm{m}\,00.400^\mathrm{s}$, and
$\delta_\mathrm{J2000} = -00^\circ\,17'\,56.74''$.
Hereinafter, the parameters corresponding to the inner orbit Aa1+Aa2 and
to the outer orbit (Aa1+Aa2)+Ab are denoted by indices 1 and 2, respectively.
The parameters of the components Aa1, Aa2, and Ab are denoted by indices 1, 2, and 3, respectively.

Many researchers have studied the system since the end of the 19th century.
For a detailed summary of the early investigation of $\delta$~Ori,
we refer readers to our earlier study of the system \citep{mayer2010}.
As far as studies of the 21st century are concerned, 
\citet{harvin2002} carried out a tomographic separation of 
the ultraviolet and optical spectra into two systems of spectral lines, 
interpreted them as the lines of the primary and secondary of the eclipsing subsystem, 
and concluded that the components have unexpectedly low masses ($m_1 = 11.2\,\Mnom$ and $m_2 = 5.6\,\Mnom$). 
However, \citet{mayer2010}, 
showed that the optical spectra are dominated by the spectral lines of 
the O9.5\,II primary (Aa1; \citealt{wal72}) 
and the similarly hot tertiary (Ab), and
that the system has normal masses for O and early-B stars \citep{Harmanec1988BAICz..39..329H}.
The previous solution of the light curves (LCs) led \citet{mayer2010} to the conclusion that 
the faint secondary (Aa2) contributes only a few percent to the total flux. 
Although they carried out disentangling of the spectra, 
they were unable to find its spectral lines convincingly, 
and could only rely on an indirect estimate of the mass ratio~$m_2/m_1$.

Five in-depth studies of \de were published in 2015 
(the first four are a series): \citet{corcoran2015} presented an overview of 
deep Chandra HETGS X-ray observations 
that covered nearly the entire binary (Aa1+Aa2) orbit. 
The observed X-ray emission was dominated by wind shocks from the primary (Aa1).
\citet{nichols2015} discussed the time-resolved and 
phase-resolved variability seen in the Chandra spectra. 
For the first time, they found phase-dependent variability in the X-ray emission line widths. 
They identified two periods in the total X-ray flux: $4.76 \pm 0.30$ and $2.04 \pm 0.05$ days.
\citet{pablo2015} carried out a detailed analysis of space-based photometry from Microvariability and Oscillations of STars (MOST) and 
simultaneously secured ground-based spectroscopy 
in the residuals of the orbital LC,
with periods ranging from 0.7 to 29~days.
\citet{shenar2015} carried out a multi-wavelength
non-local thermodynamic equilibrium (NLTE) analysis of spectra. 
The determined parameters led to a
O9.5\,II, B1\,V, and B0\,IV spectral classification for Aa1, Aa2, and Ab, respectively,
with evolved primary (Aa1) and tertiary (Ab) components.
They also found wind-driven mass loss by the Aa1 component at 
$4\cdot 10^{-7}\,\Mnom\,{\rm yr}^{-1}$.
\citet{richardson2015} used cross-correlation of the ultraviolet spectra from HST 
to obtain stellar parameter estimates for the primary, secondary, and the tertiary 
that was angularly resolved in the observations.

In this work, 
we continue our earlier analysis \citep{hec2013}, which
was devoted to the detection of very weak \ion{He}{i}~6678~\AA\ lines of the secondary
in the red spectral region.
Hereinafter, we focus on the blue spectral region.
This study was also motivated by the tentative evidence of the secondary
reported by \citet{richardson2015}, namely in the ultraviolet region,
observed by the Hubble Space Telescope (Space Telescope Imaging Spectrograph).

However, a robust detection of the secondary (Aa2) spectrum is still lacking.
Now, we have a larger set of spectra in the blue part of the optical spectrum and
procedures to successfully detect the secondary's spectrum.
Moreover, new Gaia DR3 parallax measurements have been published.
This provides the possibility to estimate the distance of bright stars,
saturated in the Gaia images, 
from the measured distances of their faint companions.
We also have new high-resolution astrometric measurements at our disposal,
which enables us to constrain the long-period orbit of (Aa1+Aa2)+Ab.

\section{Observational data}

In this section, only the spectroscopic and photometric data sets are described
as these data sets are new and fundamental to our analysis. 
Details of other data sets (astrometry, spectral energy distribution SED, speckle interferometry, etc.)
are described in the following sections 
(Sects. \ref{xitau_orbit}, \ref{sed}, \ref{3body}).

\subsection{Spectroscopy}\label{spec}

\begin{figure}
\centering
\includegraphics[width=9cm]{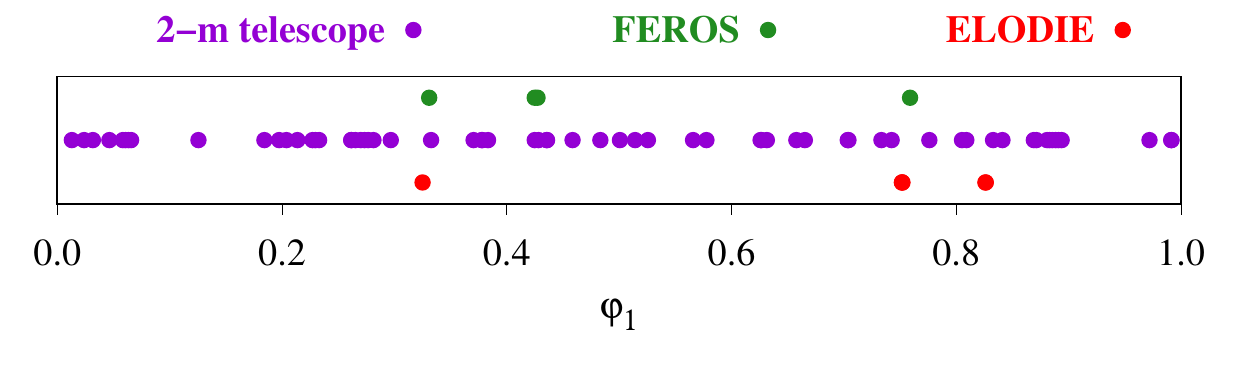}
\caption{Coverage of the orbital period $P_1$ by blue spectra.
Phases are determined with respect to time $T_0 = {\rm HJD}\,2454002.8735$
(time of periastron passage determined by \korele) for the eclipsing binary.}
\label{spectra_phase}
\end{figure}

We used digital spectra covering the blue spectral region 
secured at the coud\'e focus of the Ond\v{r}ejov 2m reflector \citep{Skoda2002PAICz..90.....S}. 
We supplemented these data sets with spectra from the public archives of 
the ELODIE echelle spectrograph \citep{moultaka2004} at the Haute Provence Observatory,
and the FEROS echelle spectrograph \citep{kaufer99} at the ESO La Silla Observatory.
The journal of the observations is presented in Table~\ref{jou} 
(see Table~\ref{oes} for more details). 
The coverage of orbital phase $\varphi_1$ is illustrated in Fig.~\ref{spectra_phase}. 
The short period $P_1$ of 5.732436\,d is well covered.
The mean signal-to-noise ratio (S/N) is 208.5 
(S/N values of individual spectra are given in Tables \ref{oes} and \ref{el_fer}), which  
was sufficient for spectral disentangling.
We normalised the spectra using polynomials of degree at least 4,
with the program \texttt{reSPEFO2}%
\footnote{\url{https://astro.troja.mff.cuni.cz/projects/respefo/}}
(written by Adam Harmanec).

\begin{table}
\caption[]{Journal of digital spectra covering the blue spectral region.
}
\label{jou}
\centering
\small
\begin{tabular*}{\hsize}{@{\hspace{0.2cm}}@{\extracolsep{\fill}}crcc}
\hline\hline\noalign{\smallskip}
Time interval & No. of  & Detector & Resolution \\
$\mathrm{[HJD-2400000]}$ & spectra &          & [\AA] \\
\noalign{\smallskip}\hline\noalign{\smallskip}
50031.68--50435.40 &  4 & ELODIE     & 0.05 \\
54136.58--54953.46 &  6 & FEROS      & 0.03 \\
55836.57--58405.57 & 65 & Site-5 CCD      & 0.13 \\
\noalign{\smallskip}\hline\noalign{\smallskip}
\end{tabular*}
\tablefoot{For more details, see Tables~\ref{oes} and \ref{el_fer}.}
\end{table}


\subsection{Photometry}

\begin{figure}
\centering
\includegraphics[width=9cm]{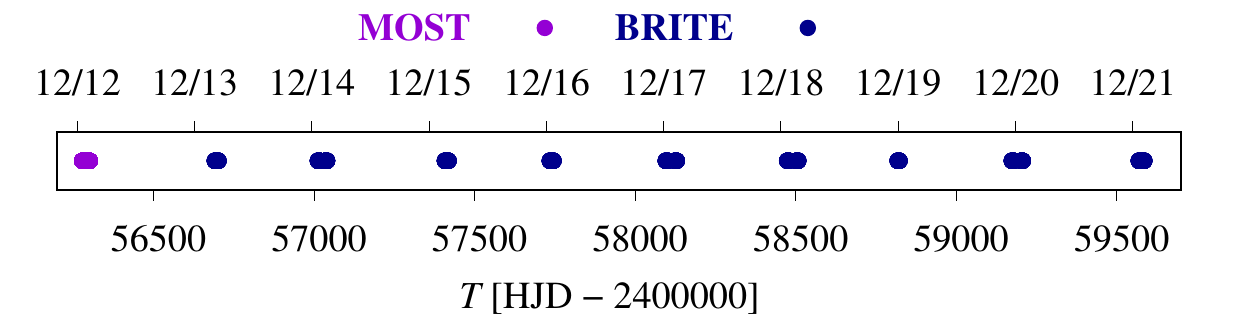}
\caption{Photometric data from MOST and BRITE
displayed with respect to time. 
The BRITE data covers six consecutive seasons between 2013 and 2021.}
\label{data-time}
\end{figure}

We used space-based photometric data from instruments on board
the BRITE (BRIght Target Explorer; \citealt{pablo2016}) and
the MOST (\citealt{carroll1998})
satellites and ground-based photometric data obtained at the Hvar Observatory
with the 0.65m telescope.
The time coverage is illustrated in Fig.~\ref{data-time}.
We did not use the saturated photometry from 
the Transiting Exoplanet Survey Satellite (TESS).

Each BRITE nanosatellite hosts a telescope,
which has a 3\,cm aperture.
The BTr, BHr, and UBr satellites are equipped with a red filter (with effective wavelength 620\,nm);
BAb and BLb have a blue filter (420\,nm).
We have eliminated instrumental effects from the raw BRITE data by 
removing outliers and worst orbits, 
and by decorrelations. 
For more information on BRITE data processing,
see \citet{Pigulski2018pas8.conf..175P}.

The MOST passband covers the visible range of the spectrum (350–750 nm).
The satellite performs high-precision optical photometry of single bright stars. 
It is equipped with a Maksutov telescope with an aperture of 15 cm and a custom broadband filter.
It can point with an error of less than 1 arcsec.
Other information can be found in Table~\ref{insa}.

The \de LC from MOST continuously covers 3 weeks of observation.
During calibration, we numerically shifted the measured magnitude
to the $V$ magnitude from the differential photometry at the Hvar Observatory.
Then, we constructed normal points by centring the errors on the satellite orbital periods
from Table~\ref{insa},
omitting the points with larger than the average uncertainty (0.5\,mmag). 

The Cassegrain 0.65m $f/11$ telescope
at the Hvar observatory
is equipped with a photoelectric detector \citep{Bozic1998HvaOB..22....1B}.
This telescope was constructed at the Ond\v{r}ejov Observatory
of the Czechoslovak Academy of Sciences and 
brought to the Hvar Observatory at the beginning of 1972.
A monitoring programme of bright variable stars has continued until today.
The Hvar all-sky photometry provides accurate UBVR magnitudes in the Johnson system.
For $\delta$~Ori~A, we used UBV differential magnitudes 
obtained between October 2006 and October 2008 and
UBVR between January 2019 and March 2021.

\begin{table}
\caption[]{Information on satellites.}\label{insa}
\centering
\small
\begin{tabular*}{\hsize}{@{\hspace{0.2cm}}@{\extracolsep{\fill}}lccc}
\hline\hline\noalign{\smallskip}
Satellite                   & Height  & Inclination & Period \\
                            & [km]    & [\st]       & [d] \\
\noalign{\smallskip}\hline\noalign{\smallskip}
MOST                      & 825--840 & 98.7        & 0.07042 \\
UBr (UniBRITE)            & 775--790 & 98.6        & 0.06972 \\
BAb (BRITE-Austria)       & 775--790 & 98.6        & 0.06972 \\
BLb (BRITE-Lem)           & 600--890 & 97.7        & 0.06917 \\
BTr (BRITE-Toronto)       & 620--643 & 97.9        & 0.06819 \\
BHr (BRITE-Heweliusz)     & 612--640 & 98.0        & 0.06743 \\
\noalign{\smallskip}\hline\noalign{\smallskip}
\end{tabular*}
\tablefoot{Sources \citet{pablo2016}, \citet{carroll1998}, \citet{webb2006}.}
\end{table}

\section{Parallax and distance of $\delta$~Ori}\label{parallax}

In Gaia DR3 \citep{Gaia2016A&A...595A...1G, Gaia2021A&A...649A...1G, Vallenari_2022},
the parallaxes of the faint components of bright stars in the Orion OB1 association were measured
(see Table~\ref{moduli}).
The parallax of $\delta$~Ori~Ca+Cb, $\pi = (2.6244\pm 0.0538)\,{\rm mas}$
implies a distance $d = (381\pm 8)\,{\rm pc}$ and 
a distance modulus $\mu = (7.90\pm 0.04)\,{\rm mag}$.
Hereinafter, we assume that the components (Aa1+Aa2)+Ab as well as Ca+Cb
are located at the same distance.
Statistically, they are located close to each other.
The number of stars brighter than Ca+Cb ($6.62\,{\rm mag}$) is limited,
there is only 15 of them within $7200''$.
Given the separation of $52''$,
the probability that stars are physically unrelated is low,
$p < 10^{-3}$.   

To the contrary, $\delta$~Ori~B, which is also a formal member of the multiple visual system ADS 4134,
is located at a substantially smaller distance (by almost 100\,pc).
It is therefore not physically related to $\delta$~Ori~A.
Either way, it is too faint (14\,mag) to affect our results.

\begin{figure}
\centering
\resizebox{\hsize}{!}{\includegraphics{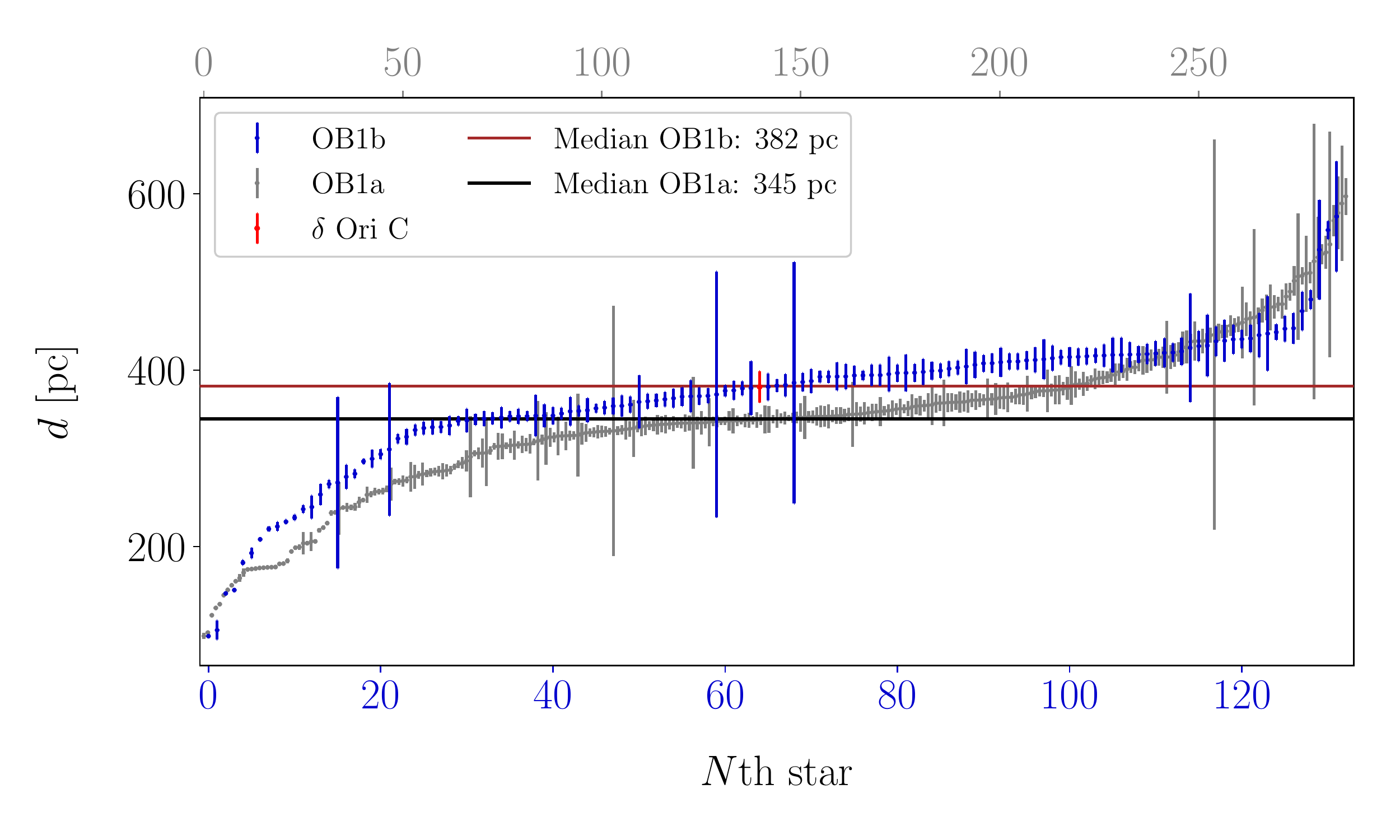}}
\caption{Sorted distances of objects from the Orion OB1b (\color{pymediumblue}blue\color{black}) 
and OB1a (\color{pygrey}grey\color{black}) associations \citep{Brown_1994A&A...289..101B}.
Parallax data were taken from the Gaia DR3 catalogue \citep{Vallenari_2022}.
The median values of distance for OB1b and OB1a are 
$(382\pm1)\,{\rm pc}$ (\color{pybrown}brown\color{black}) and $(345\pm 1\,{\rm pc})$ (black), respectively.
The distance of $\delta$~Ori~C (\color{red}red\color{black}), $(381\pm 8)\,{\rm pc}$, 
is very close to the median distance of the OB1b distribution.
}
\label{distance_Orion}
\end{figure}

The Orion OB1 stellar association is usually divided into four subgroups,
OB1a, OB1b, OB1c, and OB1d \citep{Brown_1994A&A...289..101B}.
The system $\delta$~Ori belongs to OB1b.
We used the distances of 131 members from the Gaia DR3 catalogue
and estimated the median distance to be
$(382\pm 1)\,{\rm pc}$,
using a cumulative distribution function 
that is sensitive to the local number density of stars
(see Fig.~\ref{distance_Orion}).
We obtained the same distance as the distance of $\delta$~Ori~Ca+Cb, 
within the respective intervals. 
We consider this to be an independent estimate for the $\delta$~Ori~A system
since massive stars are often located in the centre of the given association.

Other bright stars in the Orion belt are also located at very similar distances
(Table~\ref{moduli}).
For instance, the faint components of
$\zeta$~Ori~C,
$\sigma$~Ori~C, D, and~E
all have precise parallaxes.
Moreover, the single star $\varepsilon$~Ori has a similar spectroscopic distance modulus.
Again, this is an independent confirmation for the $\delta$~Ori system.

For comparison, the dispersion of distance in the radial direction (1$\sigma$) of the OB1b subgroup is only 15\,pc,
as seen in Fig.~\ref{distance_Orion},
while the angular dispersion (1$\sigma$) is about $0.5^\circ$,
which corresponds to 3\,pc, at the distance of 382\,pc.
In other words,
$1'$ corresponds to $0.11\,{\rm pc}$, and
$1''$ to $0.0018\,{\rm pc}$;
this is a range of separations for the faint components discussed above.

The age of the OB1b association is estimated between 4 and 5\,Myr \citep{Mauco_2018ApJ...859....1M}.%
\footnote{Some of the outliers seen in Fig.~\ref{distance_Orion}
might actually be former members of the OB1b association.
If they were ejected at the typical speed of $10\,{\rm km}\,{\rm s}^{-1}$,
they may travel $50\,{\rm pc}$ or $7.5^\circ$
in the radial or tangential directions.
The same is true for $\delta$~Ori~B.}
The OB1a subgroup (north-west) is older and at a smaller distance (by approximately 37\,pc),
while the OB1c and OB1d subgroups (north-east), including the Trapezium, 
are younger and at larger distances.

\begin{table*}
\caption[]{Information about bright stars and their companions in Orion.}
\label{moduli}
\small
\begin{center}
\begin{tabular*}{\textwidth}{@{\hspace{0.2cm}}@{\extracolsep{\fill}}l@{\ \ }l@{\ \ }l@{\ \ }l@{\ \ }l@{\ \ \ }l@{\ \ }l@{\ \ }l@{\ \ }l@{\ \ }l@{\ \ \ }l@{}}
\hline\hline\noalign{\smallskip}
HD        & Name             & $V$  & Spectral   & $A_V$  & $V_0$  & $M_V$ & $V_0\!-\!M_V$ & Gaia DR3 parallax      & Notes \\
          &                  & [mag]  & type     & [mag] & [mag]   &   &             &    [mas]        &         \\
\noalign{\smallskip}\hline\noalign{\smallskip}
36486     & $\delta$ Ori     & 2.22*& O9.5\,II  & 0.13 & 2.09 & $-5.81$  & 7.90 &                    & OB1b association, multiple         \\
37128     & $\varepsilon$ Ori& 1.68*& B0\,Ia    & 0.14 & 1.54 & $-6.25$**& 7.79 &                    & OB1b, single, variable 0.05\,mag        \\
37742     & $\zeta$ Ori      & 1.75*& O9.5\,Ib  & 0.17 & 1.58 & $-6.28$**& 7.92 &                    & OB1b, multiple                     \\
37468     & $\sigma$ Ori     & 3.82*& O9.5\,V   & 0.17 & 3.65 & $-4.14$  & 7.79 &                    & OB1b, multiple                     \\
37043     & $\iota$ Ori      & 2.75 & O8.5\,III & 0.09 & 2.66 & $-5.13$  & 7.79 &                    & OB1d (Trapezium), multiple         \\
\hline\noalign{\smallskip}
36486 Aa1 & $\delta$ Ori Aa1 & 2.55 & O9.5\,II  &      & 2.42 & $-5.7$** & 8.12 &                    & cf. this work                           \\
36486 Aa2 & $\delta$ Ori Aa2 & 5.5? & B2\,V     &      & 5.4? & $-2.5$?  &      &                    & $0.00052''$ from Aa1, \cite{shenar2015}          \\
36486 Ab  & $\delta$ Ori Ab  & 3.83 & B0\,IV    &      & 3.70 & $-4.0$***& 7.70 &                    & $0.32''$                                \\
36486 B   & $\delta$ Ori B   & 14.0 & K?        &      & 13.9 & $+6.6$   &      & $3.5002\pm 0.0119$ & $33''$, UCAC3 180-24383            \\
36485 Ca  & $\delta$ Ori Ca  & 6.62 & B3\,V     &      & 6.49 & $-1.6$***& 8.09 & $2.6244\pm 0.0538$ & $52''$, helium star, \cite{leone10}          \\
36485 Cb  & $\delta$ Ori Cb  & 9.8? & A0\,V     &      & 9.7? & $+1.8$?  &      &                    & $0.0012''$ from Ca                      \\
\\
37742 Aa  & $\zeta$ Ori Aa   & 2.1  & O9.5\,Ib  &      &      &          &      &                    & \cite{Hummel_2000ApJ...540L..91H}       \\
37742 Ab  & $\zeta$ Ori Ab   & 4.3  & B0.5\,IV  &      &      &          &      &                    & $0.042''$                               \\
37743     & $\zeta$ Ori B    & 4.0  & B0\,III   &      &      &          &      &                    & $2.4''$                                 \\
37742 C   & $\zeta$ Ori C    & 9.54 & A?        &      &      &          &      & $2.5876\pm 0.0387$ & $57''$                                  \\
\\
37468 Aa  & $\sigma$ Ori Aa  & 4.61 & O9.5\,V   &      &      &          &      &                    & \cite{Simon_2015ApJ...799..169S}        \\
37468 Ab  & $\sigma$ Ori Ab  & 5.20 & B0.5\,V   &      &      &          &      &                    & $0.00042''$                             \\
37468 B   & $\sigma$ Ori B   & 5.31 & B?        &      &      &          &      &                    & $0.25''$                                \\
37468 C   & $\sigma$ Ori C   & 8.79 & B0.5\,V   &      &      &          &      & $2.4720\pm 0.0292$ & $11''$                                  \\
37468 D   & $\sigma$ Ori D   & 6.62 & B2\,V     &      &      &          &      & $2.4744\pm 0.0621$ & $13''$                                  \\
37468 E   & $\sigma$ Ori E   & 6.66 & B2\,V     &      &      &          &      & $2.3077\pm 0.0646$ & $42''$, helium star                \\
\\
37043 Aa1 & $\iota$ Ori Aa1  & 2.8? & O8.5\,III &      &      &          &      &                    & \cite{Bagnuolo_2001ApJ...554..362B}     \\
37043 Aa2 & $\iota$ Ori Aa2  &      & B0.8\,III &      &      &          &      &                    & $0.0015''$, eccentric                   \\
37043 Ab  & $\iota$ Ori Ab   &      & B2\,IV    &      &      &          &      &                    & $0.15''$                                \\
37043 B   & $\iota$ Ori B    & 7.00 & B8\,III   &      &      &          &      & $2.7869\pm 0.0476$ & $11''$                                  \\
37043 C   & $\iota$ Ori C    & 9.76 & A0\,V     &      &      &          &      & $2.6057\pm 0.0241$ & $49''$, \cite{Parenago_1954TrSht..25....1P}, Brun 731 \\
\noalign{\smallskip}\hline
\end{tabular*}
\tablefoot{
* Hvar all-sky photometry;
** \citet{Martins_2005A&A...436.1049M};
*** \citet{Schmidt_1982BICDS..23....2S};
spectral types from \citet{Maiz_2019A&A...626A..20M,Burssens_2020A&A...639A..81B} and
absorption from \citet{Lallement_2019A&A...625A.135L}.
The visual magnitude including absorption is denoted by $V$;
the absorption, by $A_V$;
the visual magnitude without absorption, by $V_0$; and
the distance modulus computed from visual magnitude, by $M_V$.
}
\end{center}
\end{table*}

\section{Visual orbit of (Aa1+Aa2)+Ab}\label{xitau_orbit}

To determine the parameters of the long-period ($P_2$) orbit of the (Aa1+Aa2)+Ab system, 
we constructed a simplified two-body model in the \texttt{Xitau} program
\citep{Broz_2017ApJS..230...19B,Broz_2021A&A...653A..56B,Broz_2022A&A...657A..76B,Broz_2022A&A...666A..24B}\footnote{\url{http://sirrah.troja.mff.cuni.cz/~mira/xitau/}}.
We used astrometric data
from the Washington Double Star (WDS) Catalogue (\citealt{WDS2001AJ....122.3466M}).
If uncertainties were not available, we assumed the uncertainties of the separation $\rho$ and 
the position angle measured from $v$, +DE direction as follows:
$\sigma_\theta = 1.0^\circ$, 
$\sigma_\rho = 0.01\,\mathrm{mas}$, or
$\sigma_\theta = 0.2^\circ$, 
$\sigma_\rho = 0.005\,\mathrm{mas}$
for measurements before and after 2013, respectively. 
After removing 8 outliers due to poor resolution, incorrect plate scale, or calibration
(from 1879.12%
\footnote{
\cite{Niesten_1904AnOBN...8a...1N} reported micrométriques measures
by 15cm Merz refractor,
with two position angles $162^\circ$ and a note `en contact'.
However, it is unlikely that it corresponds to $\delta$~Ori~Ab,
because its separation at that epoch was only $0.11''$;
separations in other binaries were $1''$ or more.
This observation is not compatible with our model,
which indicates $\theta \doteq 127^\circ$.
}
, 1978.10, 1979.06, 1980.02, 1981.01, 1985.74, 1995.05, 1999.78), 
we used $N = 74$ data points (both $\rho$, $\theta$). 

Another data set incorporated into the model
included values of the systemic velocities $\gamma_1$ of $\delta$~Ori \citep{harvey87,harvin2002},
which vary between approximately $12$ and $23\,{\rm km}\,{\rm s}^{-1}$ 
(see Table~\ref{gamma_points}). 
This should correspond to the radial velocity of the (Aa1+Aa2) component. 
We did not take into account data points with possible systematic errors in $\gamma_1$, that is,
blending with Ab (1910, 1948),
low amplitude of RV curve $K_1$ (1951, 1969, 1981), and
different $\gamma_1$ for Aa2 (1987, 1997).
In some cases, also the RV of Ab was measured.

In total, we had $N = 88$ data points and
$M = 8$ free parameters, 
which means $N-M = 80$ degrees of freedom. 
The model resulted in the best-fit with $\chi^2 = 95$,
with contributions
$\chi_{\rm SKY}^2 = 60$ for astrometry and
$\chi_{\rm RV}^2 = 35$ for RVs.
Although $\chi^2 > N-M$, the fit is still acceptable.
The RV amplitude is in agreement,
as well as directly measured RV values of Ab,
which is lower than $\gamma_1$.

\begin{figure}
\centering
\leavevmode\kern-0.3cm\includegraphics[width=9.5cm]{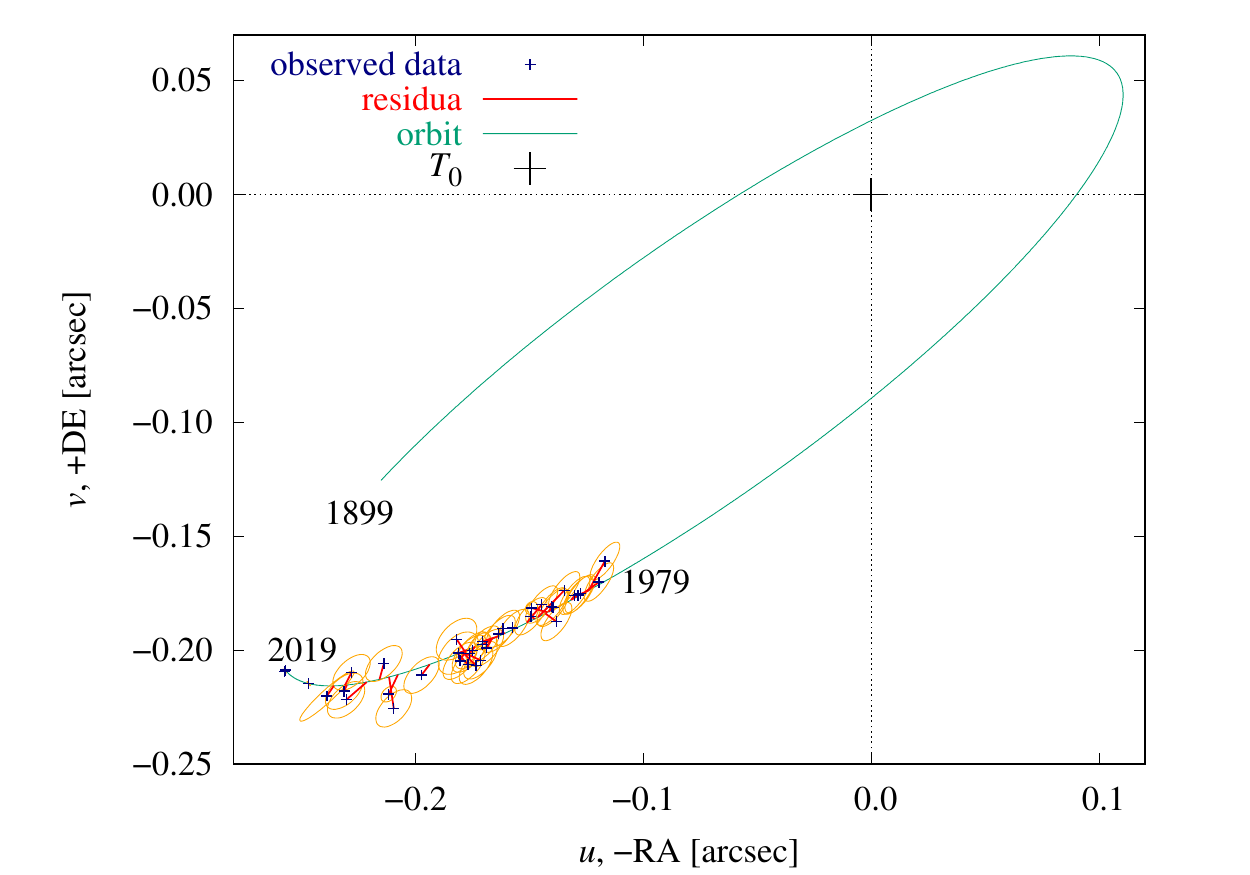}
\vskip-1cm
\includegraphics[width=9cm]{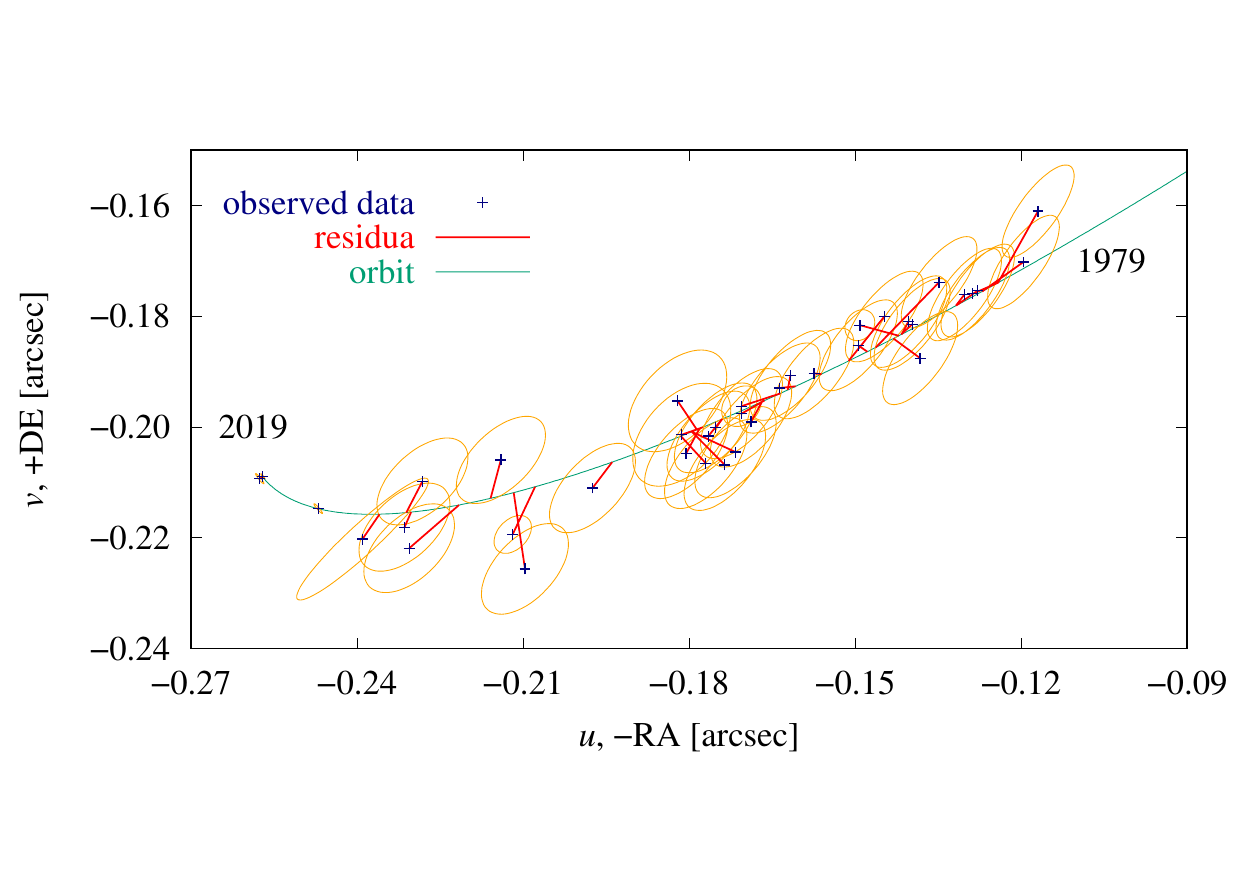}
\vskip-1cm
\caption{
Orbit of 2-body model.
Top: Orbit of (Aa1+Aa2)+Ab components in the ($u$-$v$) plane
(\color{gnu_green}green\color{black}),
calculated using the two-body model and the simplex method.
Observations are shown with \color{gnu_navy}blue\color{black}\ symbols
and uncertainty ellipses (\color{gnu_orange}orange\color{black}).
The residuals are plotted in \color{red}red\color{black};
the value of $\chi^2 = 95$.
The astrometry used for the fit is from the WDS.
The radial velocities of Aa1, from Table~\ref{gamma_points}
were also used for the fit,
extending the time span to 44\,000~days.
Bottom: Detail of the observed arc.
The most precise astrometric measurements from 2013 and 2019
constrain the orbital period $P_2$.
}
\label{orbit_delori}
\end{figure}

\begin{figure}
\centering
\resizebox{\hsize}{!}{\includegraphics{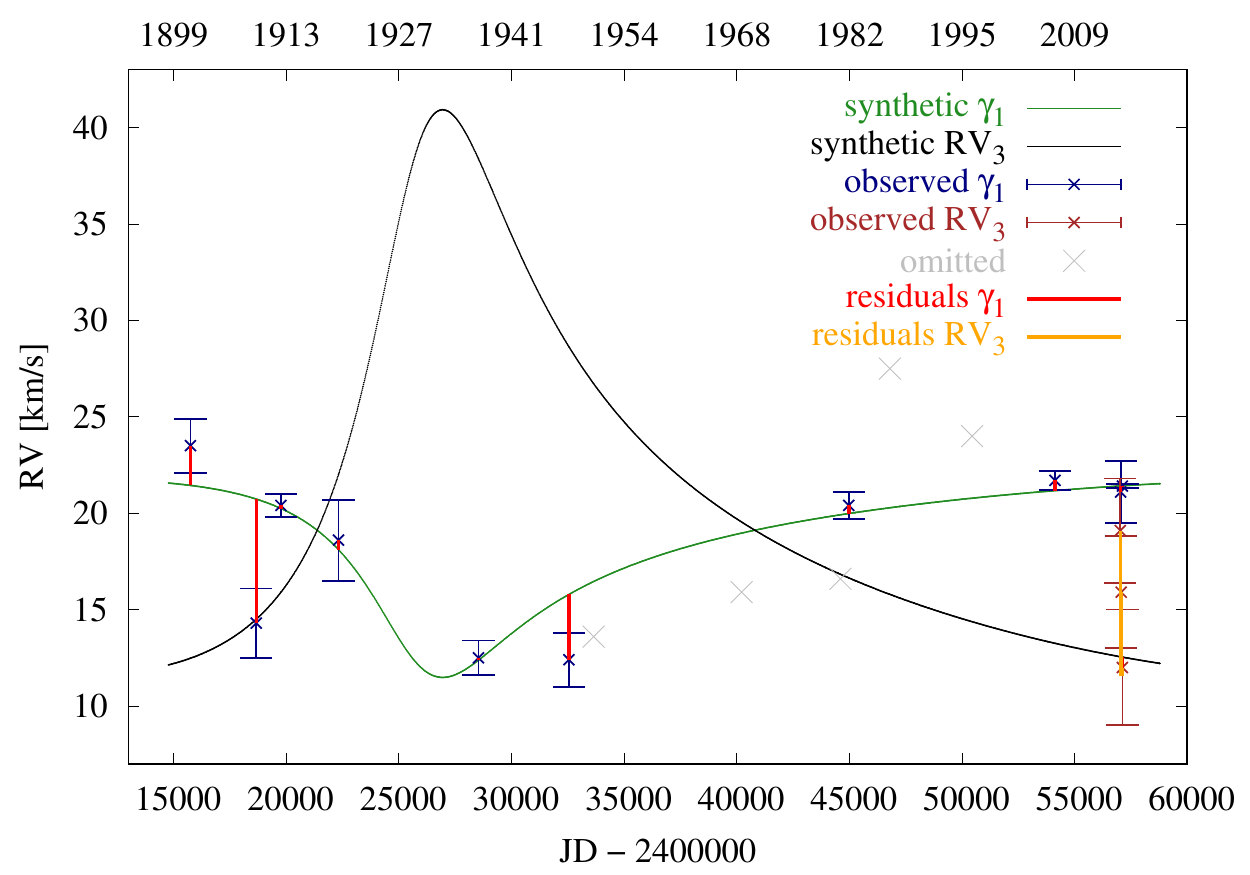}}

\caption{
Synthetic RVs of the Aa1+Aa2 (\color{forestgreen}green\color{black})
and Ab (black) components,
relative to the barycentre of the (Aa1+Aa2)+Ab system.
We used a simplified two-body model
and fitted data from Table~\ref{gamma_points},
plotted with \color{gnu_navy}blue\color{black}\ symbols.
The residuals are plotted in \color{red}red\color{black};
the value of $\chi^2 = 95$.
The last most precise point is from this work.
}
\label{orbit_delori_RV}
\end{figure}

The resulting parameters and parameters that were fixed 
are shown in Table~\ref{Xitau_param}.
We fixed the mass of the (Aa1+Aa2) components 
based on the phoebe2 model (Sect.~\ref{phoebe2})
and the distance $d$ based on the parallax (Sect.~\ref{parallax}).
The orbit is illustrated in Fig.~\ref{orbit_delori}. 
The fit of RVs is shown in Fig.~\ref{orbit_delori_RV}.

We estimated the uncertainties of parameters using $\chi^2$ mapping and verified the MCMC method.
According to the $\chi^2$ statistics,
a 1$\sigma$ level corresponds to $\chi^2 \doteq 101$.
Since the distance was fixed,
the uncertainties are relatively small ($1\%$ for the period, $10\%$ for the mass).
We determined the mass of the Ab component to be $11.0$\,\Mnom. 
Therefore, the total mass of the (Aa1+Aa2)+Ab system is around $37.5$\,\Mnom.

\paragraph{Mirror solution.}
We are aware of the existence of a mirror solution,
with the opposite sign of inclination~$i_2$.
It exhibits
higher total mass (up to $52\,\Mnom$),
higher eccentricity ($0.95$),
shorter period ($40000\,{\rm d}$),
closer periastron passage.
The RV curve of (Aa1+Aa2) component is also opposite,
with a `spike' due to the eccentricity.
According to our throughout testing, it always has a worse best-fit $\chi^2$,
especially the $\chi^2_{\rm rv}$ contribution.
Moreover, in the mirror solution, the RVs of Ab are larger than~$\gamma_1$,
which is incorrect.
A more complex model is needed to test other constraints
(see Sect.~\ref{3body}).

\begin{table}
\caption{Parameters of the orbit (Aa1+Aa2)+Ab, for the model with unreduced $\chi^2 = 95$.}
\label{Xitau_param}
\centering
\small
\sisetup{separate-uncertainty}
\begin{tabular*}{\hsize}{l @{\extracolsep{\fill}} r @{\,\( \pm \)\,} @{\extracolsep{0cm}}l  @{\extracolsep{\fill}} c}
\hline\hline\noalign{\smallskip}
{Parameter}                  & \multicolumn{2}{c}{Value}        & Unit \\
\noalign{\smallskip}\hline\noalign{\smallskip} 
$T_0$                        & 2458773.1886         & 0.1       & HJD   \\
$(m_1+m_2)$\,$^{\rm f}$      & 26.5                 & 2.0       &  \Mnom\\
$m_\mathrm{3}$               & 11.102               & 1.2       &  \Mnom\\
$P_2$                        & 53839                & 550       &  d    \\
$e_2$                        & 0.5886               & 0.016     &  1    \\
$i_2$                        & 104.710              & 0.4       &$^\circ$\\
$\Omega_2$                   & 122.416              & 0.5       &$^\circ$\\
$\varpi_2$                   & 258.944              & 2.0       &$^\circ$\\
$\lambda_2$                  & 134.307              & 1.0       &$^\circ$\\
$\gamma_1$                   & 19.1                 & 0.5       &${\rm km}\,{\rm s}^{-1}$\\
$d$\,$^{\rm f}$              & 382                  & 8         & pc\\
\noalign{\smallskip}\hline\noalign{\smallskip}
\end{tabular*}
\tablefoot{
$T_0$ denotes the time of periastron passage;
$(m_1+m_2)$, the mass of Aa1+Aa2 (the primary in this model);
$m_3$, the mass of the Ab component (the secondary in this model);
$P_2$, the orbital period of Aa and Ab;
$e_2$, the eccentricity;
$i_2$, the inclination;
$\Omega_2$, the longitude of the ascending node;
$\varpi_2$, the longitude of periastron; 
$\lambda_2$, the true longitude;
$\gamma_1$, the systemic velocity attributed to Aa1+Aa2; and
$d$, distance.
$^{\rm f}$~indicates the respective parameter was fixed.
}
\end{table}

\begin{table}
\sisetup{separate-uncertainty}
\small
\caption[]{
Observed values of $\gamma_1$ velocity
of the Aa1+Aa2 components.
It is variable due to the presence of the third (Ab) component. 
In some cases, also the RV of Ab was measured.
If the reference is not provided, the value is taken from the list of \citet{harvey87}
or \citet{mayer2010}, where more information about RV observations is provided
(in their App.~A).
}
\label{gamma_points}
\begin{flushleft}
\begin{tabular*}{\hsize}{l@{\extracolsep{\fill}}ccl}
\hline\hline\noalign{\smallskip}
$T$ (TDB) & $T_\mathrm{Bessel}$ & RV & Observatory or Ref. \\
{}[JD] & [BY] & [${\rm km}\,{\rm s}^{-1}$] & \\
\noalign{\smallskip}\hline\noalign{\smallskip}
2415793.70     & 1902.1174 & $23.5 \pm 1.4$  & Potsdam \\
2418981.13     & 1910.8443 & $14.3 \pm 1.8$  & Allegheny \\ 
2420024.23     & 1913.7002 & $20.4 \pm 0.6$  & Ann Arbor \\ 
2422391.71     & 1920.1822 & $18.6 \pm 2.1$  & Vienna \\ 
2428382.90     & 1936.5855 & $12.5 \pm 0.9$  & Yerkes \\ 
2432499.19     & 1947.8555 & $12.4 \pm 1.4$  & McDonald \\ 
2433656.03$^*$ & 1951.0229 & $13.6 \pm 1.9$  & Heidelberg \\ 
2440410.36$^*$ & 1969.5156 & $15.9 \pm 2.1$  & Kodaikanal \\ 
2444922.09$^*$ & 1981.8683 & $16.6 \pm 3.0$  & Kavalur \\ 
2445139.84     & 1982.4645 & $20.4 \pm 0.7$  & IUE \\ 
2446865.19$^*$ & 1987.1883 & $27.5 \pm 0.7$  & \citet{harvin2002} \\ 
2450535.00$^*$ & 1997.2359 & $24.0 \pm 3.0$  & \citet{harvin2002} \\ 
2454125.40     & 2007.0661 & $21.7 \pm 0.5$  & \citet{mayer2010} \\
2457040.9380   & 2015.0486 & $21.1 \pm 1.6$  & \citet{richardson2015} \\
2457121.0974   & 2015.2681 & $21.4 \pm 0.1$  & this work\\
\noalign{\smallskip}\hline\noalign{\smallskip}
Ab component: & & & \\
\noalign{\smallskip}
2457022.1757   & 2014.9972 & $19.1 \pm 2.7$ & \citet{richardson2015} \\
2457064.0764   & 2015.1120 & $15.9 \pm 2.9$ & \citet{richardson2015} \\
2457121.0974   & 2015.2681 & $12.0 \pm 3.0$ & this work \\
\noalign{\smallskip}\hline\noalign{\smallskip}
\end{tabular*}
\tablefoot{$^*$ denotes the data that were not included in the fit
due to systematic errors (see text).}
\end{flushleft}
\end{table}


\section{Spectral disentangling of residuals}\label{korel}

After obtaining the reliable value of the long-period ($P_2$) of $\delta$~Ori Aa and Ab, 
we searched for the secondary's lines in the spectra.
Our experience with the disentangling technique is that 
the result is often sensitive to the choice of initial values of the parameters.
This is understandable since the $\chi^2$ sum based on all data points
of all spectra is a complicated function of the orbital elements,
and it is easy to end up in a local minimum.

Moreover, the rotationally broadened spectral lines of the primary (Aa1)
and tertiary (Ab) blend with each other at all orbital phases
and altogether dominate the spectrum. 
Consequently, the contribution of the faint secondary (Aa2) to the $\chi^2$ sum 
is almost comparable to the noise.
The mass ratio $q_1$ of the Aa2 and Aa1 components is therefore poorly constrained. 
Nevertheless, the lines of the secondary can be detected in the residuals by
a procedure called a two-step disentangling.

To disentangle the spectra, we used the \korel program
developed by \citet{korel1, korel2, korel3, korel4}.
Rebinning of the spectra to a linear scale in RV,
needed as input for \korele, was carried out using 
the {\tt HEC35D} program written by P.H.
\footnote{The program {\tt HEC35D} with User's Manual is available at
\url{http://astro.troja.mff.cuni.cz/hec/HEC35}\,.}
The relative fluxes for the new wavelength points were derived using
the {\tt INTEP} program \citep{hill82}, 
which is a modification of the Hermite interpolation formula. 
It is possible to choose both boundaries of the desired spectral region, 
and the program smoothly interpolates the rebinned spectra 
with the continuum values of 1.0 at both edges.

To account for the variable quality of the individual spectra,
we measured their S/N ratios in the line-free regions and
assigned each spectrum a weight $w$ according to the formula:
\begin{equation}\label{wk}
w={(\mathrm{S/N})^2\over{(\mathrm{S/N})_{\rm mean}^2}},
\end{equation}
where $(\mathrm{S/N})_{\rm mean}$ denotes the root mean square of S/N ratio of all spectra.

Fitting with \korel was performed with the following
measurement equation:
\begin{equation}
{\cal F}I_i(y, t) = \sum_j s_{ij}\, {\cal F}I_{ij}\cdot\exp({\rm i}yv_{ij}) \quad\hbox{for }\forall i\,,\label{measurement_eq}
\end{equation}
where
$j$~denotes the component;
$i$,~the spectrum;
$I$,~the normalised intensity;
${\cal F}I$,~its Fourier transform;
$y$,~the Fourier-transformed quantity $x\equiv\ln\lambda/\lambda_0$, 
related to the wavelength~$\lambda$;
$s_{ij}$,~the intensity factors (constant or variable);
$v_{ij}$,~the radial velocity.

\paragraph{Two-step method.}

We used the period derived by \citet{mayer2010}, pericentre rate derived by \citet{pablo2015}, and parameters from 
Table~\ref{Xitau_param} as the initial conditions. 
With the method of spectral disentangling, 
we needed to detect a line spectrum of the secondary
in the blue spectral region 4275--4509\,\AA.

In the first step, we fitted the orbit of the close pair (Aa1+Aa2) and 
converged $q_1$, $e_1$, $\omega_1$, $K_1$, $T_0$, 
while $P_1$, $\dot{\omega}$ were fixed as well as the outer orbit (Sect.~\ref{xitau_orbit}). 
We set the same and constant intensity of lines of Aa1 and Ab ($s_1 = 1$, $s_3 = 1$) constant
and assigned zero intensity to Aa2 ($s_2 = 0$). 
The result of the first step was the disentangled spectra of only the primary (Aa1) 
and tertiary (Ab), 
and the residuals for all individual spectra after disentangling (O$-$C).

\begin{figure}
\centering
\includegraphics[width=0.49\textwidth]{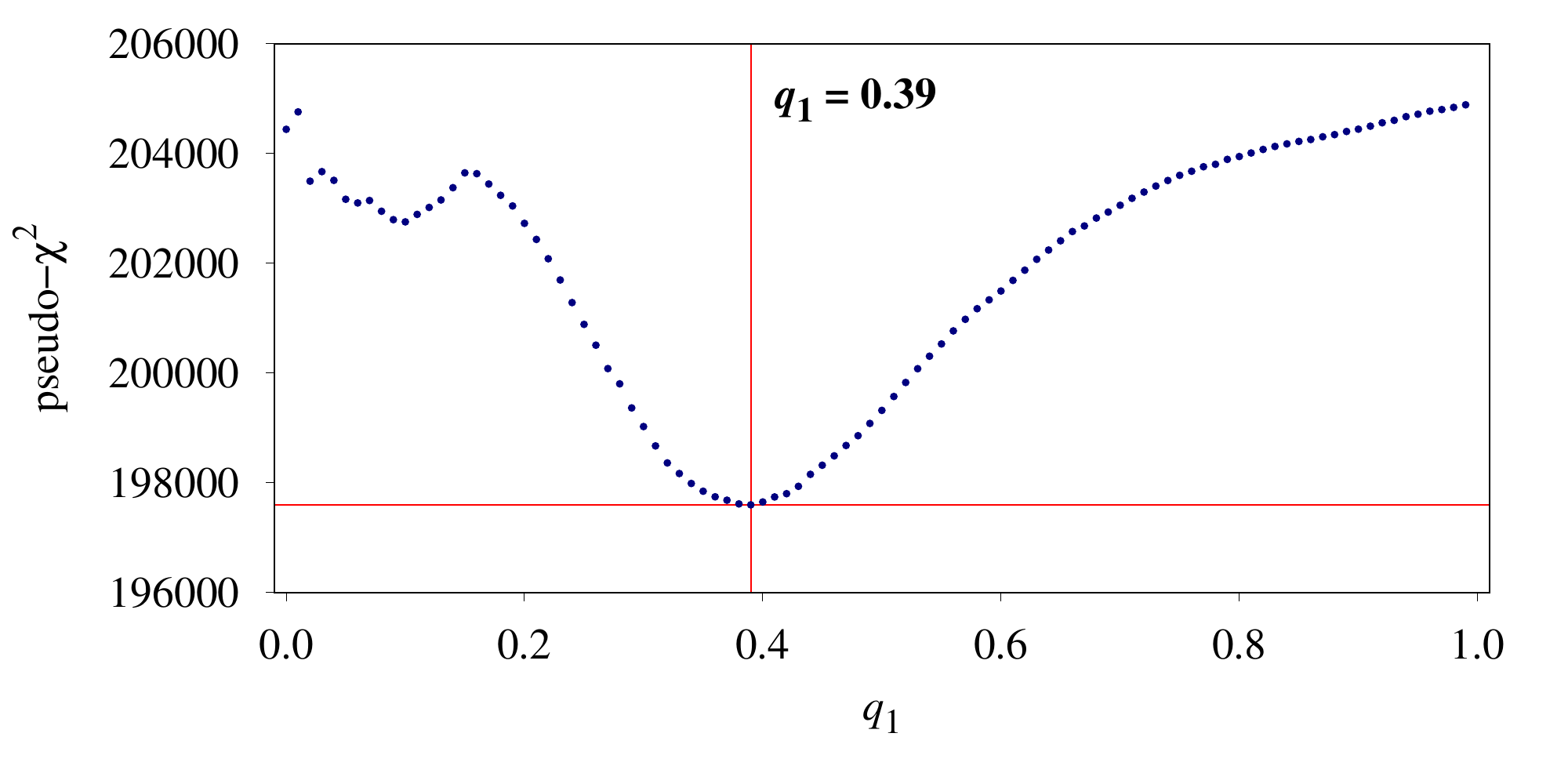}
\caption{
Plot of the pseudo-$\chi^2$ in Fourier space vs. the mass ratio $q_1 = m_2/m_1$.
This is related to the two-step disentangling method, to its second step,
when the signal of the secondary component (Aa2) was sought for in the residuals.
}
\label{chi2_q}
\end{figure}

In the second step, 
we added a value of 1.0 to the residuals and 
reran \korel on this `residual' data set. 
Now, the intensity factors of Aa1, Ab were zero 
($s_1 = 0$, $s_3 = 0$) and 
the intensity factor of Aa2 was constant $s_2 = 1$. 
We fitted the spectrum of the Aa2 component by converging the mass ratio $q$ and fixing 
$T_0$, $e_1$, $\omega_1$, $K_1$, $P_1$, and $\dot\omega_1$.
We successfully detected the desired spectrum of the Aa2 component. 
The determined parameters are summarised in Table~\ref{koreltab}. 
This method gave higher $e_1$, lower $q_1$, precise $K_1$ and $\omega_1$, 
which were well constrained. 
All disentangled spectra in these two steps have a flat continuum, not wavy. 
To confirm the detection, 
we created a pseudo-$\chi^2$ map (see Fig.~\ref{chi2_q}).%
\footnote{
We had some success using only a one-step method 
and setting Aa1, Aa2 to have fixed intensity factors $s_1 = 1$, $s_2 = 1$
and Ab to have free intensity factor $s_3$. 
Parameters $T_0$, $e_1$, $\omega_1$, $K_1$, $q_1$ were converged 
and $P_1$, $\dot{\omega_1}$ were fixed.  
This setting prevents fluctuations related to variable $s_1$, $s_2$ factors, 
making the model much more stable. 
Otherwise, the continuum is wavy when $s_1$ is free.
The results from this approach are
$e_1 = 0.0804$,
$\omega_1 = 153.9^\circ$, 
$K_1 = 108.3\,{\rm km}\,{\rm s}^{-1}$, 
$q_1 = 0.4893$};
uncertainties as in Table~\ref{koreltab}.

\paragraph{Three-step method.}
A more precise orbital solution can be obtained by using \korel in a sequence (three-step method). 
We started the process by fitting the primary and 
tertiary (Aa1+Ab) with variable intensities $s_1$, $s_3$. 
We fixed $P_1$ and $\dot{\omega_1}$ of 
the close orbit and converged $T_0$, $e_1$, $\omega_1$, $K_1$. 
The outer parameters of the orbit were fixed.

We continued by fitting Aa1, Ab with the constant $s$-factors and Aa2 with the variable one. 
Except for $q_1$, all parameters were fixed. 
Finally, we found the solution for all three components with 
constant $s$-factors, 
free $T_0$, $e_1$, $\omega_1$, $K_1$, $q_1$, and 
fixed $P_1$, $\dot{\omega_1}$.

The resulting $s_1(t)$ is variable with time and 
should correspond to the LC; however, 
the amplitude of the eclipses (without reflection) is too low (0.04\,mag) to be seen.
The resulting value of $q_1 = 0.4517$ is higher, compared to the two-step disentangling, while
$e_1 = 0.0761$ is close to that found from the LC, and 
$K_2 = 239.7\,{\rm km}\,{\rm s}^{-1}$.

In the three-step method, 
which we considered to be more reliable, 
we also computed the radial velocities of all three components
(see Tables~\ref{oes} and \ref{el_fer}).
We estimated the uncertainties as a standard deviation weighted by S/N.

\begin{table}
\caption[]{Solution to the disentangling of 75 blue spectra in \korele.
We prefer the solution from the three-step method ({\bf bold}).}
\label{koreltab}
\centering
\small
  \begin{tabular*}{\hsize}{l@{\extracolsep{\fill}}rrr}
\hline\hline\noalign{\smallskip}
Parameter & Two-step & {\bf Three-step} & $\sigma$ \\
\noalign{\smallskip}\hline\noalign{\smallskip}
$P_\mathrm{anom,1}\,[{\rm d}]$                  & \multicolumn{3}{c}{5.732821 \citep{mayer2010}}   \\ 
$\dot\omega_1\,[^\circ\,{\rm y}^{-1}]$ & \multicolumn{3}{c}{1.45 \citep{pablo2015}}\\

\noalign{\smallskip}\hline\noalign{\smallskip}
$T_0$                           & 2454002.8737 & 2454002.8735 & 0.02 \\
$e_1$                           & 0.0833       & 0.0761       & 0.01 \\    
$\omega_1\,[^\circ]$            & 153.6        & 153.5        & 3.0  \\     
$K_1\,[{\rm km}\,{\rm s}^{-1}]$ & 110.1        & 108.3        & 0.5  \\
$q_1$                           & 0.3996       & 0.4517       & 0.02 \\
$\sigma_{s_1}$                  & 0            & 0.1          & --   \\
$\sigma_{s_2}$                  & 0            & 0.4!         & --   \\
$\sigma_{s_3}$                  & 0            & 0.1          & --   \\
\noalign{\smallskip}\hline\noalign{\smallskip}
$K_2\,[{\rm km}\,{\rm s}^{-1}]$ & 295          & 239          & 10 \\
\noalign{\smallskip}\hline\noalign{\smallskip}
pseudo-$\chi^2$                 & 197588       & 172575       & --   \\
\noalign{\smallskip}\hline
\end{tabular*}
\tablefoot{
The anomalistic period $P_\mathrm{anom,1}$ and the pericentre rate $\dot\omega_1$ were fixed.
Free parameters were
the time of periastron passage $T_0$,
eccentricity $e_1$,
argument of periastron $\omega_1$,
semi-amplitude of the primary $K_1$,
mass ratio $q_1$,
standard deviations of the intensity factors for the primary, secondary, and tertiary,
$\sigma_{s_1}$,
$\sigma_{s_2}$,
$\sigma_{s_3}$, respectively.
The dependent parameter is the
semi-amplitude of the secondary $K_2$.
The models are quantified by pseudo-$\chi^2$ in Fourier space.}
\end{table}

\section{Atmospheric parameters of Aa1, Aa2, and Ab}\label{pyterpol}

We used the disentangled blue spectra to 
estimate the atmospheric parameters,
namely, \tef, \lgg, \vsi, and the relative luminosities
of the three components using
the program {\tt PYTERPOL} \citep{Nemravova_2016A&A...594A..55N}%
\footnote{\url{https://github.com/miroslavbroz/pyterpol3}}.
The program uses the simplex minimisation technique to fit 
the synthetic spectra to the observed ones.
As model spectra, OSTAR and BSTAR grids \citep{Lanz_2003ApJS..146..417L,Lanz_2007ApJS..169...83L} 
were used.

\begin{figure}
\centering
\resizebox{\hsize}{!}{\includegraphics{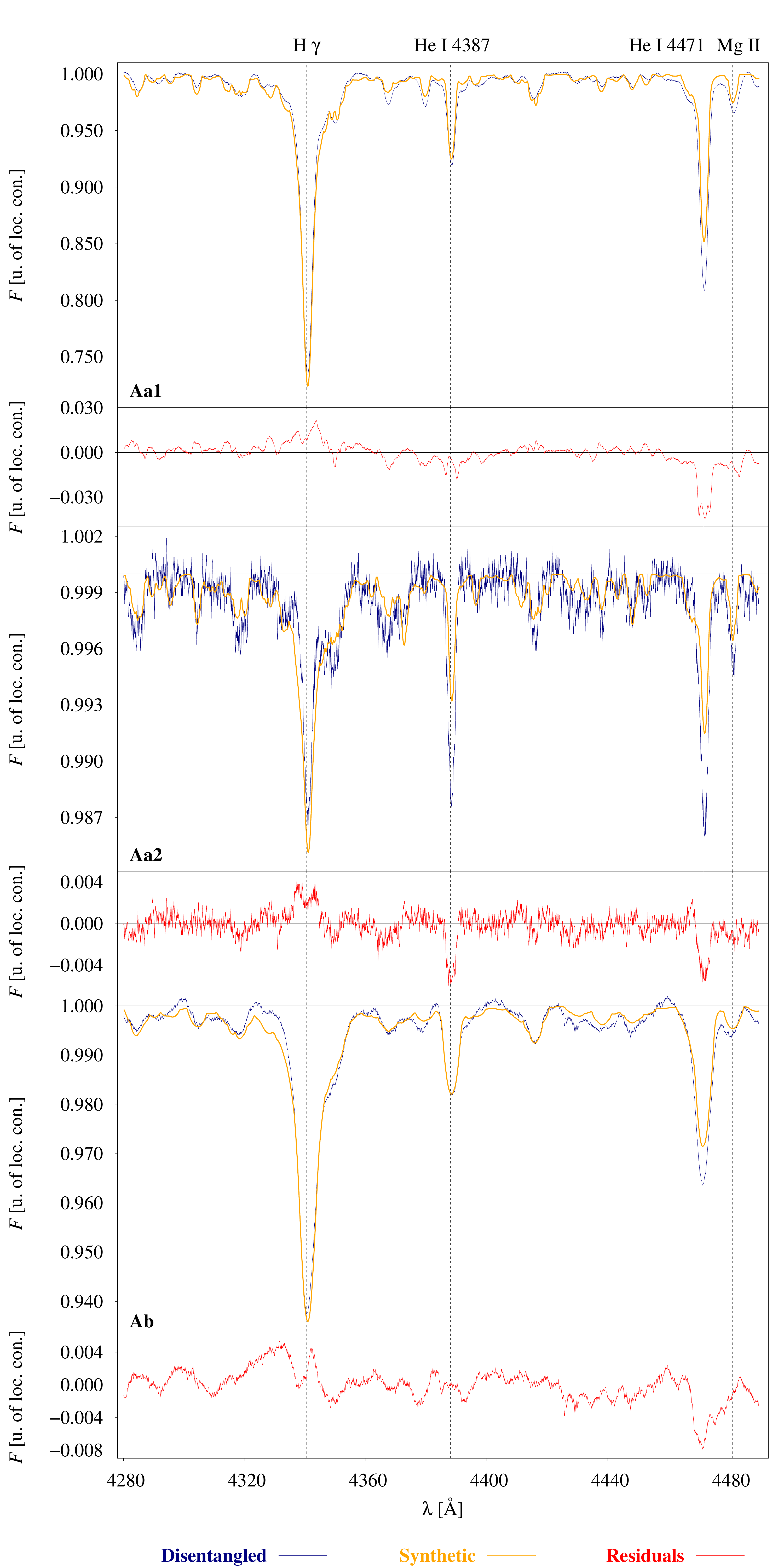}}
\caption{
Comparison of the disentangled spectra (\color{gnu_navy}blue\color{black})
of the Aa1, Aa2, and Ab components
with the best-fit synthetic spectra (\color{gnu_orange}orange\color{black}) found by \pyte.
The range from 428 to 449\,nm was used for disentangling.
The flux is normalised to the local continuum.
The small panels show the residuals (\color{red}red\color{black}).
The relative luminosities of the Aa1 and Ab components
significantly exceed that of the Aa2 component
(see also Table~\ref{pyttab}).
}
\label{profiles}
\end{figure}

The results are summarised in Table~\ref{pyttab}, where
the uncertainties were estimated from several independent trials.
The fitted spectral line profiles of all components are shown in Fig.~\ref{profiles}.
Most of the lines are fitted reasonably well, except for \ion{He}{I}\,4471. 
The value of $\log g$ was determined primarily from the H$\gamma$ wings. 
The metallicity $Z$ is not well constrained.

The sum of relative luminosities that were fitted independently 
(0.692 + 0.035 + 0.194 = 0.921) is close to 1, 
which is an independent verification of the correctness of the KOREL disentangling.
The effective temperatures agree with the spectral classifications of Aa1 (O9.5\,II) and Ab (B0\,IV), although
the effective temperature of the Aa2 component is significantly lower (around 25000\,K) than that of the other components,
corresponding to B1\,V, 
according to \cite{Harmanec1988BAICz..39..329H} calibrations.
The metallicities were fixed to the solar value since they are not well constrained by the blue spectra containing only one strong magnesium line.
The values from Table~\ref{pyttab} are the initial parameters for the phoebe2 model.

\begin{table}
\caption[]{Atmospheric parameters derived with \pyt from the blue spectral region
4271--4513\,\AA, with the mean resolution of 0.0144\,\AA.}
\label{pyttab}
\small
\centering
  \begin{tabular*}{\hsize}{l@{\extracolsep{\fill}}rrr}
\hline\hline\noalign{\smallskip}
Parameter       & Aa1 &  Aa2 &  Ab \\
\noalign{\smallskip}\hline\noalign{\smallskip}
\multirow{2}{*}{\tef[K]}         & 31400(1000)       & 25442(1500)                   & 30250(1000)    \\
                                 & \color{pygrey}29500\color{black}       & \color{pygrey}25600\color{black}                   & \color{pygrey}28400\color{black}    \\
& \\[-8pt]
\multirow{2}{*}{\lgg \,[cgs]}   & 3.55(5)       & 3.48(6)                  & 3.64(5)    \\
                                &  \color{pygrey}3.37\color{black}        & \color{pygrey}3.9\color{black}                   &  \color{pygrey}3.5\color{black}    \\
& \\[-8pt]
\multirow{2}{*}{\vsi \:[\ks]}    & 114(20)         &   89(15)                    & 216(25)        \\
                                & \color{pygrey}130\color{black}           & \color{pygrey}150\color{black}                   &  \color{pygrey}220\color{black}       \\
& \\[-8pt]
\multirow{2}{*}{$L_R$}           & 0.692(34)       & 0.035(15)                 & 0.194(10)   \\ 
                                & \color{pygrey}0.707\color{black}         & \color{pygrey}0.059\color{black}                 & \color{pygrey}0.234\color{black}   \\
\noalign{\smallskip}\hline\noalign{\smallskip}
$\chi^2_R$      & 2.562       & 1.769                 & 0.2384    \\
\noalign{\smallskip}\hline\noalign{\smallskip}
\end{tabular*}
\tablefoot{
\tef \ denotes the effective temperature;
\lgg, \ logarithm of surface gravity;
\vsi, \ projected rotational velocity;
$Z$, metallicity;
$L_R$,~relative luminosity;
$\chi^2_R$, the reduced value of $\chi^2$ 
(divided by the degrees of freedom), and
$^{\mathrm{f}}$ indicates the fixed parameter. 
For comparison with previous results,
the values from \citet{shenar2015} are shown in 
\color{pygrey}grey\color{black}.
In the case of Aa1 and Ab components, the results are usually in agreement 
within the uncertainties. 
More significant differences are for the Aa2 component; however, our values for the secondary are constrained by the disentangled spectra and mass ratio from \texttt{KOREL}.
The uncertainties of the parameters are given in concise form in brackets.
}

\end{table}

\section{Orbit of eclipsing binary Aa1+Aa2}\label{phoebe2}

For the eclipsing binary Aa1+Aa2, 
we solved the inverse problem 
using \texttt{PHOEBE2} \citep{Conroy2020ApJS..250...34C}, 
obtaining a more precise model than with \texttt{PHOEBE1} \citep{prsa2005} 
in our preliminary analysis \citep{Oplistilova_2020CoSka..50..585O}. 
The initial values of parameters for \texttt{PHOEBE2} were inferred from the analysis performed with \texttt{PHOEBE1}. 
We had three photometric data sets available for analysis, SMEI, MOST, and BRITE \citep{Oplistilova_2020CoSka..50..585O}, 
but we preferred to use only BRITE data to have a homogeneous
data set spanning nine seasons.
MOST was used in Sect. \ref{3body}.
We did not use SMEI data since they suffer from a contamination problem
because stellar images in SMEI image have angular sizes of the order of 1 degree.

\texttt{PHOEBE2}\footnote{\url{http://phoebe-project.org}}, a Python module, 
is software for modelling eclipsing binaries. 
To achieve the smallest possible discretisation error, 
the software uses a mesh of triangular elements. 
Each element of the mesh is assigned local properties (e.g.~temperature, intensity), 
and the eclipse algorithm determines which elements are visible, 
which are partially visible, and which are not visible at all. 
The total flux is obtained by integrating over all visible elements. 

We implemented a custom object-oriented Python wrapper 
to construct a model of the eclipsing binary
and combine different data types. 
Each model was quantified by the $\chi^2$ value. 
First, we fitted the stellar parameters using the simplex method \citep{Nelder_1965} or the supblex method \citep{Rowan_1990}. 
Second, we used the Markov chain Monte Carlo method (MCMC; \citealt{Robert2011, Tierney1994}),
which was originally invented by Slanis\l aw Ulam alongside with the atomic bomb. 
This method uses a sequence of random samples 
and provides a straightforward algorithm for the numerical estimation of parameters and their uncertainties. 
In other words, it describes the topology of the parameter space in the vicinity of the local/global minimum.
The MCMC method was run using the API OpenMP (application program interface), 
which allows our code to run in parallel on multiple CPUs. 

The MCMC method relies on Bayes' Theorem, 
which relates four probabilities as follows:
\begin{equation}
P(\bm{\Theta}_\mathrm{M}|\bm{D}) = \frac{P(\bm{D}|\bm{\Theta}_\mathrm{M})P(\bm{\Theta}_\mathrm{M})}{P(\bm{D})},
\end{equation}
where
$\bm{D}$~denotes the vector of data; $\bm{\Theta}_\mathrm{M}$,~the vector of parameters of our model;
$P(\bm{D})$, the probability of obtaining the data (normalisation); 
$P(\bm{\Theta}_\mathrm{M})$, the prior, a priori knowledge of parameters (we used uniform, uninformative priors); $P(\bm{D}|\bm{\Theta}_\mathrm{M})$, the likelihood function, which is equivalent to the forward model or $\chi^2$; and
$P(\bm{\Theta}_\mathrm{M}|\bm{D})$, the posterior distribution, 
which quantifies our belief in the parameters after combining our prior distribution with the current observations and 
normalising by the overall evidence.

The input data for the script are 
the RV curves of the primary and secondary, and the
LCs in the blue and red filters. 
The synthetic fluxes were normalised by two free parameters $S_\mathrm{red}$ and $S_\mathrm{blue}$ satisfying:
\begin{equation}\label{flux_norm}
F_{c,\mathrm{norm}} = S_\mathrm{c} \cdot \frac{F_{c,i}}{\mathrm{max}(F_{\mathrm{c},i})},
\end{equation}
where $c$ denotes the colour of the filter (blue or red), and $i$ is the point number. 

We set the algorithm parameters as follows: 
for the spatial discretisation, 
we used 1500 triangles covering the surface of the primary and 
500 triangles for the small secondary surface. 
As a sampler, we used \texttt{emcee} \citep{Foreman_2013PASP..125..306F} with 30 walkers and 2000 iterations. 
After some initial tests, we set the number of initial steps (burn-in) to 300. 
These are not taken into account as 
they are irrelevant and randomly distributed within the prior. 
The program ran on 30 CPUs. 

In our modelling, we fixed the orbital sidereal period to $5.732436\,{\rm d}$ following \citet{mayer2010}, 
the pericentre rate $\dot{\omega} = 1.45^\circ\,{\rm y}^{-1}$ \citep{pablo2015}, and 
in some models also the effective temperature of the primary $T_\mathrm{1} = 31\,000\,{\rm K}$, 
and the third light, additional to the components Aa1 and Aa2, $l_3 = 0.26685$ calculated from Table~\ref{pyttab} (Sect.~\ref{pyterpol}). 

We used the following parameters in our model:
\begin{itemize}
    \item \textit{atmosphere}, black-body (approximation),
    \item \textit{limb darkening}, linear, 
    \item \textit{limb darkening coefficients}, interpolated based on \citet{vanHamme_1993AJ....106.2096V},
    \item \textit{gravity brightening}, 1.0 (corresponding to the $\beta$ coefficient for gravity darkening corrections),
    \item \textit{reflection and heating fraction}, 1.0,
    \item \textit{distortion method}, Roche,
    \item \textit{irradiation method}, \citet{Wilson1990ApJ...356..613W}, Wilson's original reflection effect scheme incorporates all irradiation effects, including reflection and redistribution, 
    \item \textit{radial velocity method}, flux-weighted (i.e. radial velocities are determined by the radial velocity of each element of visible surface, weighted by its intensity). Consequently, the RV curve includes the Rossiter-McLaughlin effect.
\end{itemize}
We did not take into account either the effects of light travel time or gravitational redshift. 
This setting of the phoebe2 model is used for all models in Sects. \ref{m2016} and \ref{m_all}.


\subsection{Model for season 2016}\label{m2016}

We had the BRITE LCs from 9 seasons at our disposal (Table~\ref{seasons}).
First, we selected well-covered season 2016 and 
fitted several models with some parameters free or fixed,
namely the effective temperature of the primary and 
the third light (for both blue and red filters).
The results are presented in Table~\ref{phoebe2_result}.
We prefer the model with the fixed effective temperature of the primary,
which also has the lowest value of $\chi^2$.
The data, the model, and the residuals are shown in Fig.~\ref{final_fixT}.

Then, we used the MCMC method to estimate the uncertainties of the parameters.
Figs.~\ref{corner_fixT} and~\ref{walkers_fixT} show the corner plot and the paths of walkers.
In particular, masses $M_\mathrm{i}$ and radii $R_\mathrm{i}$ show strong positive correlations.
In contrast, the inclination $i_1$ and $R_\mathrm{i}$ show negative correlations due to geometrical reasons.
In binaries, the sum of masses is inversely proportional to the third power of $\sin i$; 
thus, $i_1$ and $m_\mathrm{i}$ show negative correlations.
The value of the systemic velocity~$\gamma$ is a little problematic
as the value of $21.96\,{\rm km}\,{\rm s}^{-1}$ was assumed and subtracted, 
then our model drifted to about $-2.5\,{\rm km}\,{\rm s}^{-1}$,
so that the resulting value is $18.5\,{\rm km}\,{\rm s}^{-1}$.

\begin{figure}
\centering
\includegraphics[width=9cm]{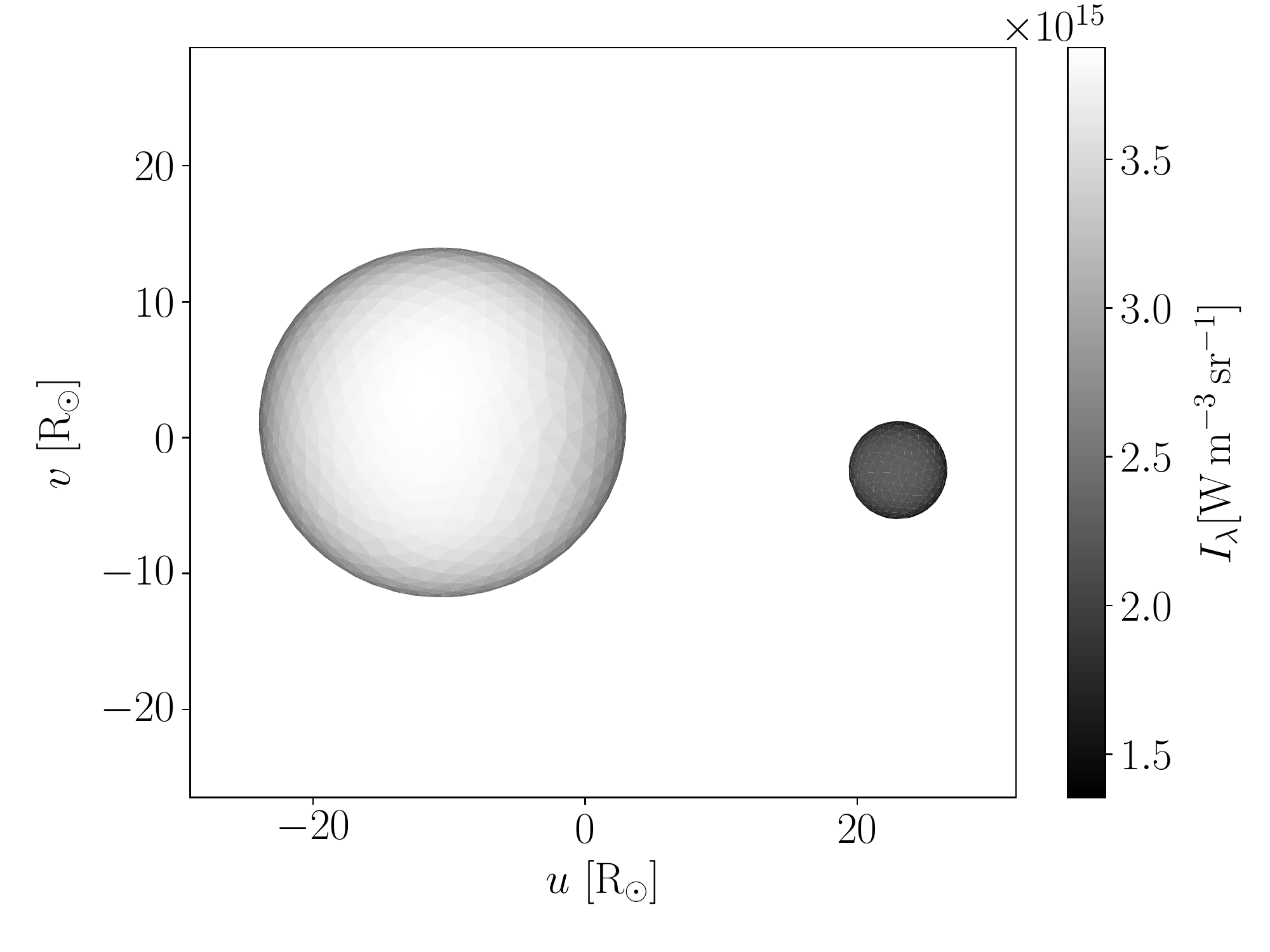}
\caption{
Mesh plot of $\delta$~Ori Aa1+Aa2 binary from the phoebe2 model (with $\chi^2 = 604$).
This is a $(u, v)$ plane projection with scale in \Rnom, at phase $\varphi = 0.75$.
The grey scale corresponds to the monochromatic intensity $I_\lambda\,[{\rm W}\,{\rm m}^{-3}\,{\rm sr}^{-1}]$
for the effective wavelength of the BRITE blue filter (420\,{\rm nm}).
}
\label{Fig8}
\end{figure}

\begin{figure*}
\centering
\resizebox{\hsize}{!}{\includegraphics{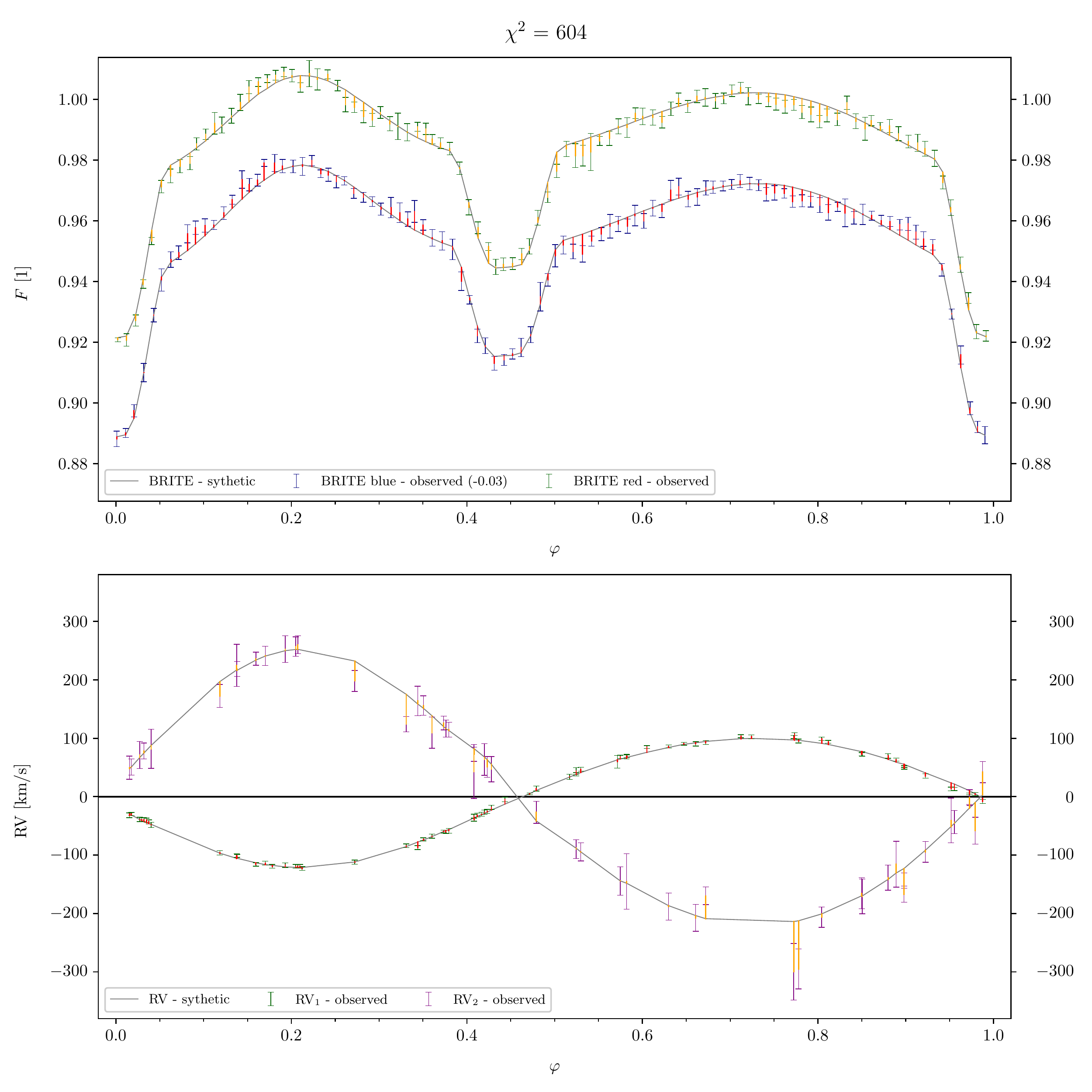}}
\caption{
Comparison of observations and the phoebe2 model of $\delta$~Ori with $\chi^2 = 604$.
The values of the effective temperature of the primary~$T_1$ and the third light~$l_3$ were fixed.
The upper panel shows the phased LCs in the blue and red BRITE filters.
The lower panel shows the RV curves for the primary Aa1 (\color{forestgreen} green\color{black}) and the secondary Aa2 (\color{pypurple} purple\color{black}).
The grey points correspond to our model, the red lines to the residuals, or contributions to~$\chi^2$.
}
\label{final_fixT}
\end{figure*}

\begin{figure*}
\centering
\resizebox{\hsize}{!}{\includegraphics{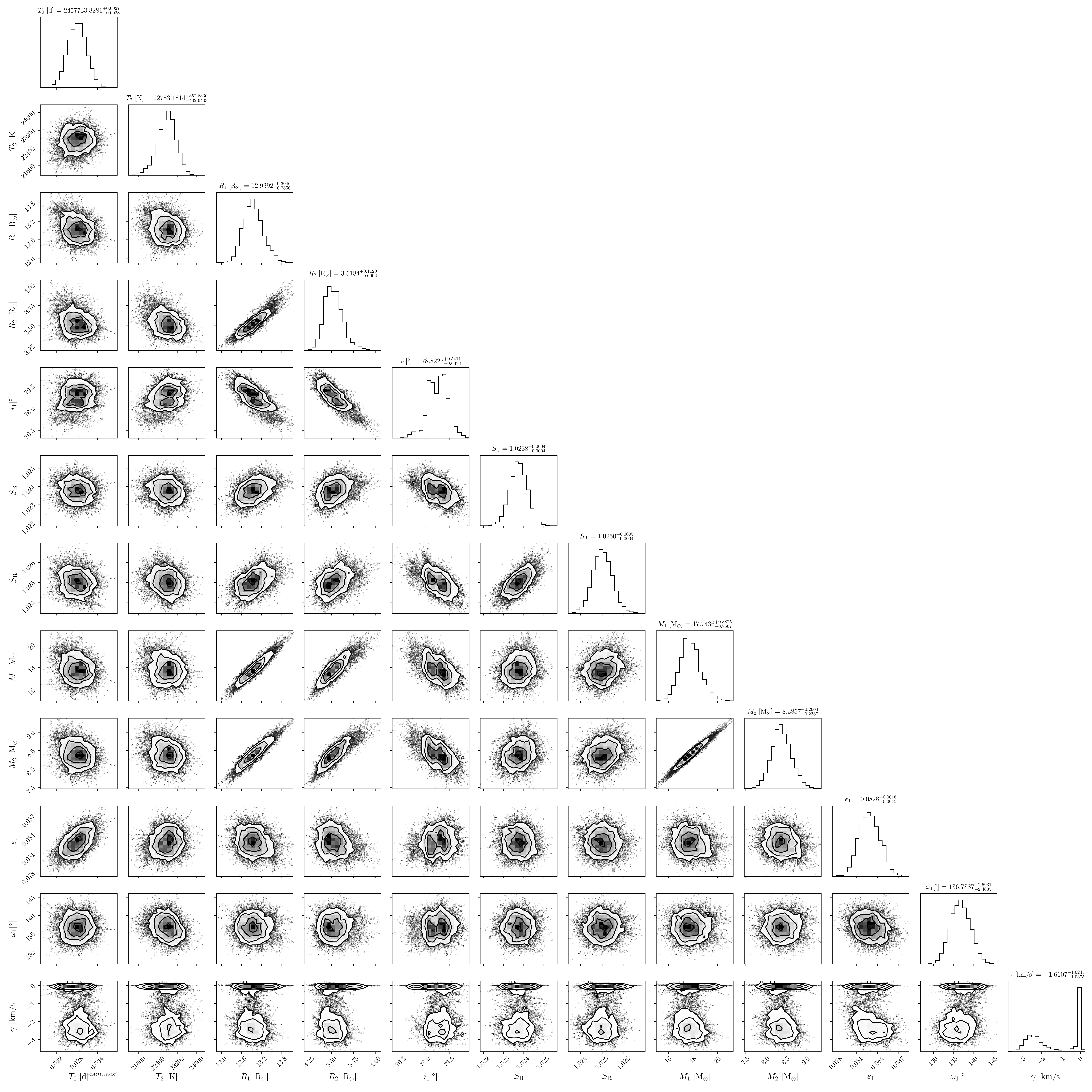}}
\caption{
Corner plot (a full covariance matrix) for the $\delta$~Ori model,
as derived by the MCMC analysis.
The model is the same as in Fig.~\ref{final_fixT}.
Each diagonal panel shows a 1-D histogram (posterior distribution) for one parameter
(explained in Table~\ref{phoebe2_result}).
Each sub-diagonal panel shows a 2-D histogram,
the isolines corresponding to the confidence intervals,
and the correlations between parameters.
}
\label{corner_fixT}
\end{figure*}

\begin{figure*}
\centering
\resizebox{\hsize}{!}{\includegraphics{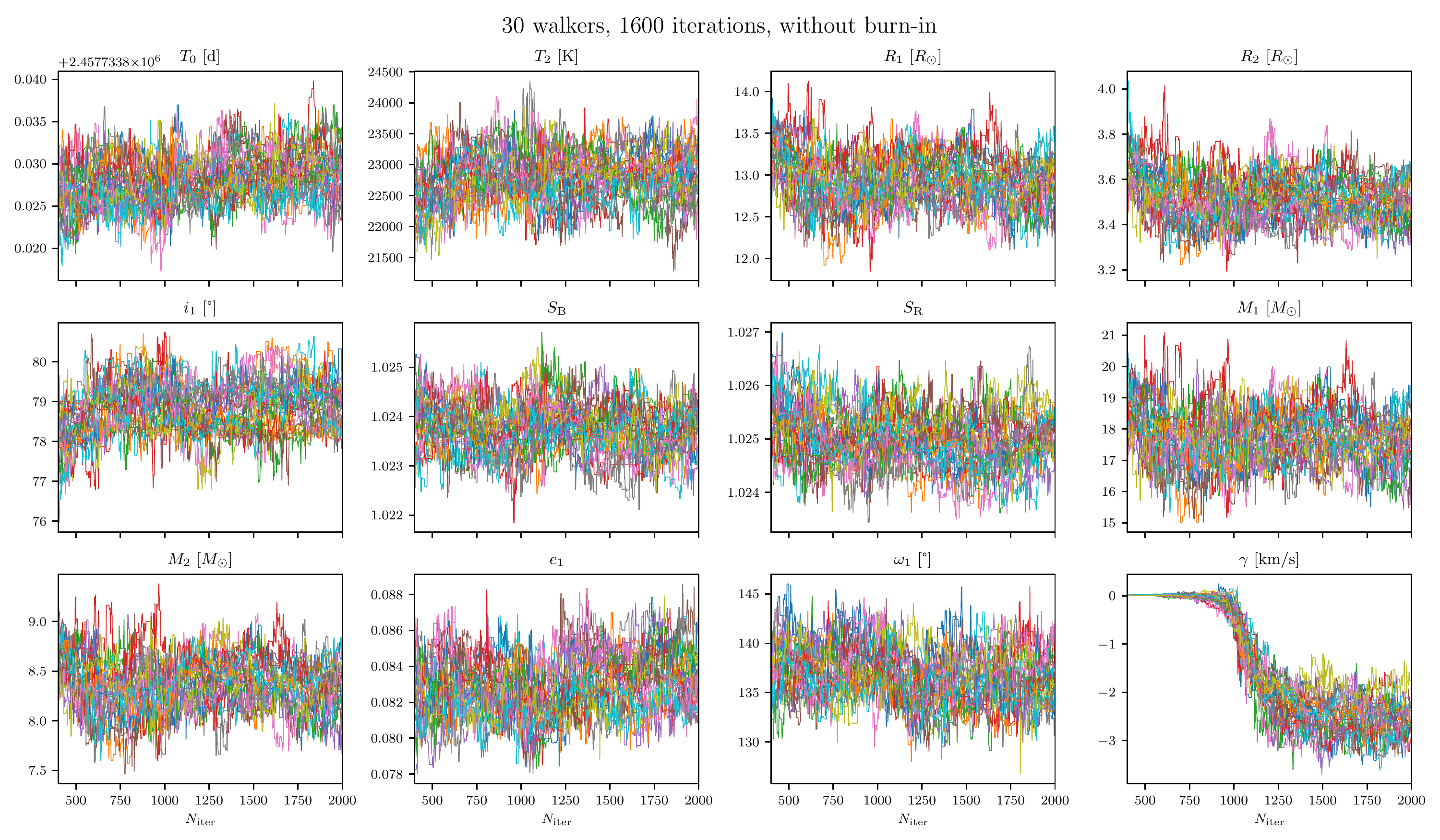}}
\caption{
Parameter values versus iterations during the MCMC analysis (as in Fig.~\ref{corner_fixT}).
We used 30 walkers (one walker corresponds to one colour) for the computation and 
300 steps for the burn-in.
For most parameters, the distribution of walkers is already stationary.
The systemic velocity~$\gamma$ is stationary only after 1000 iterations.
The value of $21.96\,{\rm km}\,{\rm s}^{-1}$ should be added to $\gamma$.
}
\label{walkers_fixT}
\end{figure*}

The detached binary system Aa1+Aa2 is shown in Fig.~\ref{Fig8}.
In addition, we derived several parameters from the nominal phoebe2 model
($\chi^2 = 604$); see Table~\ref{derived}. 
We estimated the synthetic apparent brightness of $\delta$~Ori~A as follows.
The passband flux in Johnson V at the observer location is (in $[{\rm W}\,{\rm m^{-2}}]$):
\begin{equation}
\Phi_{\mathrm{V}} = \Delta_{\rm eff}\kern.3pt\omega\!\sum_k I_{\lambda k}\kern.2pt S_{\!k\,}\mu_k\,\eta_k\,,
\end{equation}
where
$\Delta_{\rm eff}\,[{\rm m}^{-1}]$ stands for the effective wavelength range;
$\omega = 1\,{\rm m}^2/d^2\,[{\rm sr}^{-1}]$, for~the solid angle;
$d$, for~the distance of the system;
$\sum_k$, for~the summation over the triangular elements (grid);
$I_\lambda\,[{\rm W}\,{\rm m}^{-3}\,{\rm sr}^{-1}]$,~the monochromatic intensity on the stellar surface;
$S\,[{\rm m}^2]$,~the surface area of the element;
$\mu = \cos\theta$, where
$\theta$ is~the angle between the normal and the line of sight;
$\eta$,~the visibility in the range from 0 (visible element) to 1 (hidden or eclipsed element). 

We assumed the monochromatic calibration flux of \citet{Bessell2000eaa..bookE1939B}:
\begin{equation}
\Phi_{\lambda,{\rm cal}} = \Phi_{\nu,{\rm cal}} {c\over\lambda_{\rm eff}^2} = 0.03669877\,{\rm W}\,{\rm m}^{-3}
\end{equation}
and the Johnson V passband flux:
\begin{equation}
\Phi_{\mathrm{V,cal}} = \int_\lambda f\Phi_{\lambda,{\rm cal}}{\rm d}\lambda
\doteq \Delta_{\rm eff}\Phi_{\lambda,{\rm cal}}
\doteq 3.119396\cdot 10^{-9}\,{\rm W}\,{\rm m}^{-2}\,,
\end{equation}
where
$f(\lambda)$~denotes the filter transmission;
$\lambda_\mathrm{eff} = 545$\,nm, the effective wavelength; and
$\Delta_\mathrm{eff} = 85$\,nm, the effective range.

The apparent magnitude $V_0$ (without absorption) of the primary Aa1 and Aa2 is then:
\begin{equation}\label{V0}
V_0 = 0\,{\rm [mag]} - 2.5 \log\frac{\Phi_{\mathrm{V}}}{\Phi_{\mathrm{V,cal}}}.
\end{equation}
For the tertiary, we used the value of the third light:
\begin{equation}
\Phi_{\rm V,2} = (\Phi_{\rm V,1} + \Phi_{\rm V,2})\, l_{\rm 3V}\,.
\end{equation}

Comparing $V_0$ of the Aa1 + Aa2 + Ab system with Table~\ref{moduli},
we get synthetic values $2.65 + 5.91 + 4.02 = 2.34\,{\rm mag}$ and 
observed values $2.42 + 5.4 + 3.70 = 2.08\,{\rm mag}$.
Thus, the total synthetic magnitude $V_0$ is about 
$0.26\,{\rm mag}$ fainter than observed.
This result is acceptable, especially because
the phoebe2 model was constrained only by the relative BRITE photometry
(see also Sects.~\ref{sed}, \ref{3body}). 
In other words, the result can be considered to be an independent confirmation of the distance.

\begin{table}
\caption[]{
Time intervals of the BRITE LCs (red and blue filters)
when $\delta$~Ori was observed.
The whole time span was divided into 9~seasons.
Time is given in ${\rm HJD}-2400000$.
Season 2019 denoted by $^*$ was omitted since 
red-filter data were not available.
In total, BRITE nanosatellites measured 499\,656 raw data points.
The numbers of points are shown for each season and filter.}
\label{seasons}
\small
\begin{center}
  \begin{tabular*}{\hsize}{l@{\extracolsep{\fill}}crcr}
\hline\hline\noalign{\smallskip}
Season &  Blue filter & $N_{\rm B}$& Red filter & $N_{\rm R}$\\
\noalign{\smallskip}\hline\noalign{\smallskip}
2013         & $56628\mbox{--}56734$  & 22822     & $56603\mbox{--}56733$    & 33357     \\ 
2014         & $56926\mbox{--}57098$  & 38126     & $56924\mbox{--}57095$    & 61672     \\ 
2015         & $57370\mbox{--}57434$  & 6472      & $57374\mbox{--}57443$    & 9665      \\ 
2016         & $57645\mbox{--}57810$  & 22800     & $57729\mbox{--}57734$    & 22996     \\ 
2017         & $58011\mbox{--}58178$  & 17055     & $57645\mbox{--}58178$    & 12812     \\ 
2018         & $58375\mbox{--}58561$  & 61777     & $58430\mbox{--}58556$    & 28001     \\ 
2019$^*$     & $58745\mbox{--}58917$  & 10291     & $-$                      & 39888     \\ 
2020         & $59112\mbox{--}59256$  & 4727      & $59129\mbox{--}59286$    & 44131     \\ 
2021         & $59469\mbox{--}59638$  & 5065      & $59504\mbox{--}59646$    & 58109     \\ 

\noalign{\smallskip}\hline\noalign{\smallskip}
\end{tabular*}
\end{center}
\end{table}

\begingroup
\renewcommand{\arraystretch}{1.3}
\begin{table}
  \caption{
  Results of fitting three phoebe2 models for $\delta$~Ori Aa1+Aa2.
  LC from season~2016 and all RVs were used to constrain the models.
  We fixed or released effective temperature~$T_1$ of the primary and the third light~$l_3$.
  We prefer the model with fixed $T_1$ and $l_3$ ({\bf bold}).
  The uncertainty $\sigma$ is estimated the same for all models.
  The following numbers of data points were used: 
  $N_{\mathrm{total}} = 321$, 
  $N_{\mathrm{LCB}} = N_{\mathrm{LCR}} = 100$, 
  $N_{\mathrm{RV1}} = 71$, and 
  $N_{\mathrm{RV2}} = 50$.
  }
  
  \label{phoebe2_result}
  \small
  \centering
  \renewcommand{\arraystretch}{1.3}
  \begin{tabular*}{\hsize}{l@{\extracolsep{\fill}}rrrc}
      \hline\hline
      {Parameter} & \bf{Fixed }\bm{$T_\mathrm{1}$}& {Fixed $T_\mathrm{1}$, free $l_3$} & {$\sigma$} & {Unit} \\
      \hline
      $T_0$                     & 0.8303	         & 0.8277	      &  0.0030        & HJD$^*$           \\
      $T_\mathrm{1}$            & 31000 	 	     & 31000		  &  300           &  K                \\
      $T_\mathrm{2}$            & 22709		     & 22778		  &  300           &  K                \\
      $R_\mathrm{1}$            & 13.27	 	     & 13.01		  &  0.55          & \Rnom             \\
      $R_\mathrm{2}$            & 3.70			 & 3.66		      &  0.35	       & \Rnom             \\
      $i_1$        			  & 77.671		     & 78.45          &  1.1           & \st               \\
      $S_\mathrm{B}$			  & 1.0079		     & 1.0236 	      &  0.0008        & 1                 \\
      $S_\mathrm{R}$			  & 1.0084	 	     & 1.0250	      &  0.0009        & 1                 \\
      $m_\mathrm{1}$            & 18.07			 & 17.39	 	  &  1.5	       & \Mnom             \\
      $m_\mathrm{2}$            & 8.47			 & 8.26	  	      &  0.4	       & \Mnom             \\
      $e_1$					  & 0.0825		     & 0.0826 	      &  0.0031        &                   \\
      $\omega_1$       		  & 129.98	  	     & 137.50	      &  2.5           & \st               \\
      $\dot{\omega_1}$  	      & 1.45$^{\rm f}$   & 1.45$^{\rm f}$ & 1.45$^{\rm f}$ & \st \, yr$^{-1}$  \\
      $\gamma$         		  & 19.43        	 & 21.96 	      &  0.5           & \ks               \\
      $l_{3\mathrm{B}}$         & 0.26685     	 & 0.308	 	  &  0.05	       &                   \\  
      $l_{3\mathrm{R}}$         & 0.26685 		 & 0.316	 	  &  0.05          &                   \\
      \hline
       $\chi^2_\mathrm{sum}$	 & 604        	    & 490		     &                &                   \\  
	   $\chi^2_\mathrm{LCB}$ 	 & 145         	    & 136		     &                &                   \\
	   $\chi^2_\mathrm{LCR}$ 	 & 113         	    & 113		     &                &                   \\
	   $\chi^2_\mathrm{RV1}$	 & 171        	    & 216	         &                &                   \\
	   $\chi^2_\mathrm{RV2}$     & 70			    & 72		     &                &                   \\
  \noalign{\smallskip}\hline\noalign{\smallskip}
  \end{tabular*}
  \tablefoot{
  $T_0$ denotes the time of periastron passage ($^*$~means HJD$-2457733$);
  $T_\mathrm{1}$ and $T_\mathrm{2}$, the effective temperatures of the primary and secondary, respectively;
  $R_\mathrm{1}$ and $R_\mathrm{2}$, the radii of the primary and secondary, respectively;
  $i$,~inclination;
  $S_\mathrm{B}$ and $S_\mathrm{R}$, the coefficients adjusting normalisation of the flux defined in Eq. (\ref{flux_norm});
  $m_\mathrm{1}$ and $m_\mathrm{2}$, the masses of the primary and secondary, respectively;
  $e_1$,~eccentricity;
  $\omega_1$, the argument of periastron;
  $\gamma$, the systemic velocity;
  $l_{3\mathrm{R}}$ and $l_{3\mathrm{B}}$, the third light in the blue and red filters, respectively;
  $\chi^2_\mathrm{sum}$, the total value of $\chi^2$;
  $\chi^2_\mathrm{LCB}$,
  $\chi^2_\mathrm{LCR}$,
  $\chi^2_\mathrm{RV1}$, and
  $\chi^2_\mathrm{RV2}$, the contributions to $\chi^2$
  from the LCs in the red and blue filters and radial velocities of the primary and secondary;
  $N_{\mathrm{total}}$, the total number of data points;
  $N_{\mathrm{LCB}}$,
  $N_{\mathrm{LCR}}$,
  $N_{\mathrm{RV1}}$, and
  $N_{\mathrm{RV2}}$, the corresponding numbers of data points.
  Uncertainties were estimated from Figs.~\ref{walkers_fixT} and \ref{seasons_comparison}.
  }
\end{table}
\endgroup

\begin{table}
\caption{
Derived parameters corresponding to the $\delta$~Ori Aa1+Aa2 model with $\chi^2 = 604$.
}
\label{derived}
\centering
\small
\renewcommand{\arraystretch}{1.1}
\begin{tabular*}{\hsize}{l@{\extracolsep{\fill}}ll}
\hline\hline\noalign{\smallskip}
Derived parameter & Value & Unit\\
\hline\noalign{\smallskip}
$\log g_1$                         & $3.46 \pm 0.01$       & cgs \\
$\log g_2$                         & $4.27 \pm 0.01$       & cgs  \\
$\Phi_{V,1}$                       & $2.714\cdot 10^{-10}$ & ${\rm W}\,{\rm m}^{-2}$  \\
$\Phi_{V,2}$                       & $1.342\cdot 10^{-11}$ & ${\rm W}\,{\rm m}^{-2}$ \\
$\Phi_{V,1}+\Phi_{V,2}+\Phi_{V,3}$ & $3.618\cdot 10^{-10}$ & ${\rm W}\,{\rm m}^{-2}$ \\
$V_{0,1}$                          & 2.65                  & mag \\
$V_{0,2}$                          & 5.91                  & mag\\
$V_{0,1}+V_{0,2}+V_{0,3}$          & 2.34                  & mag\\
\noalign{\smallskip}\hline\noalign{\smallskip}
\end{tabular*}
\tablefoot{
$\log g$ denotes the logarithm of surface gravity,
$\Phi_V$ the passband flux at Earth,
$V_0$ the corresponding apparent magnitude (without absorption, at a distance of 382\,pc).
}
\end{table}


\subsection{Model for all observing seasons}\label{m_all}

\begin{figure}
\centering
\includegraphics[width=8cm]{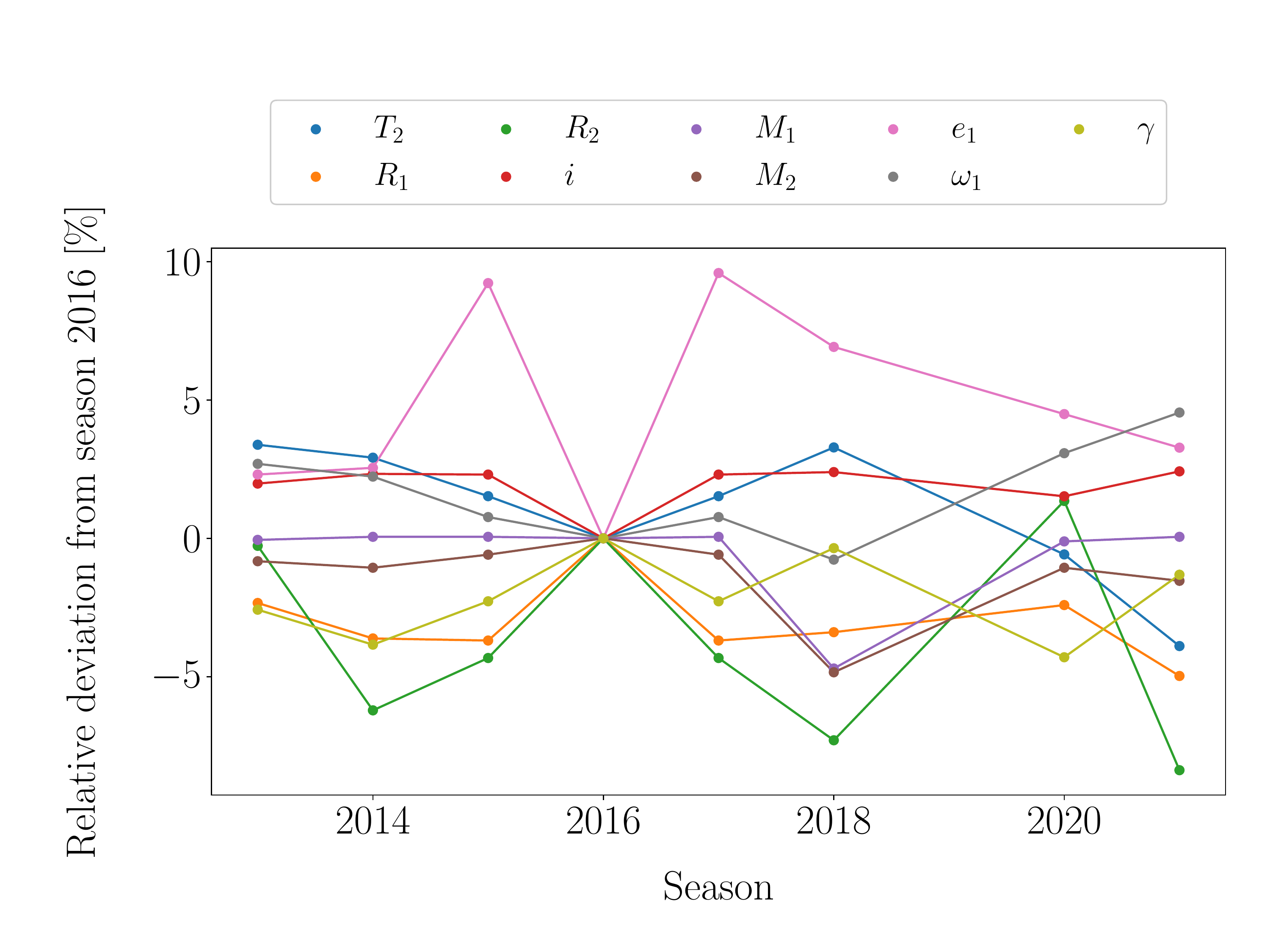}
\caption{
Relative differences of parameters derived for $\delta$~Ori for eight seasons (Table~\ref{phoebe2_seasons}).
In all models, the effective temperature $T_1$ of the primary (Aa1) and 
the third light $l_3$ were fixed.
Season 2016 was taken as a reference. 
Explanations of the parameters can be found in Table~\ref{phoebe2_result}.
}
\label{seasons_comparison}
\end{figure}

Then, we made fits for all seasons.
We kept the effective temperature of the primary $T_1$ and the third light $l_3$ fixed
(as for the preferred model).
The detailed results are presented in Table~\ref{phoebe2_seasons}.
We took season~2016 as a reference.
In Fig.~\ref{seasons_comparison}, we show the deviations from the solution for season~2016.

In season 2018, the red filter measurements had greater uncertainties,
however, it did not lead to any artefacts.
We also omitted season 2019 since only measurements in the blue filter are available.
We cannot confirm that variations of the parameters are intrinsic.
Since the masses of the components must be constant, 
the mean values over all seasons should be preferred.
The variations are most likely due to the oscillations.

\section{Spectral energy distribution (SED)}\label{sed}

The absolute flux is an additional observational constraint.
In the case of $\delta$~Ori, the absorption is low
because the star is not located behind the Orion molecular clouds.
According to the reddening maps of
\cite{Lallement_2019A&A...625A.135L}%
\footnote{\url{https://stilism.obspm.fr/}},
$E(B-V) = 0.042\,{\rm mag}$,
with a substantial scatter of individual samples
(due to the Orion clouds).
Therefore, the total absorption
$A_V \doteq 3.1\,E(B-V) = 0.130\,{\rm mag}$,
if extinction is not anomalous.
There are not enough line-of-sight samples in the maps of
\cite{Green_2019ApJ...887...93G}%
\footnote{\url{http://argonaut.skymaps.info/}}.

From the photometric catalogues in 
\href{http://vizier.u-strasbg.fr/vizier/sed/doc/}{VizieR tool}
\citep{2014ASPC..485..219A},
we took the standard Johnson system photometry \citep{Ducati_2002yCat.2237....0D}
and measurements from
Hipparcos \citep{Anderson_2012AstL...38..331A},
Gaia DR3 \citep{Gaia_2020yCat.1350....0G},
2MASS \citep{Cutri_2003yCat.2246....0C},
WISE \citep{Cutri_2012yCat.2311....0C},
MSX \citep{Egan_2003yCat.5114....0E},
Akari \citep{Ishihara_2010A&A...514A...1I}, and
IRAS \citep{Neugebauer_1984ApJ...278L...1N}.
The data covered the spectral range from $0.35$ to $100\,\mu{\rm m}$.

We have removed clear outliers, multiple entries, and points without uncertainties.
The IRAS 60 and $100\,\mu{\rm m}$ measurements show an excess,
probably due to the far-infrared emission behind $\delta$~Ori;
thus, they have been removed too.
In the end, our photometric data set contained 31 data points (see Fig.~\ref{chi2_SED_FLUX}).

For the Hvar photometry \citep{Bozic1998HvaOB..22....1B},
we performed a removal of eclipse phases (around 0.0, 0.45, 1.0),
and computed average values at the maximum light.
In this case, the absolute photometry is more reliable.
The magnitudes transformed to the Johnson system are as follows:
$U = 0.940\,{\rm mag}$,
$B = 1.977\,{\rm mag}$,
$V = 2.221\,{\rm mag}$,
$R = 2.334\,{\rm mag}$,
with uncertainties less than $0.010\,{\rm mag}$.
The comparison star used was HD\,36591 (HR\,1861):
$V = 5.341\,{\rm mag}$,
$B-V = -0.190\,{\rm mag}$,
$U-B = -0.927\,{\rm mag}$,
$V-R = -0.050\,{\rm mag}$.
In order to compare with the absolute flux, we used the calibrations from
\cite{Bessell2000eaa..bookE1939B} (see also Fig.~\ref{chi2_SED_FLUX}).


\section{Three-body model with all observables}\label{3body}

In order to account for additional observables in {\tt Xitau},
we replaced the two-body model (Aa1+Aa2)+Ab
with a three-body model Aa1+Aa2+Ab.
Thus, the equations of motion were:
\begin{equation}
\vec f_i = \sum_{j\ne i} {Gm_j\over r_{ij}^3}\vec r_{ij} + \vec f_{\rm oblat} + \vec f_{\rm ppn}\quad\hbox{for }\forall\,i\,,
\end{equation}
where the first term is Newtonian gravitational interactions;
the second, oblateness;
and the third, relativistic effects.
This model includes all relevant $N$-body perturbations
(e.g.
the radial velocities with respect to the common centre of mass,
the light-time effect,
precession of $\Omega_1$, $\varpi_1$, $\Omega_2$, $\varpi_2$,
variation,
evection;
see also Appendix~\ref{secular}),
even though some of them are of minor importance for $\delta$~Ori.
We included the oblateness of the bodies,
parametrised with the Love number $k_{\rm F10} \simeq 0.015$ \citep{Fabrycky_2010exop.book..217F}%
\footnote{
In the \cite{Fabrycky_2010exop.book..217F} model, only the radial force component is included.
In the multipole model \citep{Broz_2021A&A...653A..56B}, containing all components,
the value of $J_2 = -C_{20} \simeq 1.5\cdot 10^{-4}$,
or equivalently $k_2 = J_2(\Omega_0/n_0)^2 \simeq 2.5\cdot 10^{-3}$.
},
which results in the observed value of precession
$\dot\omega \simeq 1.45^\circ\,{\rm y}^{-1}$.
Finally, we also included the parametrised post-Newtonian (PPN) approximation
of relativistic effects \citep{Standish_2006,Broz_2022A&A...666A..24B}
since the stars are both massive and close.
The motion was integrated numerically using a Bulirsch--Stoer integrator,
with a precision parameter $\varepsilon = 10^{-8}$, and
output every $0.2\,{\rm d}$
(plus exact times of observations).

Our model was constrained by
astrometry (as in Sect.~\ref{xitau_orbit}),
RVs of all components (Aa1, Aa2, Ab),
eclipse timings,
eclipse duration,
LCs,
synthetic spectra, and
the SED.
Individual contributions to the $\chi^2$ metrics
were multiplied by weights:
\begin{eqnarray}
\chi^2 &=& w_{\rm \scaleto{\rm SKY}{4pt}}\chi^2_{\rm \scaleto{\rm SKY}{4pt}} + w_{\rm \scaleto{\rm RV}{4pt}}\chi^2_{\rm \scaleto{\rm RV}{4pt}} + w_{\rm \scaleto{\rm ETV}{4pt}}\chi^2_{\rm \scaleto{\rm ETV}{4pt}} + w_{\rm \scaleto{\rm ECL}{4pt}}\kern.5pt\chi^2_{\rm \scaleto{\rm ECL}{4pt}} + w_{\rm \scaleto{\rm LC}{4pt}}\kern.5pt\chi^2_{\rm \scaleto{\rm LC}{4pt}}+ \nonumber\\
&&+\,w_{\rm \scaleto{\rm SYN}{4pt}}\kern.5pt\chi^2_{\rm \scaleto{\rm SYN}{4pt}} + w_{\rm \scaleto{\rm SED}{4pt}}\kern.5pt\chi^2_{\rm \scaleto{\rm SED}{4pt}}\,,
\end{eqnarray}
where subscripts denote the data sets mentioned above, respectively.
We used
$w_{\rm \scaleto{\rm SKY}{4pt}} = w_{\rm \scaleto{\rm ECL}{4pt}} = 10$,
due to the limited number of points, and
$w_{\rm \scaleto{\rm SYN}{4pt}} = w_{\rm \scaleto{\rm SED}{4pt}} = 0.1$,
due to the remaining systematics
of rectification of spectra and absolute flux measurements.

Our model had 27 free parameters (see Table~\ref{deltaori_test27_LCECC_free}).
The osculating elements are referenced to the epoch 2458773.188651 (TDB),
corresponding to the most precise speckle interferometry measurement.
They are defined in the Jacobi coordinates,
suitable for a system with hierarchical geometry.
In this particular case,
the distance $d_{\rm pc}$ (Sect.~\ref{parallax}) was fixed.

We used the MOST LC \citep{pablo2015} to derive
three times of the primary eclipse timings:
2456283.521,
2456289.277, and
2456294.994 (TDB, barycentric).
Additional timings were obtained from the TESS \citep{Ricker_2014SPIE.9143E..20R}:
2458473.344,
2458479.080,
2458484.830 (TDB, barycentric).
Due to large-amplitude oscillations, 
the uncertainty is degraded to $0.005\,{\rm d}$.
The primary eclipse duration is
$0.667\,{\rm d}$,
with an uncertainty of $0.010\,{\rm d}$,
again due to the oscillations.
We used a simplified eclipse algorithm for spherical stars.
  
At the same time, we computed the synthetic LC
with the modified version of the Wilson-Devinney program \citep{Wilson_1971ApJ...166..605W,Wilson_1979ApJ...234.1054W,vanHamme_2007ApJ...661.1129V,Wilson_2010ApJ...723.1469W,Broz_2017ApJS..230...19B};
similarly as in Sect.~\ref{phoebe2}.
In this case, however, the instantaneous true phase and distance
were determined by the N-body integration.
The third light is no longer an independent parameter;
instead, it is determined by the third component
($m_3$, $T_3$, $\log g_3$).
This allowed us to constrain our model by eclipse depths.
Other improvements included:
a correction of computations for highly eccentric binaries,
precise computations of the Roche potential from the volume-equivalent radius \citep{Leahy_2015ComAC...2....4L}, 
and more photometric filters \citep{Prsa_2005ApJ...628..426P},
including MOST.
As the oscillations were not accounted for in the synthetic LC,
uncertainties of 0.01\,mag were assigned to all data points
(see Fig.~\ref{chi2_LC}).
The observed spectra cover the blue region (430 to 450\,nm).
The synthetic spectra were interpolated by \texttt{Pyterpol} \citep{Nemravova_2016A&A...594A..55N}
from the BSTAR and OSTAR grids \citep{Lanz_2003ApJS..146..417L,Lanz_2007ApJS..169...83L}.

\begin{figure}
\centering
\includegraphics[width=9cm]{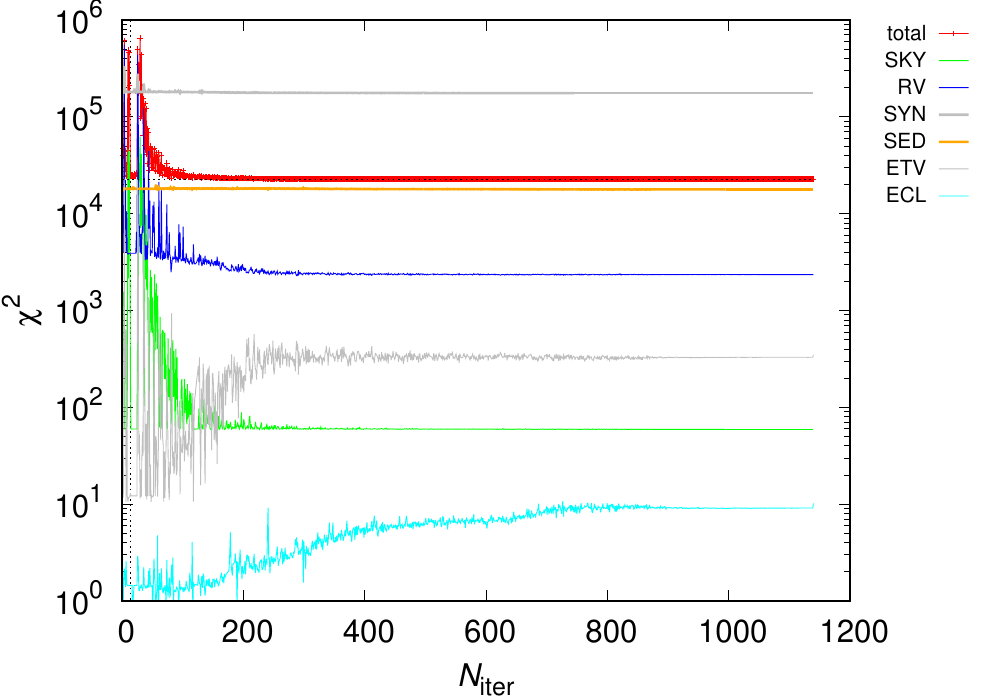}
\caption{
Example of a convergence of the three-body model of $\delta$~Ori.
The individual contributions to $\chi^2$ correspond to
astrometry (SKY),
radial velocities (RV),
eclipse timings (ETV),
eclipse duration (ECL),
synthetic spectra (SYN), and
the spectral energy distribution (SED).
The total $\chi^2$ is summed with non-unit weights (see values in the main text).
}
\label{chi2_iter}
\end{figure}

We used the Planck (black body) approximation for the whole range of wavelengths,
or absolute synthetic spectra for the limited range
of the respective grids.
The fit was performed with the simplex algorithm
(see Fig.~\ref{chi2_iter}).
We consider the best-fit model to be a compromise
because it exhibits a tension between
i) the synthetic spectra (in particular, $\log g_2$ or $R_2$)
and the duration of eclipses,
ii) the minima timings, RVs, and oblateness (see also Fig.~\ref{logg1_logg3_ALL}).
The best-fit parameters are summarised in Table~\ref{deltaori_test27_LCECC_free},
and the derived parameters in Table~\ref{deltaori_test27_LCECC_derived}.
Uncertainties were estimated by the $\chi^2$ mapping and by the MCMC method.
Actually, for the Aa1, Aa2 components, they seem to be comparable
to the phoebe2 model (Sect.~\ref{phoebe2}),
but here we use a different and more extensive set of observations,
in order to constrain all components at the same time.

\begin{figure}
\centering
\includegraphics[width=9cm]{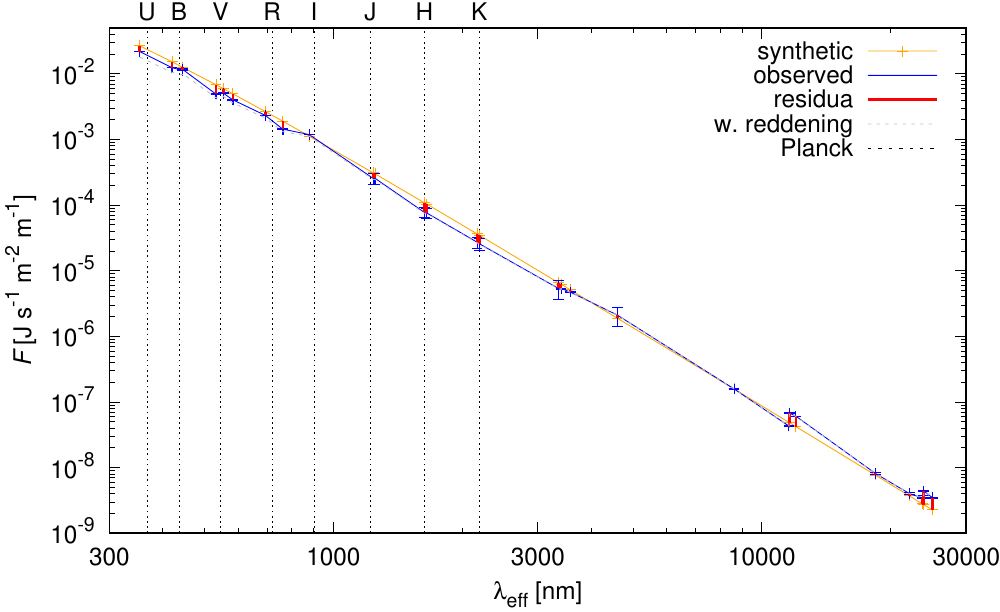}
\includegraphics[width=9cm]{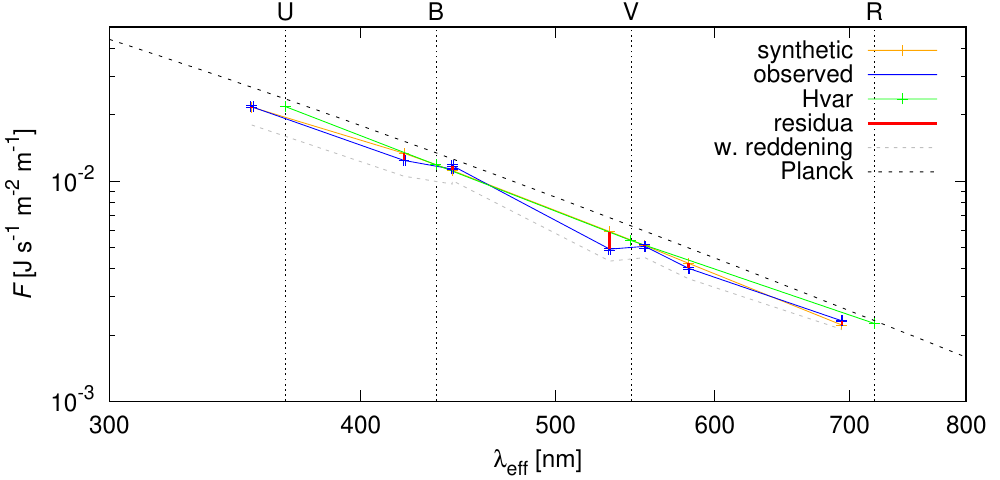}
\caption{
Model from SED data.
Top: Comparison of the observed (\color{blue}blue\color{black}) and
synthetic (\color{gnu_orange}orange\color{black}) SED of $\delta$~Ori.
The residuals are plotted in \color{red}red\color{black}.
The wavelength range is from $350\,{\rm nm}$ (ultraviolet) to $25\,\mu{\rm m}$ (far-infrared).
Bottom: The same for the limited wavelength range of synthetic spectra.
The Hvar differential UBVR photometry with removed eclipses is plotted in \color{green}green\color{black}.}
\label{chi2_SED_FLUX}
\end{figure}

The observed and synthetic SEDs are compared in Fig.~\ref{chi2_SED_FLUX}.
Even though the corresponding contribution $\chi^2_{\rm \scaleto{\rm LC}{4pt}}$
is larger than the number of data points $N_{\rm \scaleto{\rm SED}{4pt}}$,
we consider the fit to be acceptable
as there are several multiple (but independent) measurements
of the same band
that are not consistent with each other.
At the same time, there is no systematic offset of the SED.
In other words, our model provides independent confirmation
of the parallax distance.

\begin{figure*}
\centering
\includegraphics[width=18cm]{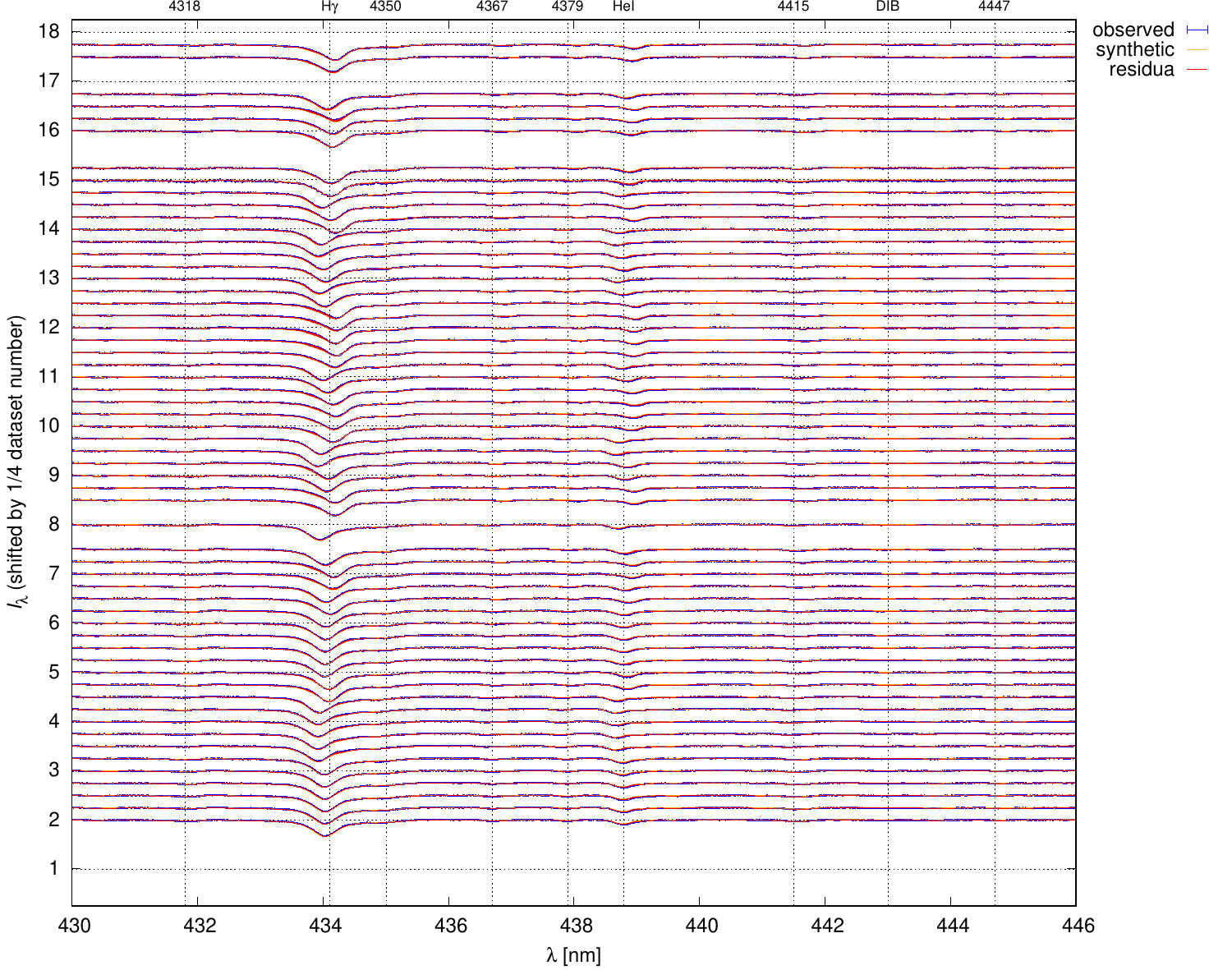}
\caption{
Comparison of the observed (\color{blue}blue\color{black}) and
synthetic (\color{gnu_orange}orange\color{black}) rectified spectra.
The residuals are plotted in \color{red}red\color{black}.
All components Aa1, Aa2, Ab of $\delta$~Ori contribute to the total flux.
The wavelength range included the spectral lines:
H$\gamma$ 4341,
\ion{He}{I} 4378,
and numerous weaker lines.
}
\label{chi2_SYN}
\end{figure*}

All blue spectra are shown in Fig.~\ref{chi2_SYN}.
There were remaining systematics between the observations and the model 
related to the rectification procedure,
especially close to the \ion{He}{I}\,4387 line.
While the H$\gamma$ was fitted without problems,
the synthetic \ion{He}{I} 4471 line was much shallower than the observed one.
These problematic regions were removed from the fitting.
These spectra constrain not only the RVs
but also the relative luminosities~$L$, $\log g$, or radii $R$ of all components.

\begin{figure}
\centering
\includegraphics[width=9cm]{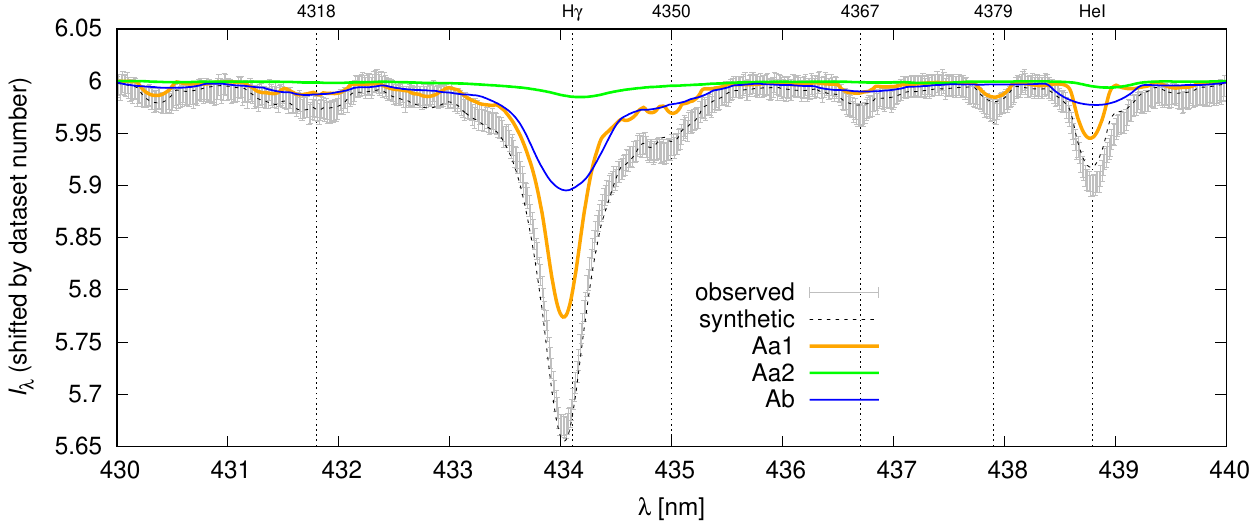}
\caption{
Same as Fig.~\ref{chi2_SYN},
with the distinguished contributions of individual components Aa1, Aa2, Ab of $\delta$~Ori.
The wavelength range is limited to 430 to 440\,nm.
Only the sixth spectrum, 2455836.582700 is shown.
}
\label{synthetic2}
\end{figure}

Contributions of individual components are demonstrated in Fig.~\ref{synthetic2}.
Indeed, the secondary (Aa2) is faint (relative $L_2 = 0.038$).
Unfortunately, its contribution is comparable to the systematics mentioned above.
Consequently, some of the parameters are not very stable
(in particular, $\log g_2$, $v_{\rm rot2}$).
Nevertheless, our fitting in the direct space is independent and complementary
to the fitting in Fourier space (Sect.~\ref{korel}).
Moreover, the secondary is constrained by other observables
(e.g.
eclipse duration,
eclipse depth,
RVs of the primary,
the 3rd-body orbit,
or the total mass $m_1+m_2+m_3$).

\paragraph{Mirror solution.}
Eventually, we explored the mirror solution (Sec.~\ref{xitau_orbit}).
We fixed the total mass $m_1+m_2+m_3 = 52.0\,\Mnom$
and performed a similar testing as in Fig.~\ref{logg1_logg3_ALL}.
The overall best-fit has $\chi^2 = 25468$,
which is worse than the nominal model.
It exhibits a~strong tension between the synthetic spectra and the SED.
Especially the H$\gamma$ line profiles are fitted poorly 
($\chi^2_{\rm syn} = 69441$ vs. $44795$).
This is directly related to the $\log g_3$ value,
which is very low (around 3.2)
according to our model, 
as well as modelling of the disentangled spectrum of the tertiary component
(Sects.~\ref{pyterpol}).
Consequently, we exclude the mirror solution and prefer the nominal model
presented above.

\begin{table}
\caption{
Free parameters of the three-body model of $\delta$~Ori.
The best-fit model has non-reduced $\chi^2 = 23739$.
}
\label{deltaori_test27_LCECC_free}
\centering
\small
\renewcommand{\arraystretch}{1.2}
\begin{tabular*}{\hsize}{@{\hspace{0.2cm}}@{\extracolsep{\fill}}llrrr@{\hspace{0.2cm}}}
\hline\hline\noalign{\smallskip}
Component & Parameter & Value & Unit  & $\sigma$ \\
\noalign{\smallskip}\hline\noalign{\smallskip}
Aa+Ab   & $m_{\rm tot}$  & 35.108              & \Mnom        & 2.0 \\
Aa1+Aa2 & $q_1$          &  0.4785             & 1            & 0.03 \\
Aa+Ab   & $q_2$          &  0.3338             & 1            & 0.03 \\
Aa1+Aa2 & $P_1$          & 5.733121            & d            & 0.000001 \\
        & $\log e_1$     & $-1.089$            & 1            & 0.01 \\
        & $i_1$          &  79.124             & deg          & 1.0 \\
        & $\Omega_1$     & 224.294             & deg          & 1.0 \\
        & $\varpi_1$     &  43.451             & deg          & 1.0 \\
        & $\lambda_1$    &  72.695             & deg          & 1.0 \\
Aa+Ab   & $P_2$          & 55453               & d            & 1000.0 \\
        & $\log e_2$     & $-0.248$            & 1            & 0.01 \\
        & $i_2$          & 105.170             & deg          & 1.0 \\
        & $\Omega_2$     & 122.735             & deg          & 1.0 \\
        & $\varpi_2$     & 261.255             & deg          & 1.0 \\
        & $\lambda_2$    & 133.391             & deg          & 1.0 \\
Aa1     & $T_1$          & 31385               & K            & 1000 \\
Aa2     & $T_2$          & 24515               & K            & 1000 \\
Ab      & $T_3$          & 27906               & K            & 1000 \\
Aa1     & $\log g_1$     & 3.452               & cgs          & 0.1 \\
Aa2     & $\log g_2$     & 4.128               & cgs          & 0.1 \\
Ab      & $\log g_3$     & 3.220               & cgs          & 0.1 \\
Aa1     & $v_{\rm rot1}$ & 114                 & ${\rm km}\,{\rm s}^{-1}$ & 10 \\
Aa2     & $v_{\rm rot2}$ & 184                 & ${\rm km}\,{\rm s}^{-1}$ & 100 \\
Ab      & $v_{\rm rot3}$ & 279                 & ${\rm km}\,{\rm s}^{-1}$ & 10 \\
Aa1     & $C_{20}$       & $-9.2\times10^{-4}$ & 1            & $0.1\times10^{-4}$ \\
        & $z_0$          & 2.221               & mag          & 0.01 \\
        & $\gamma$       & 18.455              & ${\rm km}\,{\rm s}^{-1}$ & 1 \\
        & $d_{\rm pc}$   & 382$^{\rm f}$       & pc           & $-$ \\
\noalign{\smallskip}\hline\noalign{\smallskip}
        & $N_{\rm \scaleto{\rm SKY}{4pt}}$      &     74 & \\
        & $N_{\rm \scaleto{\rm RV}{4pt}}$       &    105 & \\
        & $N_{\rm \scaleto{\rm ETV}{4pt}}$      &      6 & \\
        & $N_{\rm \scaleto{\rm ECL}{4pt}}$      &      1 & \\
        & $N_{\rm \scaleto{\rm LC}{4pt}}$       &  22136 & \\
        & $N_{\rm \scaleto{\rm SYN}{4pt}}$      &  76794 & \\
        & $N_{\rm \scaleto{\rm SED}{4pt}}$      &     31 & \\
        & $N$                                        & 147147 & \\
        & $\chi^2_{\rm \scaleto{\rm SKY}{4pt}}$ &     59 & \\
        & $\chi^2_{\rm \scaleto{\rm RV}{4pt}}$  &   2991 & \\
        & $\chi^2_{\rm \scaleto{\rm ETV}{4pt}}$ &     82 & \\
        & $\chi^2_{\rm \scaleto{\rm ECL}{4pt}}$ &     10 & \\
        & $\chi^2_{\rm \scaleto{\rm LC}{4pt}}$  &  13704 & \\
        & $\chi^2_{\rm \scaleto{\rm SYN}{4pt}}$ &  44795 & \\
        & $\chi^2_{\rm \scaleto{\rm SED}{4pt}}$ &  17872 & \\
        & $\chi^2$                              &  23739 & \\
\noalign{\smallskip}\hline\noalign{\smallskip}
\end{tabular*}
\tablefoot{
$m_{\rm tot}$~denotes the total mass;
$q_1 = m_2/m_1$, $q_2 = m_3/(m_1+m_2)$~the respective mass ratios;
$P$,~osclulating period;
$e$,~eccentricity;
$i$,~inclination;
$\Omega$,~longitude of node;
$\varpi$,~longitude of pericentre;
$\lambda$,~true longitude;
$T$,~effective temperature;
$g$,~gravitational acceleration;
$v_{\rm rot}$,~rotational velocity;
$C_{20}$, quadrupole moment;
$z_{\rm 0}$,~magnitude zero point;
$\gamma$,~systemic velocity; and
$d_{\rm pc}$,~distance.
$^{\rm f}$~indicates the respective parameter was fixed.
Orbital elements are osculating for the epoch $T_0 = 2458773.188651$ (TDB).
}
\end{table}

\begin{table}
\caption{
Derived parameters of the three-body model of $\delta$~Ori.
}
\label{deltaori_test27_LCECC_derived}
\centering
\small
\renewcommand{\arraystretch}{1.2}
\begin{tabular*}{\hsize}{l@{\extracolsep{\fill}}lrrr}
\hline\hline\noalign{\smallskip}
Component & Parameter & Value & Unit & $\sigma$ \\
\noalign{\smallskip}\hline\noalign{\smallskip}
Aa1         & $m_1$          & 17.803     & \Mnom & 1.0 \\
Aa2         & $m_2$          & 8.518      & \Mnom & 1.0 \\
Ab          & $m_3$          & 8.787      & \Mnom & 1.0 \\
Aa1+Aa2     & $a_1$          & 40.099     & \Rnom & 1.0  \\
Aa+Ab       & $a_2$          & 20038      & \Rnom & 100  \\
Aa1+Aa2     & $e_1$          & 0.081      & 1     & 0.02 \\
Aa+Ab       & $e_2$          & 0.565      & 1     & 0.01 \\
Aa1         & $R_1$          & 13.119     & \Rnom & 1.0  \\
Aa2         & $R_2$          &  4.168     & \Rnom & 0.5  \\
Ab          & $R_3$          & 12.045     & \Rnom & 1.0  \\
Aa1         & $L_1$          &  0.563     & 1     & 0.1  \\
Aa2         & $L_2$          &  0.038     & 1     & 0.01 \\
Ab          & $L_3$          &  0.397     & 1     & 0.1  \\
\noalign{\smallskip}\hline\noalign{\smallskip}
\end{tabular*}
\tablefoot{
$m$~denotes the mass;
$a$,~the semi-major axis;
$e$,~the eccentricity;
$R$,~the stellar radius;
$L$,~the relative luminosity (in~V).
}
\end{table}

\section{Pulsations}\label{pulsations}

\begin{figure}
\centering
\includegraphics[width=9cm]{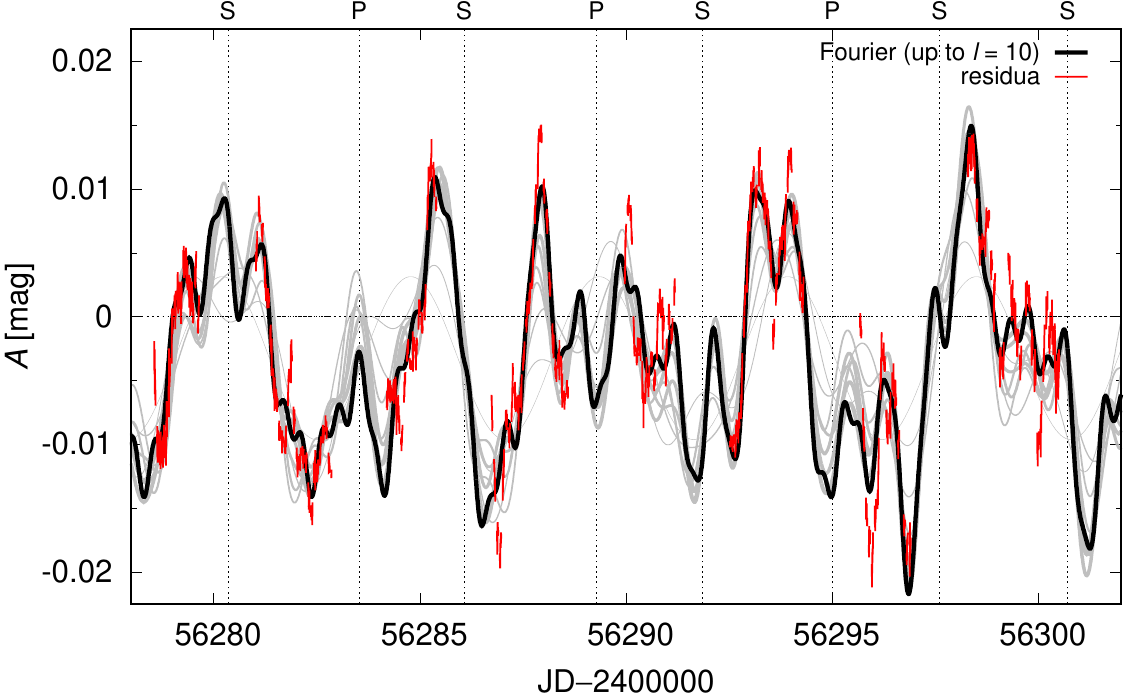}
\caption{
Residuals of the MOST LC (\color{red}red\color{black}),
and the Fourier series up to the order of $\ell = 10$
(black; Table~\ref{tab15}).
Individual orders as they are summed up are shown in \color{pygrey}grey\color{black}.
The times of primary (P) and secondary (S) eclipses are indicated on top.
}
\label{noeclipse}
\end{figure}

After obtaining a well-constrained model,
we analysed the residual MOST LC (Fig.~\ref{noeclipse}),
in order to address the large-amplitude oscillations (BRITE data were not used because of instrumental issues).
Our analysis is similar to that in \cite{pablo2015},
albeit it is different in several aspects:
i)~our model (from Sect.~\ref{3body}) is constrained by all observables,
ii)~not only did we subtract the synthetic LC
but we also removed the eclipse intervals (both primary and secondary),
to suppress the binary signal.
Consequently, the remaining frequencies should be preferentially related
to rotation or pulsations,
even though `gaps' also create spurious signals.

\begin{figure}
\centering
\includegraphics[width=9cm]{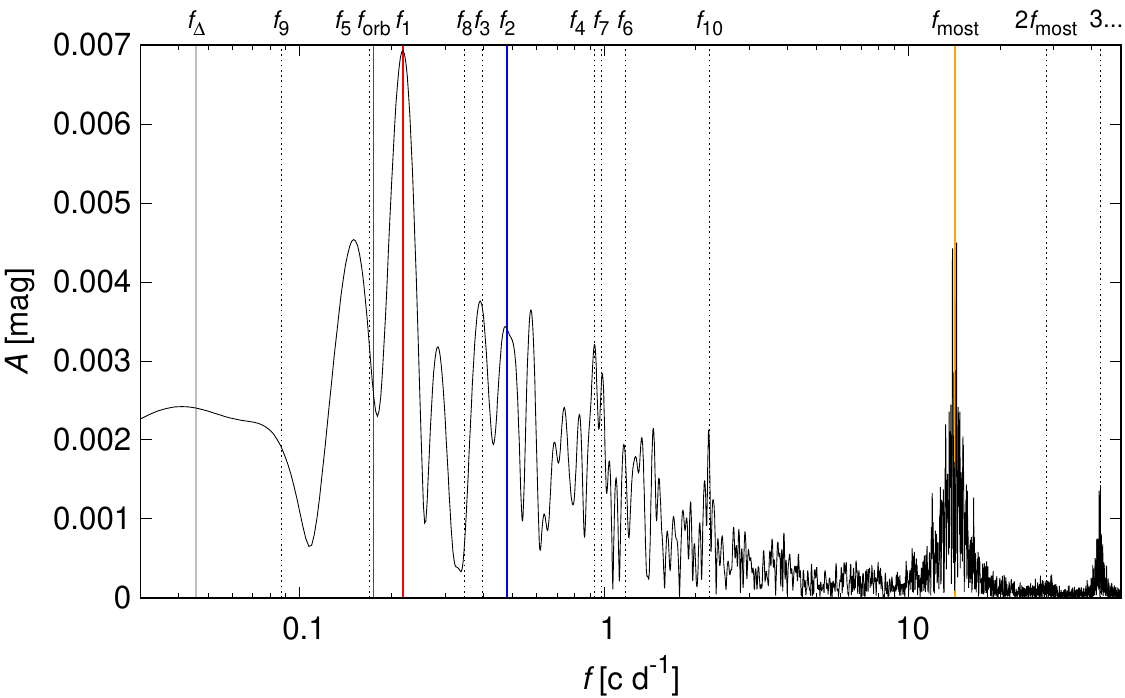}
\caption{
Periodogram of the residual MOST LC
with the 10 principal frequencies indicated on top.
The first periodogram is shown;
it is subsequently modified by subtraction (prewhitening).
The broad peak between 10~to $20\,{\rm c}\,{\rm d}^{-1}$
corresponds to the satellite frequency~$f_{\rm most}$
and its combinations with $f_1$, etc.
}
\label{fourier}
\end{figure}

We used the \texttt{Period04} program \citep{Lenz_2004IAUS..224..786L}
to compute the Fourier spectrum (Fig.~\ref{fourier}),
subtract the dominant term (prewhitening),
recompute the spectrum again,
and repeat these steps ten times.
Our result is shown in Table~\ref{tab15}.

For reference, the minimum frequency (also spacing)
is given by the time span of observations,
$f_\Delta = 1/\Delta = 0.045\,{\rm c}\,{\rm d}^{-1}$,
which means that frequencies differing by $f_\Delta$
are certainly possible to distinguish.
The maximum frequency (also Nyquist) is given by the sampling,
$f_{\rm Ny} = 1/(2\delta) \doteq 1000\,{\rm c}\,{\rm d}^{-1}$.
The orbital frequency of the binary Aa1+Aa2 is
$f_{\rm orb} = 1/P_1 = 0.174\,{\rm c}\,{\rm d}^{-1}$.

In the MOST data, the first two frequencies are
$f_{\rm 1} = 0.218\,{\rm c}\,{\rm d}^{-1}$,
$f_{\rm 2} = 0.478\,{\rm c}\,{\rm d}^{-1}$.
These correspond to the Chandra data \citep{nichols2015},
$f_{\rm X1} = 0.210\,{\rm c}\,{\rm d}^{-1}$,
$f_{\rm X2} = 0.490\,{\rm c}\,{\rm d}^{-1}$,
within uncertainties.
The first one was identified as the rotation frequency of the primary (Aa1)
by \cite{nichols2015}.
Here, we interpret the second one as the rotation of the tertiary (Ab).
This is confirmed by the rotational broadening.
For parameters from Tables~\ref{deltaori_test27_LCECC_free}, \ref{deltaori_test27_LCECC_derived},
$f_{\rm rot3} = v_{\rm rot3}/(2\pi R_3\sin i_2) = 0.474\,{\rm c}\,{\rm d}^{-1}$,
where we assumed an alignment.

Interestingly,
$f_{\rm rot1} = v_{\rm rot1}/(2\pi R_1\sin i_1) = 0.174\,{\rm c}\,{\rm d}^{-1}$
derived from rotational broadening
is not equal to $f_1$,
which differs by $f_\Delta$.
It corresponds to the orbital frequency,
which would indicate a synchronous binary.

Alternatively, when we assume an asynchronous primary,
the synchronicity is
$F_1 = f_1/f_{\rm orb} = 1.250$.
For comparison, the periastron-synchronised value is
$F_1 = [(1+e_1)/(1-e_1)^3]^{1/2} = 1.181$.
In both cases, it means a minor modification of the LC.

The remaining frequencies ($f_3$, $f_4$, $f_{10}$)
are likely associated with pulsations,
namely low-order modes $\ell = 0,1,2$, or $3$,
typical for $\beta$~Cep or SPB stars
\citep{Paxton_2015ApJS..220...15P}.
They can be present either on the primary (Aa1),
or the tertiary (Ab),
which contributes up to 40\% of the light.

\begin{table}
\centering
\small
\renewcommand{\arraystretch}{1.1}
\caption{
Fourier analysis of the residual MOST LC (from Fig.~\ref{noeclipse}).
The synthetic LC was subtracted and
eclipses were removed 
to suppress the binary signal.
The first two frequencies are then identified
as the rotation frequency of the primary (Aa1; \citealt{nichols2015})
and the tertiary (Ab; this work).
}
\label{tab15}
\begin{tabular*}{\hsize}{l@{\extracolsep{\fill}}rrrl@{\hspace{0.2cm}}}
\hline\hline\noalign{\smallskip}
& Frequency & Period & Amplitude & Notes \\
& [${\rm c}\,{\rm d}^{-1}$] & [${\rm d}$] & [mmag] & \\
\noalign{\smallskip}\hline\noalign{\smallskip}
$f_1$    & 0.218 &  4.572 & 6.37 & rotation of Aa1                        \\
$f_2$    & 0.478 &  2.090 & 3.39 & rotation of Ab                         \\
$f_3$    & 0.396 &  2.522 & 4.04 & pulsation?                             \\
$f_4$    & 0.929 &  1.075 & 3.05 & pulsation?                             \\
$f_5$    & 0.168 &  5.931 & 2.95 & orbital (Aa1+Aa2)                      \\
$f_6$    & 1.168 &  0.855 & 1.95 & orbital$\,+\,1\,{\rm c}\,{\rm d}^{-1}$ \\
$f_7$    & 0.975 &  1.025 & 1.96 & approx. $1\,{\rm c}\,{\rm d}^{-1}$     \\
$f_8$    & 0.346 &  2.887 & 2.54 & $2f_{\rm orb}$?                        \\
$f_9$    & 0.086 & 11.551 & 1.67 & $0.5f_{\rm orb}$?                      \\
$f_{10}$ & 2.214 &  0.451 & 1.29 & pulsation?                             \\
\noalign{\smallskip}\hline\noalign{\smallskip}
\end{tabular*}
\tablefoot{
The uncertainty is determined by the time span,
$f_\Delta = 0.045\,{\rm c}\,{\rm d}^{-1}$.
For reference, the orbital frequency of the Aa1+Aa2 binary
$f_{\rm orb} = 0.174\,{\rm c}\,{\rm d}^{-1}$, and
the orbital frequency of the satellite
$f_{\rm most} = 14.19\,{\rm c}\,{\rm d}^{-1}$.
}
\end{table}

\section{Conclusions}

In this paper, 
we studied the triple star Aa1+Aa2+Ab 
in the multiple system $\delta$~Ori. 
The close eclipsing binary Aa1+Aa2 contains an O star,
is non-interacting, and has a negligible mass transfer.
Consequently, it represents a target suitable for defining 
the intrinsic parameters of evolved O stars.
Our main results are as follows:
\begin{enumerate}
    \item The distance of the system was estimated from the new Gaia DR3 parallax of the faint component $\delta$ Ori (Ca+Cb).
    \item The outer orbit (Aa1+Aa2)+Ab was constrained by new speckle interferometric measurements from the WDS (the period of approximately 152 years and eccentricity $0.58$) and by $\gamma$ velocities.
    \item The secondary (Aa2) spectrum in the blue spectral region was detected by the two- and three-step disentangling.
    \item The RV curve of the secondary was obtained by cross-correlation with a disentangled template spectrum.
    \item The two-body model of the eclipsing binary was constructed in \texttt{PHOEBE2}.
    \item The three-body model in \texttt{Xitau} was constrained by all observables.
\end{enumerate}

Compared to previous studies, 
we obtained significantly lower masses than \citet{pablo2015}, 
in their low-mass model ($23.81+8.54$)\,\Mnom.
In contrast to the study of \cite{shenar2015}, 
where models were calculated for two distances, 212 and 380\,pc, 
we adopted the latter.
Our results give lower radii of Aa1, Aa2;
\citet{shenar2015} have [($16.5\pm 1)+(6.5\pm2$)]\,\Rnom.
Nevertheless, the radius of Ab is in agreement with
\citet{shenar2015} who reported $(10.4 \pm 2)\,$\Rnom.
We also obtained low $\log g_3$ similarly as \citet{shenar2015}.

\begin{figure}
\centering
\includegraphics[width=8cm]{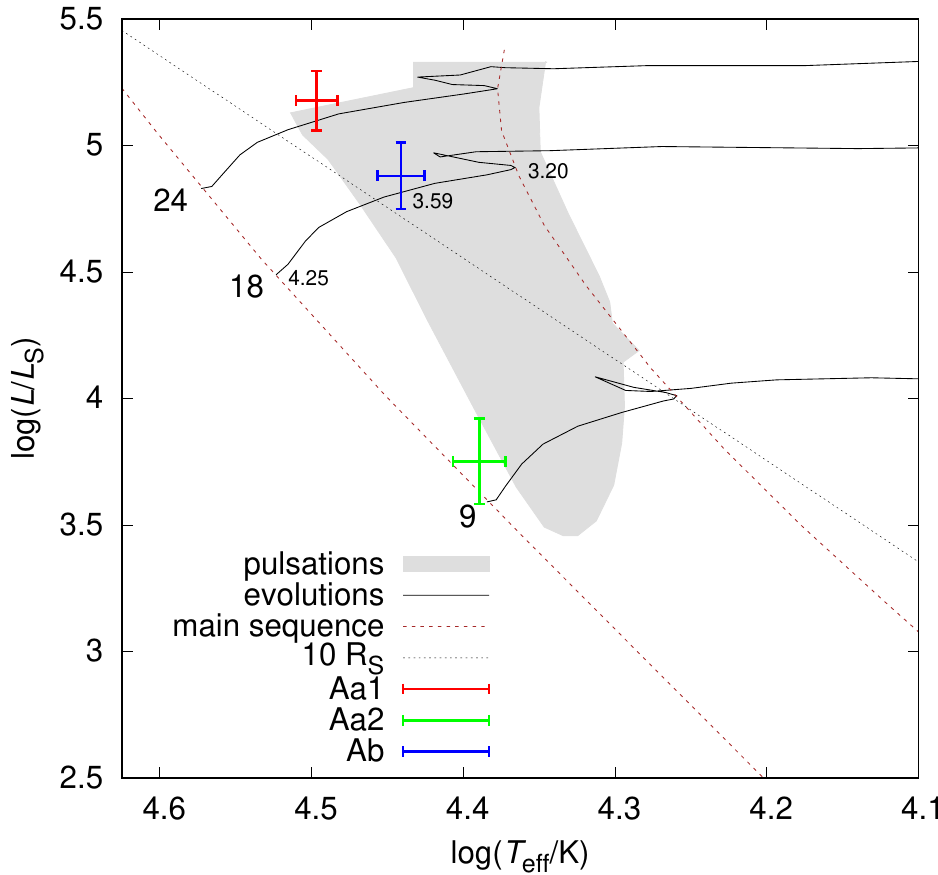}
\caption{
Hertzsprung--Russel diagram
with the positions of the Aa1, Aa2, and Ab components of $\delta$~Ori
and evolutionary tracks from \cite{Paxton_2015ApJS..220...15P}.
Numbers next to the main sequence indicate the theoretical masses,
small numbers the theoretical gravitational acceleration~$\log g$.
According to the 3-body model,
the masses are $17.1$, $8.5$, $8.7\,\Mnom$.
For the tertiary (Ab), it is in disagreement with the evolutionary track,
but in agreement with the value of~$\log g_3$
inferred from normalised spectra (H$\gamma$).
The instability region of pulsations ($\beta$~Cep type, order $\ell=0$) is indicated as grey area.
Other modes ($\ell = 1, 2, 3$) can be found in a very similar region \citep{Paxton_2015ApJS..220...15P}.
Both Aa1, Ab components are located here, 
and they can exhibit photometric variability 
attributed to pulsations.}

\label{HRD}
\end{figure}

\paragraph{Hertzprung--Russel diagram.}
Given the spectral types of Aa1 + Aa2 + Ab,
O9.5\,II + B2\,V + B0\,IV
(\citealt{pablo2015} and this work),
the primary has evolved from the main sequence;    
however, it has not reached the overflow yet.
The Hertzsprung--Russel diagram (Fig.~\ref{HRD})
with the positions of Aa1, Aa2, Ab components
indicates an interesting problem
--- the Ab component is very offset from a normal position.
This offset is either related to its $\log g_3$ value (3.2)
or to its $m_3$ value ($8.7\,\Mnom$).
However, it is not easy to modify these values,
because they are well constrained by observations.
To put Ab on the evolutionary track,
either $\log g_3 \simeq 3.7$,
or $m_3 \simeq 18\,\Mnom$.
If all components were normal, the sum of masses should be about
$(24+18+10)\,\Mnom = 52\,\Mnom$.
Interestingly, this is similar to the mirror solution,
which was excluded (see the discussion in Sects.~\ref{xitau_orbit} and \ref{3body}).
Consequently, we are left with an unusual stellar component.
Actually, it is not unusual ---
see for example
$\delta$~Ori~Ca, or
$\sigma$~Ori~E
which are both helium-rich, with H$\alpha$ emission (Tab.~\ref{moduli}).
Detailed stellar-evolution models with possible mass transfer
between (some of) the components
shall be computed in future work.
Additionally, long-baseline optical or near-infrared interferometry 
may be able to measure precisely the angular diameters of the component stars 
(e.g. \citep{Shabun_2008IAUS..248..118S}) 
and the separation of the inner orbit in the sky, 
giving direct constraints on the size of the orbit, 
helping to resolve any discrepancies in the masses measured 
between this study and other similar studies of $\delta$ Ori.

\vskip\baselineskip
A comparison of $\delta$~Ori with other bright stars in the Orion belt
(see Table~\ref{moduli})
shows that $\sigma$~Ori has a similar architecture
(((Aa+Ab)+B)+C)+D+E,
and even a very similar angular scale \citep{Simon_2015ApJ...799..169S}.
All of its components seem to be less evolved.%
\footnote{$\iota$~Ori is located farther away, in the Trapezium region.}
On the other hand, $\zeta$~Ori exhibits
an angular scale about 10 times larger and has 
the primary evolved in an O supergiant \citep{Hummel_2000ApJ...540L..91H}.
In this sense, $\varepsilon$~Ori,
which seems to be a single variable B supergiant \citep{Puebla_2016MNRAS.456.2907P},
may represent an even more evolved object.

Given the fact that all these stars ($\delta$, $\varepsilon$, $\zeta$, $\sigma$)
are the most massive within the Orion OB1b association,
they might have encountered and perturbed (destabilised) each other.
Again, a possible convergence of their proper motions
will be analysed in future work.

\begin{acknowledgements}
This research was supported by the grants P209/10/0715, GA15-02112S, and
GA19-01995S of the Czech Science Foundation.
M.B. was supported by the Czech Science Foundation grant GA21-11058S.
We acknowledge the support from the Research Programmes MSM0021620860
``Physical study of objects and processes in the solar system and in astrophysics''
of the Ministry of Education of the Czech Republic and
the grant AV0Z10030501 of the Academy of Sciences of the Czech Republic.
Thanks to Dr. Rachel Matson for providing the data from Washington Double Star Catalogue.
H.B. acknowledges financial support from the Croatian Science Foundation 
under the project 6212 ``Solar and Stellar Variability''.
We gratefully acknowledge the use of the program {\tt KOREL} written by Petr Hadrava,
{\tt PYTERPOL} by Jana Nemravov\'a, and {\tt reSPEFO2} by Adam Harmanec.
This research has made use of the \texttt{AMBER data reduction package} of the
Jean-Marie Mariotti Center\footnote{\url{http://www.jmmc.fr/amberdrs}}.
Our thanks are due to \v{S}.~Dvo\v{r}\'akov\'a, D.~Kor\v{c}\'akov\'a,
J.~Kub\'at, J.~Nemravov\'a, M.~Oksala, P.~\v{S}koda, and V.~Votruba
who obtained some of the Ond\v{r}ejov blue spectra used in this study.
We also acknowledge the use of the electronic database from CDS Strasbourg and
the electronic bibliography maintained by the NASA/ADS system.
CCL and GAW acknowledge Discovery Grant support by the Natural Science 
and Engineering Research Council (NSERC) of Canada.
Last but not least, we would like to thank our anonymous 
referee for the constructive useful comments that 
helped to improve the paper.

\end{acknowledgements}


\bibliographystyle{aa}
\bibliography{delori}


\newpage
\begin{appendix}

\section{Secular rates for $\delta$~Ori~A}\label{secular}

We used the standard Gauss equations to estimate the secular rates
for the $\delta$~Ori~A system.
A perturbation due to the quadrupole moment $J_2 \equiv -C_{20}$
of the primary induces a precession of
the argument of pericentre:
\begin{equation}
\dot\omega_1 = +3n_1J_2\left({R_1\over a_1}\right)^2{5\cos^2 \tilde i_1-1\over 4\eta_1^4}\,,
\end{equation}
and the longitude of the node:
\begin{equation}
\dot\Omega_1 = -{3\over 2}n_1J_2\left({R_1\over a_1}\right)^2{\cos\tilde i_1\over\eta_1^4}\,,
\end{equation}
where
$n_1 = \sqrt{G(m_1{+}m_2)/a_1^3}$,
$\eta_1 \equiv \sqrt{1-e_1^2}$, and
$\tilde i_1$ is the inclination with respect to the primary equator.
Otherwise, the notation is the same as in Table~\ref{deltaori_test27_LCECC_free}.
For $\tilde i_1\to 0$, $e_1\to 0$, $J_2 > 0$, we would have
$\dot\omega_1 > 0$,
$\dot\Omega_1 < 0$,
$\dot\omega_1 = -2\dot\Omega_1$.

The quadrupole moment is determined by the internal structure:
\begin{equation}
J_2 = -{1\over m_1R_1^2}\int_V \rho |\vec r|^2P_2(\cos\theta)\,{\rm d}V\,,
\end{equation}
where
the Legendre polynomial $P_2(x) = {1\over 2}(3x^2-1)$.
For a homogeneous body, it would be related to the ellipticity
\citep{Fitzpatrick_2012icm..book.....F}%
\footnote{$r(\vartheta) = R(1-2/3\varepsilon_0 P_2(\cos\vartheta))$}:
\begin{equation}
J_2 = -{2\over 5}\varepsilon_0\,.
\end{equation}
For a rotating body, it is related to the Love number%
\footnote{approximately, $0.02$ for the Sun;
$0.3$ for the Earth, Jupiter, or an M-dwarf;
up to $0.75$ for incompressible fluid}:
\begin{equation}
J_2 \simeq k_2\left({\Omega_0\over n_0}\right)^2,
\end{equation}
where $\Omega_0$ is the angular rotation frequency;
$n_0$, the mean motion at the surface.
Assuming $J_2 = 1.8\cdot 10^{-4}$ results in
$\dot\omega = 1.47\,{\rm deg}\,{\rm y}^{-1}$,
$\dot\Omega = -0.73\,{\rm deg}\,{\rm y}^{-1}$,
where the node circulates with respect to the primary equator,
but it only librates with respect to the observer plane.

For the binary Aa1+Aa2 (acting on a mass-less particle),
the equations should be modified as follows:
\begin{equation}
\dot\omega_2 = +3n_2 J_2\left({a_1\over a_2}\right)^2 {5\cos^2\tilde i_2-1\over 4\eta_2^4}\,,
\end{equation}
\begin{equation}
\dot\Omega_2 = -{3\over 2}n_2 J_2\left({a_1\over a_2}\right)^2 {\cos\tilde i_2\over\eta_2^4}\,,
\end{equation}
where the inclination~$\tilde i_2$ is with respect to the binary.
The effective quadrupole moment is:
\begin{equation}
J_2 = {1\over 2} {m_1m_2\over(m_1+m_2)^2}\,,
\end{equation}
because the respective radius ($|\vec r| = a_1$) is the same
as the reference radius ($R = a_1$).
Given that $J_2 = 0.109$,
and the ratio of $a_1/a_2 = 0.002$,
the precession rates should be of the order of $10^{-6}\,{\rm deg}\,{\rm y}^{-1}$.

However, for the massive triple system (Aa1+Aa2)+Ab,
we used the theory of \cite{Breiter_2015MNRAS.449.1691B};
for the longitudes of pericentre and node:
\begin{eqnarray}
\dot\varpi_2 &\!\!\!\!=\!\!\!\!& {3\over 8}{n_2\over\eta_2^3}{m_3\over m_1\!+\!m_2\!+\!m_3}{n_2\over n_1}\gamma\cdot \nonumber\\
&\!\!\!\!\cdot\!\!\!\!&\left[3\cos^2\!J - 1 - {\gamma\sin J\sin 2J\over 1 + \gamma\cos J +\!\sqrt{1+\gamma^2+2\gamma\cos J}}\right],
\end{eqnarray}
\begin{equation}
\dot\Omega_2 = {3\over 4}{n_2\over\eta_2^3}{m_3\over m_1\!+\!m_2\!+\!m_3}{n_2\over n_1}\cos J\sqrt{1 + \gamma^2 + 2\gamma\cos J}\,,
\end{equation}
where
$m_1'\! = m_1m_2/(m_1{+}m_2)$,
$m_2'\! = (m_1{+}m_2)\,m_3/(m_1{+}m_2{+}m_3)$ denote the reduced masses;
$L_1 = m_1'n_1a_1^2$,
$L_2 = m_2'n_2a_2^2$, the angular momenta;
$\cos J = \cos i_1\cos i_2 + \sin i_1\sin i_2\cos(\Omega_1-\Omega_2)$, the mutual inclination; and
$\gamma = L_1/(L_2\eta_2)$.
Assuming parameters from Table~\ref{deltaori_test27_LCECC_free},
$\dot\varpi_2 = 3.4\cdot 10^{-6}\,{\rm deg}\,{\rm y}^{-1}$,
$\dot\Omega_2 = 8.5\cdot 10^{-5}\,{\rm deg}\,{\rm y}^{-1}$,
which is negligible on the observational time span.
All these effects were nevertheless included in our numerical N-body model
(Sect.~\ref{3body}).


\section{Supplementary figures and tables}\label{supplement}

\begin{figure}
\centering
\includegraphics[width=9cm]{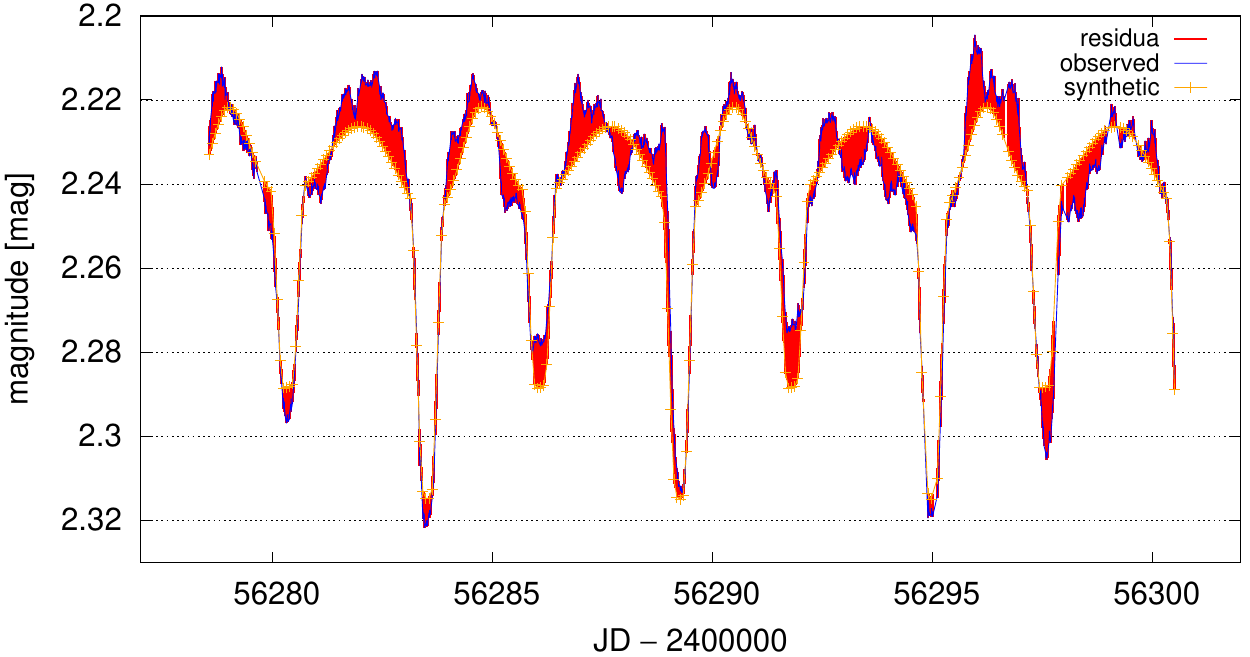}
\caption{
Comparison of the observed MOST (\color{blue}blue\color{black}) and
synthetic (\color{gnu_orange}orange\color{black}) light curve.
The residuals are plotted in \color{red}red\color{black}.
Apart from the eclipses, the light curve contains
large-amplitude oscillations (not included in our model);
uncertainties 0.01\,mag were thus assigned to all data points.
}
\label{chi2_LC}
\end{figure}

\begin{figure*}[p!]
\centering
\includegraphics[height=22cm]{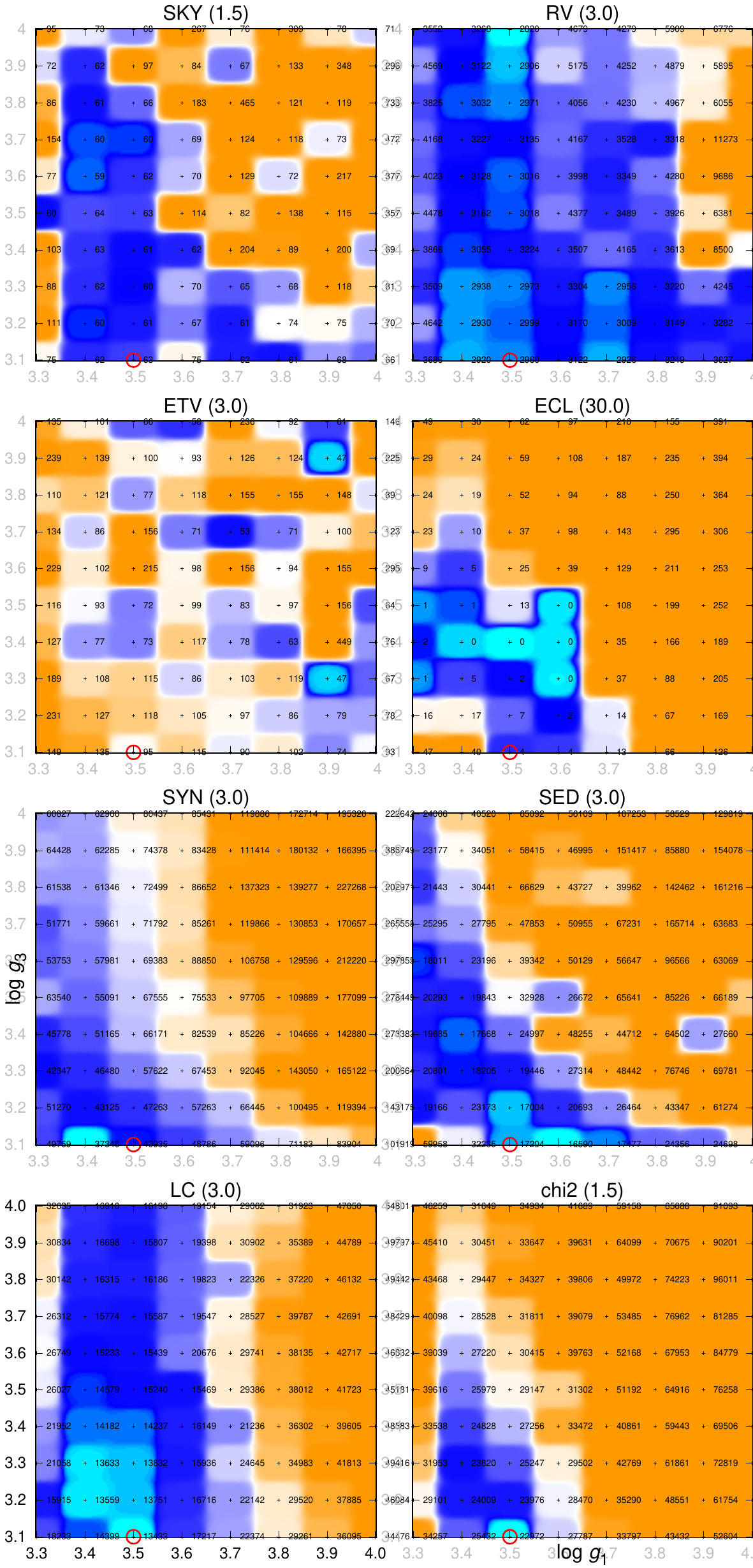}
\caption{
Contributions to $\chi^2$ for a set 100 best-fit models of $\delta$~Ori.
Individual contributions are shown in the panels (from top left):
astrometry (SKY),
RV,
eclipse timings (ETV),
eclipse duration (ECL),
normalised spectra (SYN),
SED,
light curve (LC), and
total.
Every simplex was initialised with a different combination of
the gravitational accelerations $\log g_1$, $\log g_3$,
which were kept fixed to obtain a regular grid.
All other parameters were free.
The number of convergence steps was limited to 2000,
consequently, 200000 models were computed in total.
Axes correspond to the values of $\log g_1$, $\log g_3$,
colours to $\chi^2$ (see also tiny numbers).
The colour scale was adjusted as follows:
\color{cyan}cyan\color{black}\ the very best fit for a given data set,
\color{blue}blue\color{black}\ acceptable fits (${<}3.0$ min $\chi^2$),
\color{orange}orange\color{black}\ poor fits (${\ge}3.0$ min $\chi^2$).
The factor was 1.5 for the SKY, total; and 30.0 for the ECL data set.
`Forbidden regions' can be seen,
in particular, large $\log g_1$, $\log g_3$ due to the SYN, SED,
or large $\log g_1$ due to the ECL, LC.
The weighted very best fit is denoted by the \color{red}red\color{black}\ circle.
}
\label{logg1_logg3_ALL}
\end{figure*}

In Tables~\ref{oes} and~\ref{el_fer},
we present more details on the spectral data sets
discussed in Sect.~\ref{spec}.
In Table~\ref{phoebe2_seasons}, we report parameters derived
for the 8 seasons observed by the BRITE satellites.
The LC from our three-body model (Sect.~\ref{3body})
is shown in Fig.~\ref{chi2_LC}.
Individual contributions to $\chi^2$ computed for an extensive set models of $\delta$~Ori
is shown in Fig.~\ref{logg1_logg3_ALL}.

\begin{table*}[p]
\caption[]{
Details on spectra from the coud\'e focus of the Ond\v{r}ejov 2m reflector in the blue region. 
The RVs were determined during the spectral disentangling (three-step method) in \texttt{KOREL}. 
}
\label{oes}
\centering
\small
\footnotesize
\begin{tabular*}{\textwidth}{@{\hspace{0.2cm}}@{\extracolsep{\fill}}lrrrrrrrrr@{\hspace{0.2cm}}}
\hline\hline\noalign{\smallskip}
$T$             &  Exposure time & S/N     & Heliocentric correction  & RV$_1$ & $\sigma_{\mathrm{RV}_1}$ & RV$_2$ & $\sigma_{\mathrm{RV}_2}$ & RV$_3$ & $\sigma_{\mathrm{RV}_3}$\\
{[$\mathrm{HJD-2400000}$}]   &  [s]           &           & [\ks] & [\ks] & [\ks] & [\ks] & [\ks]& [\ks]& [\ks]\\
\noalign{\smallskip}\hline\noalign{\smallskip}
55836.5692 & 1113 & 235.3 & $ 26.2335$ & $-13.03$ & 1.73 & $ 101.44$ & 1.48 & 17.25 & 1.72 \\
55836.5827 & 1131 & 268.3 & $ 26.2097$ & $-13.29$ & 1.52 & $ 100.76$ & 1.30 & 16.46 & 1.51 \\
55836.5963 & 1129 & 217.3 & $ 26.1849$ & $-13.97$ & 1.87 & $ 104.81$ & 1.60 & 17.91 & 1.86 \\
55836.6137 & 1250 & 238.0 & $ 26.1517$ & $-14.86$ & 1.71 & $ 108.90$ & 1.46 & 17.19 & 1.70 \\
55836.6287 & 1265 & 236.2 & $ 26.1222$ & $-17.11$ & 1.72 & $ 115.37$ & 1.47 & 17.70 & 1.71 \\
55836.6434 & 1174 & 254.2 & $ 26.0929$ & $-19.88$ & 1.60 & $ 113.68$ & 1.37 & 15.16 & 1.59 \\
55837.5877 & 1124 & 214.3 & $ 26.0588$ & $-94.63$ & 1.90 & $ 278.78$ & 1.63 & 16.35 & 1.88 \\
55837.6016 & 1134 & 239.5 & $ 26.0328$ & $-96.07$ & 1.70 & $ 280.05$ & 1.45 & 13.32 & 1.69 \\
55837.6152 & 1150 & 232.7 & $ 26.0064$ & $-95.62$ & 1.75 & $ 279.59$ & 1.50 & 14.24 & 1.74 \\
55837.6294 & 1218 & 237.1 & $ 25.9784$ & $-98.09$ & 1.71 & $ 278.48$ & 1.47 & 13.54 & 1.70 \\
55871.5969 & 1999 & 217.1 & $ 16.9755$ & $-79.45$ & 1.87 & $ 251.90$ & 1.60 & 17.42 & 1.86 \\
55893.4601 & 1662 & 226.5 & $  7.9171$ & $ 36.33$ & 1.79 & $ -31.61$ & 1.54 & 17.48 & 1.78 \\
55893.4820 & 2003 & 225.8 & $  7.8667$ & $ 36.04$ & 1.80 & $ -25.90$ & 1.54 & 11.27 & 1.79 \\
55953.4009 & 1916 & 224.9 & $-19.2387$ & $-11.07$ & 1.81 & $  91.43$ & 1.55 & 15.22 & 1.80 \\
55953.4226 & 1766 & 243.8 & $-19.2836$ & $ -9.87$ & 1.67 & $  87.03$ & 1.43 & 12.58 & 1.66 \\
55953.4508 & 1599 & 250.2 & $-19.3381$ & $ -7.24$ & 1.62 & $  79.34$ & 1.39 & 14.11 & 1.61 \\
55953.4692 & 1503 & 252.0 & $-19.3704$ & $ -5.12$ & 1.61 & $  75.37$ & 1.38 & 14.21 & 1.60 \\
55953.4871 & 1468 & 216.3 & $-19.3992$ & $ -1.59$ & 1.88 & $  71.30$ & 1.61 & 14.07 & 1.87 \\
55956.3498 & 3561 & 230.1 & $-20.1601$ & $ 58.08$ & 1.77 & $ -72.93$ & 1.51 & 11.98 & 1.75 \\
55959.2469 & 2134 & 230.6 & $-20.9128$ & $  4.46$ & 1.76 & $  68.90$ & 1.51 & 14.61 & 1.75 \\
55977.3276 & 2165 & 209.0 & $-25.6205$ & $ 86.76$ & 1.95 & $-121.98$ & 1.67 & 14.70 & 1.93 \\
55990.3318 & 4872 & 206.6 & $-27.3138$ & $ 96.47$ & 1.97 & $-152.94$ & 1.69 & 14.40 & 1.95 \\
55991.2779 & 1466 & 258.1 & $-27.2883$ & $ -5.95$ & 1.58 & $  77.34$ & 1.35 & 14.43 & 1.56 \\
55992.2949 & 1415 & 199.3 & $-27.3830$ & $-93.85$ & 2.04 & $ 278.69$ & 1.75 & 18.81 & 2.03 \\
56003.3328 & 1200 & 186.8 & $-27.6165$ & $-72.08$ & 2.18 & $ 231.46$ & 1.86 & 12.62 & 2.16 \\
56011.3498 & 1361 & 142.0 & $-27.1206$ & $ 54.90$ & 2.86 & $ -52.75$ & 2.45 & 17.89 & 2.84 \\
56167.6213 & 2054 & 154.9 & $ 26.1472$ & $118.69$ & 2.62 & $-195.80$ & 2.25 & 15.19 & 2.61 \\
56241.4709 & 2405 & 199.5 & $ 15.4295$ & $112.55$ & 1.86 & $-178.56$ & 1.59 &  6.27 & 1.85 \\
56257.4224 & 2287 & 219.6 & $  8.5701$ & $ 13.12$ & 1.85 & $  42.02$ & 1.59 & 16.01 & 1.84 \\
56257.6273 & 3324 & 227.7 & $  8.1164$ & $ 35.28$ & 1.79 & $  -5.17$ & 1.53 & 13.21 & 1.77 \\
56330.3789 & 1399 & 245.5 & $-22.8962$ & $-91.20$ & 1.66 & $ 271.45$ & 1.42 & 16.21 & 1.64 \\
56354.3038 & 3112 & 241.5 & $-27.1587$ & $-50.33$ & 1.54 & $ 180.31$ & 1.32 &  5.28 & 1.53 \\
56357.3761 & 1138 & 170.5 & $-27.5049$ & $ 88.62$ & 2.38 & $-124.65$ & 2.04 & 15.73 & 2.37 \\
56596.5631 & 1415 & 212.4 & $ 19.1204$ & $ 98.83$ & 1.91 & $-147.64$ & 1.64 & 17.80 & 1.90 \\
56608.6423 & 2196 & 181.0 & $ 14.3991$ & $121.15$ & 2.25 & $-201.59$ & 1.92 & 13.34 & 2.23 \\
56609.4312 & 3840 & 166.5 & $ 14.4441$ & $ 93.94$ & 2.44 & $-153.01$ & 2.09 & 19.46 & 2.43 \\
56621.6375 & 6298 & 264.2 & $  8.7784$ & $ 22.16$ & 1.54 & $  14.69$ & 1.32 & 14.10 & 1.53 \\
56629.6451 & 5211 & 232.3 & $  5.0123$ & $-30.90$ & 1.75 & $ 135.90$ & 1.50 & 15.84 & 1.74 \\
56642.5644 & 2340 & 236.0 & $ -1.1527$ & $103.79$ & 1.72 & $-163.06$ & 1.48 & 14.29 & 1.71 \\
56643.3817 & 2365 & 218.8 & $ -1.2259$ & $121.57$ & 1.86 & $-196.79$ & 1.59 & 12.93 & 1.85 \\
56643.6078 & 4554 & 246.3 & $ -1.7159$ & $110.68$ & 1.51 & $-198.65$ & 1.29 & 19.58 & 1.50 \\
56666.3166 & 2968 & 197.2 & $-11.9535$ & $120.06$ & 2.06 & $-196.17$ & 1.77 & 16.15 & 2.05 \\
56666.4940 & 5013 & 237.6 & $-12.3482$ & $120.37$ & 1.56 & $-187.88$ & 1.34 &  8.33 & 1.55 \\
56704.3508 & 2793 & 279.8 & $-24.9170$ & $-15.22$ & 1.45 & $  90.44$ & 1.25 & 13.73 & 1.44 \\
56714.2672 & 2292 & 260.6 & $-26.4964$ & $-77.91$ & 1.56 & $ 246.02$ & 1.34 & 16.17 & 1.55 \\
56719.3096 & 3214 & 240.8 & $-27.1475$ & $ -1.73$ & 1.69 & $  79.65$ & 1.45 & 18.73 & 1.68 \\
56721.3496 & 2993 & 256.5 & $-27.3779$ & $-32.32$ & 1.59 & $ 140.09$ & 1.36 & 16.81 & 1.57 \\
56737.3202 &  629 & 226.3 & $-27.4373$ & $-93.55$ & 1.80 & $ 265.23$ & 1.54 & 15.84 & 1.78 \\
56738.3020 & 2404 & 184.0 & $-27.3532$ & $-58.28$ & 2.21 & $ 194.34$ & 1.89 & 15.57 & 2.19 \\
56746.2894 & 1392 & 199.8 & $-26.5682$ & $120.01$ & 2.03 & $-201.76$ & 1.74 & 16.48 & 2.02 \\
56928.5832 & 1652 & 221.9 & $ 26.6856$ & $ 59.66$ & 1.83 & $ -64.14$ & 1.57 & 14.96 & 1.82 \\
57105.2900 & 1986 & 205.6 & $-27.2198$ & $-45.65$ & 1.98 & $ 170.70$ & 1.69 & 12.24 & 1.96 \\
57106.2877 &  565 & 118.5 & $-27.1303$ & $ 60.16$ & 3.43 & $ -62.99$ & 2.94 & 14.03 & 3.41 \\
57106.3182 & 1784 & 166.9 & $-27.1696$ & $ 63.36$ & 2.44 & $ -69.25$ & 2.09 & 16.40 & 2.42 \\
57116.3066 & 1651 & 156.0 & $-25.8607$ & $-88.80$ & 2.61 & $ 254.62$ & 2.23 & 11.35 & 2.59 \\
57128.2762 & 1434 & 158.1 & $-23.3452$ & $-39.86$ & 2.57 & $ 158.43$ & 2.20 & 14.13 & 2.55 \\
57297.5482 & 3202 & 187.7 & $ 26.2886$ & $ 83.80$ & 2.17 & $-113.55$ & 1.86 & 20.79 & 2.15 \\
57364.5793 & 2224 & 219.8 & $  2.9803$ & $ 85.92$ & 1.85 & $-124.80$ & 1.58 & 14.50 & 1.84 \\
57445.3477 & 3194 & 218.4 & $-26.6982$ & $114.68$ & 1.86 & $-185.19$ & 1.60 & 16.60 & 1.85 \\
57464.3540 & 6102 & 211.3 & $-27.6372$ & $ 18.76$ & 1.92 & $  28.03$ & 1.65 & 15.26 & 1.91 \\
58389.5221 & 2398 & 176.3 & $ 26.7622$ & $-29.27$ & 2.31 & $ 132.85$ & 1.98 & 16.06 & 2.29 \\
58390.6224 &  161 &  79.9 & $ 26.4953$ & $ 81.98$ & 5.09 & $-114.83$ & 4.36 & 15.47 & 5.05 \\
58390.6416 & 3001 & 174.5 & $ 26.4581$ & $ 85.89$ & 2.33 & $-118.96$ & 2.00 & 17.81 & 2.31 \\
58402.5225 & 2556 & 171.8 & $ 24.7883$ & $108.15$ & 2.37 & $-177.53$ & 2.03 & 16.77 & 2.35 \\
58405.5661 & 2701 & 193.3 & $ 24.0578$ & $-89.81$ & 1.92 & $ 273.70$ & 1.65 &  6.62 & 1.91 \\
\noalign{\smallskip}\hline\noalign{\smallskip}
\end{tabular*}
\end{table*}

\begin{table*}[p]
\caption[]{Details on ELODIE and FEROS spectra in the blue region. 
ELODIE is at the upper part of the table, 
FEROS at the lower part.
The RVs were determined during the spectral disentangling (three-step method) in \texttt{KOREL}. 
}
\small
\label{el_fer}
\centering
\begin{tabular*}{\textwidth}{@{\hspace{0.2cm}}@{\extracolsep{\fill}}lrrrrrrrrr@{\hspace{0.2cm}}}
\hline\hline\noalign{\smallskip}
$T$             &  Exposure time & S/N    & Heliocentric correction  & RV$_1$ & $\sigma_{\mathrm{RV}_1}$ & RV$_2$ & $\sigma_{\mathrm{RV}_2}$ & RV$_3$ & $\sigma_{\mathrm{RV}_3}$\\
{[$\mathrm{HJD-2400000}$]}   &  [s]           &           & [\ks] & [\ks]& [\ks]& [\ks]& [\ks]& [\ks]& [\ks]\\
\noalign{\smallskip}\hline\noalign{\smallskip}
50033.5779 & 187.13 &  86.9 &$  14.76 $&$ 126.27 $& 4.27 &$ -206.96 $& 3.66 &$ 18.15 $& 4.24 \\
50033.5920 & 333.78 &  99.6 &$  14.72 $&$ 123.38 $& 3.73 &$ -206.12 $& 3.19 &$ 16.55 $& 3.70 \\
50435.4013 & 120.50 & 102.9 &$  -1.87 $&$ 112.98 $& 3.61 &$ -184.64 $& 3.09 &$ 20.89 $& 3.58 \\
\noalign{\smallskip}\hline\noalign{\smallskip}
54136.5830 &  30    & 295.3 &$ -22.15 $&$  22.27 $& 1.26 &$    4.33 $& 1.08 &$ 14.51 $& 1.25 \\
54809.7226 &  20    & 216.8 &$   2.19 $&$  76.14 $& 1.71 &$ -101.90 $& 1.47 &$ 15.43 $& 1.70 \\
54809.7238 &  20    & 213.4 &$   2.19 $&$  76.02 $& 1.74 &$ -101.64 $& 1.49 &$ 15.43 $& 1.73 \\
54809.7255 &  20    & 219.8 &$   2.19 $&$  72.08 $& 1.69 &$ -100.56 $& 1.45 &$ 12.72 $& 1.68 \\
54953.4599 &  30    & 232.0 &$ -18.33 $&$  27.82 $& 1.60 &$    6.80 $& 1.37 &$ 15.67 $& 1.59 \\
54953.4615 &  30    & 216.8 &$ -18.33 $&$  27.35 $& 1.71 &$   10.92 $& 1.47 &$ 10.76 $& 1.70 \\
\noalign{\smallskip}\hline\noalign{\smallskip}
\end{tabular*}
\end{table*}

\begin{table*}[p]
  \caption{
  Results of eight phoebe2 models of $\delta$~Ori.
  LCs from eight individual seasons S. (Table~\ref{seasons}) and all RVs were used to constrain the models.
  We assumed a fixed value of the temperature~$T_1$ and the third light~$l_3$. The uncertainties are the same as in Table~\ref{phoebe2_result}.
  }
  \label{phoebe2_seasons}
  \small
  \centering
  \renewcommand{\arraystretch}{1.1}
  \begin{tabular*}{\textwidth}{l@{\extracolsep{\fill}}rrrrrrrrr}
      \hline\hline\noalign{\smallskip}
      {Parameters}              & {S. 2013} 	  & {S. 2014} 	    & {S. 2015}       & {\textbf{S. 2016}} & {S. 2017}    & {S. 2018}       & {S. 2020}       & {S. 2021}       & {All seasons}   \\
      \noalign{\smallskip}\hline\noalign{\smallskip}
      $T_0$ [HJD]               & 733.8355$^*$    & 733.8355$^*$    & 733.8412$^*$    & 773.3830$^*$    & 733.8412$^*$    & 733.8359$^*$    & 733.8343$^*$    & 733.8447$^*$    & 733.8340$^*$	  \\
      $T_\mathrm{1}$ [K]        & 31000$^{\rm f}$ & 31000$^{\rm f}$ & 31000$^{\rm f}$ & 31000$^{\rm f}$ & 31000$^{\rm f}$ & 31000$^{\rm f}$	& 31000$^{\rm f}$ & 31000$^{\rm f}$ & 31000$^{\rm f}$ \\
      $T_\mathrm{2}$ [K]        & 23477	 		  & 23371		    & 23055           & 22709		    & 23055			  & 23455	        & 22577           & 21825           & 22940           \\
      $R_\mathrm{1}$ [\Rnom]    & 12.96 		  & 12.79		    & 12.78           & 13.27		    & 12.78			  & 12.82		    & 12.95           & 12.61           & 12.87           \\
      $R_\mathrm{2}$ [\Rnom]	& 3.69	 		  & 3.47	 		& 3.54            & 3.70		    & 3.54			  & 3.43		    & 3.75            & 3.39            & 3.56            \\
      $i_1$ [\st]				& 79.21 	 	  & 79.48	 	    & 79.46           & 77.67		    & 79.46		      & 79.53		    & 78.85           & 79.55           & 79.15           \\
      $S_\mathrm{B}$			& 1.00631 		  & 1.00489 		& 1.00556         & 1.00786		    & 1.00872		  & 1.02264	        & 1.00430         & 1.00455         & var             \\
      $S_\mathrm{R}$			& 1.00375 		  & 1.00355		    & 1.00872         & 1.00841		    & 1.00361         & 1.02431         & 1.00342         & 1.00640         & var             \\
      $m_\mathrm{1}$ [\Mnom]	& 18.06	 		  & 18.08	 	    & 18.08           & 18.07  		    & 18.08			  & 17.22		    & 18.05           & 18.08           & 17.97           \\
      $m_\mathrm{2}$ [\Mnom]	& 8.40	 		  & 8.38	 		& 8.42            & 8.47		    & 8.42			  & 8.06		    & 8.38            & 8.34            & 8.35            \\
      $e_1$					    & 0.0843  		  & 0.0845  		& 0.0900          & 0.0824		    & 0.0903		  & 0.0881 	        & 0.0861          & 0.0851          & 0.0864          \\
      $\omega_1$ [\st]			& 133.5 		  & 132.9   	    & 131.0           & 130.0 		    & 131.0		      & 129.0           & 134.0           & 135.91          & 129.86          \\
      $\dot{\omega_1}$ [\st \, yr$^{-1}$] 	& 1.45$^{\rm f}$ 		  & 1.45$^{\rm f}$    	    & 1.45$^{\rm f}$           & 1.45$^{\rm f}$  		    & 1.45$^{\rm f}$ 		      & 1.45$^{\rm f}$          & 1.45$^{\rm f}$           & 1.45$^{\rm f}$          & -           \\
      $\gamma$ [\ks]			& 19.27 	 	  & 19.02 		    & 19.33           & 19.78 		    & 19.33 		  & 19.71           & 18.93           & 19.52           & 19.36           \\
      $l_{3\mathrm{B}}$         & 0.273$^{\rm f}$ & 0.273$^{\rm f}$ & 0.273$^{\rm f}$ & 0.273$^{\rm f}$ & 0.273$^{\rm f}$ & 0.273$^{\rm f}$ & 0.273$^{\rm f}$ & 0.273$^{\rm f}$ & -               \\
      $l_{3\mathrm{R}}$         & 0.273$^{\rm f}$ & 0.273$^{\rm f}$ & 0.273$^{\rm f}$ & 0.273$^{\rm f}$ & 0.273$^{\rm f}$ & 0.273$^{\rm f}$ & 0.273$^{\rm f}$ & 0.273$^{\rm f}$ & -               \\
      \noalign{\smallskip}\hline\noalign{\smallskip}
      $\chi^2_\mathrm{sum}$		& 661    		  & 683	 	        & 1051	          & 604			    & 1003		      & 877	            & 1310            & 1176            &                 \\ 
	  $\chi^2_\mathrm{LCB}$ 	& 195	 		  & 174	 	        & 598	          & 153 		    & 280			  & 392	            & 895             & 61              &                 \\
	  $\chi^2_\mathrm{LCR}$ 	& 201	 		  & 254	 	        & 180	          & 200			    & 463			  & 217	            & 160             & 284             &                 \\
	  $\chi^2_\mathrm{RV1}$	    & 197	 		  & 188	 	        & 206	          & 184			    & 192			  & 200		        & 186             & 207             &                 \\   
	  $\chi^2_\mathrm{RV2}$     & 68		      & 66	            & 67	          & 67			    & 68			  & 68 		        & 69              & 77              &                 \\
\noalign{\smallskip}\hline\noalign{\smallskip}
\end{tabular*}
\tablefoot{
$^*$ $-2457000\,$HJD. The explanation of variables is the same as in Table~\ref{phoebe2_result}.
$^{\rm f}$~indicates the respective parameter was fixed.
var denotes variable values for each season (they were between 1.004 and 1.010).
For each season, the numbers of data points were:
$N_{\mathrm{total}} = 321$, $N_{\mathrm{LCB}} = N_{\mathrm{LCR}} = 100$, $N_{\mathrm{RV1}} = 71$, and $N_{\mathrm{RV2}} = 50$.}
\end{table*}

\end{appendix}
\end{document}